\documentclass[12pt]{article}

\usepackage{amsmath}
\usepackage{amssymb}
\usepackage{amsthm}
\usepackage{natbib}
\usepackage{graphicx}
\usepackage{threeparttable}
\usepackage{bm}
\usepackage{bbm}
\usepackage{comment}
\usepackage[noabbrev]{cleveref}
\usepackage{enumitem}   
\usepackage{xcolor}
\usepackage{setspace}
\usepackage{caption}
\usepackage{subcaption}
\usepackage{soul}
\theoremstyle{plain}
\newtheorem{prop}{Proposition}
\theoremstyle{remark}
\newtheorem{example}{Example}

\graphicspath{{figs/}}

\newcommand{\x}{\bm{x}}

\newcommand{\y}{\bm{y}}

\newcommand{\bs}{\bm{s}}

\newcommand{\bx}{\bm{x}}
\newcommand{\by}{\bm{y}}

\newcommand{\bz}{\bm{z}}

\newcommand{\bV}{\bm{V}}
\newcommand{\bD}{\bm{D}}
\newcommand{\bbeta}{\bm{\beta}}
\newcommand{\bga}{\bm{\gamma}}

\newcommand{\bzeta}{\bm{\zeta}}
\newcommand{\bxi}{\bm{\xi}}

\newcommand{\bphi}{\bm{\phi}}

\newcommand{\D}{\mathcal{D}}
\newcommand{\R}{\mathbb{R}}
\newcommand{\BS}{\mathcal{S}}
\newcommand{\V}{\mathcal{V}}
\newcommand{\sv}{\bm{v}}
\newcommand{\su}{\bm{u}}
\newcommand{\U}{\mathcal{U}}

\newcommand{\tNe}{\text{Ne}}
 
\newcommand{\bmu}{\bm{\mu}}

\newcommand{\blind}{0}

\begin{document}

\def\spacingset#1{\renewcommand{\baselinestretch}
		{#1}\small\normalsize} \spacingset{1}

	%%%%%%%%%%%%%%%%%%%%%%%%%%%%%%%%%%%%%%%%%%%%%%%%%%%%%%%%
	
	\if0\blind
	{
		\title{\bf Bayesian Geostatistical Modeling for Discrete-Valued Processes}
		\author{
		Xiaotian Zheng,
		%\thanks{X. Zheng (xiaotian@ucsc.edu) is doctoral student, Department of Statistics, University of California, Santa Cruz.}
        Athanasios Kottas,
        %\thanks{A. Kottas (thanos@soe.ucsc.edu) is Professor, Department of Statistics, University of California, Santa Cruz.} 
        and
        Bruno Sans\'o
        %\thanks{B. Sans\'o (bruno@soe.ucsc.edu) is Professor, Department of Statistics, University of California, Santa Cruz.}
        \\
        Department of Statistics, University of California Santa Cruz %California, USA
        }
		\maketitle
	} \fi
	
	\if1\blind
	{
		\bigskip
		\bigskip
		\bigskip
		\begin{center}
			{\LARGE\bf Bayesian Geostatistical Modeling for Discrete-Valued Processes}
		\end{center}
		\medskip
	} \fi
	
	\bigskip
\begin{abstract}
We introduce a flexible and scalable class of Bayesian geostatistical models for discrete 
data, based on the class of nearest neighbor mixture transition distribution processes (NNMP),
referred to as discrete NNMP. The proposed class characterizes spatial variability by a weighted
combination of first-order conditional probability mass functions (pmfs) for each one of a 
given number of neighbors. The approach supports flexible modeling for multivariate dependence 
through specification of 
general bivariate discrete distributions that define the conditional pmfs. 
Moreover, the discrete NNMP allows for construction of models given a pre-specified family of 
marginal distributions that can vary in space, facilitating covariate inclusion.
In particular, we develop a modeling and inferential framework for copula-based NNMPs 
that can attain flexible dependence structures, motivating the use of bivariate copula families
for spatial processes. Compared to the traditional class of spatial generalized linear mixed 
models, where spatial dependence is introduced through a transformation of 
response means, our process-based modeling approach provides both computational and inferential 
advantages. We illustrate the benefits with synthetic data examples and an analysis of North 
American Breeding Bird Survey data.
\end{abstract}
	
\noindent
{\it Bayesian hierarchical models; 
Copula functions; 
Count data;
Mixture transition distribution;
Nearest neighbors; 
Spatial classification.} 
	
	\vfill
	
	\newpage

	\spacingset{1.5} % DON'T change the spacing!

\section{Introduction}

Discrete geostatistical data arise in many areas, such as biology, ecology, 
and forestry. Such data sets consist of observations that take discrete values and 
are indexed in a continuous spatial domain.
As an example, consider observations for counts of a species of interest,
commonly used to estimate the species distribution over a geographical domain.

The most common approach to modeling such data is through a spatial generalized 
linear mixed model (SGLMM, \cite{diggle1998model}), under which an exponential 
family distribution is specified for the response at a given location, assuming 
independence between locations, conditional on an underlying spatial process.
Such process is specified in the second stage of the SGLMM through a link function 
that associates the response mean to a set of spatial random effects.
A Gaussian process is typically used for the spatial random effects. 
Thus, SGLMMs provide a general modeling tool for geostatistical 
discrete data applications 
\citep{wikle2002spatial, recta2012two, zhang2020bayesian}.

However, SGLMMs have a number of drawbacks. First, they do not correspond to spatial 
processes for the observed data. Since the spatial random effects are incorporated 
into the transformed mean, SGLMMs model spatial structure on a function of the 
response means, not the observations directly. Thus, the model may impose a strong 
correlation between means over locations that are close, even though the corresponding 
observations may not be strongly correlated. 
In addition, the SGLMM specification poses computational challenges. 
Unlike Gaussian geostatistical models, the spatial random effects cannot be 
marginalized out. Under simulation-based inference, estimating the spatial random 
effects generally requires sampling a large number of highly correlated parameters
within a Markov Chain Monte Carlo (MCMC) algorithm, which is likely to produce
slow convergence, and a large memory footprint. Although efficient computational 
strategies have been explored in the literature 
\citep{christensen2002bayesian, christensen2006robust, sengupta2013hierarchical},
the computational challenge is unavoidable, especially for large spatial datasets.

An alternative to SGLMMs involves Gaussian copula models which construct 
random fields given a pre-specified family of marginal distributions. Here, the
joint cumulative distribution function (cdf) of the spatial responses is 
characterized by a Gaussian copula corresponding to an underlying Gaussian process;
see, e.g., \cite{madsen2009maximum}, \cite{kazianka2010copula},
and \cite{han2016correlation}. Gaussian copulas provide simplicity in
specifying spatial dependence, and flexibility in selecting discrete 
marginal distributions. However, the evaluation of the resulting  likelihood requires
efficient approximations of high-dimensional multivariate Gaussian integrals, 
limiting the applicability of this class of models.

In this paper, we introduce a new class of spatial process models for discrete 
geostatistical data. This class builds from the nearest-neighbor mixture 
transition distribution process (NNMP), proposed by \cite{zheng2021nnmp} for 
modeling large continuous geostatistical data. 
The NNMP structured mixture formulation is motivated by mixture transition 
distribution (MTD) models for non-Gaussian time series \citep{le1996modeling}. 
In particular, \cite{zheng2021construction} discuss construction of stationary 
MTDs for both continuous and discrete time series, using particular bivariate 
distributions from the literature.

The contribution of this paper is threefold. 
First, we develop a discrete analogue of the NNMP, referred to as the discrete NNMP, 
with particular focus on using bivariate copulas to define 
the spatially varying conditional probability mass functions (pmfs) for the 
structured mixture that gives rise to the joint distribution.
We show that the joint pmf of the discrete copula NNMP can be further decomposed into
a collection of bivariate copulas, providing interpretability for model construction
using different families of copulas. In fact, our approach allows for the 
use of arbitrary bivariate copula families, which enhances model flexibility and 
enables the description of complex spatial dependencies. We
demonstrate with a simulation study the impact of using different copula families, 
exploring alternatives to the traditional Gaussian copula for spatial modeling. 
Secondly, we extend the first-order strict stationarity result in \cite{zheng2021nnmp}.
The extension is key for discrete NNMPs, providing
a constructive approach to develop models with spatially varying marginal pmfs. 
This can be used, for example, to incorporate either continuous or discrete covariates, 
which is practically important in the context of regression modeling for discrete-valued
spatial responses. Finally, utilizing the stationarity extension result, 
we develop a Bayesian hierarchical framework that consists of using uniform random variables to 
transform discrete variables into continuous ones.
The proposed approach leverages the properties of copulas for continuous random vectors, 
thus facilitating the use of different copulas as well as efficient computation.
We show through a simulation study that, compared with popular
SGLMM methods, this approach yields
reliable posterior inference at a much lower computational cost.

The paper is organized as follows. In Section 2, we introduce NNMPs for 
discrete data, with copula-based discrete NNMPs developed in Section 3. Section 4 
presents the Bayesian model formulation for inference, validation and prediction, 
followed by illustration with synthetic and real datasets in Section 5. 
Finally, Section 6 concludes with a summary and discussion.

\section{NNMPs for discrete data}
\label{sec:nnmp}

\subsection{Modeling framework}\label{sec:framework}

Consider a univariate spatial process $Y(\sv)$ indexed by $\bm{v} \in\D\subset\R^p$, 
for $p\geq1$. Let $\by_{\BS} = (y(\bs_1),\dots,y(\bs_n))^\top$ be
a realization of the process $Y(\sv)$, where $\BS = (\bs_1,\dots,\bs_n)$
denotes the reference set. Using a directed acyclic graph (DAG) with vertexes 
given by $y(\bs_i)$ for the locations in $\BS$, the joint density 
$p(\by_{\BS})$ can be expressed as:
\begin{equation}\label{eq:dag}
    p(\y_{\BS}) = p(y(\bs_1))\prod_{i=2}^np(y(\bs_i)\mid y(\bs_{i-1}),\dots,y(\bs_1)),
\end{equation}
where the conditional distributions depend on the set of 
parents of each vertex in the DAG. 

Reducing the size of the conditioning set to be at most $L$, 
we obtain a valid joint density for $\y_{\BS}$ that approximates 
\eqref{eq:dag} as
\begin{equation}\label{eq:approx_dag}
    \tilde{p}(\y_{\BS}) = p(y(\bs_1)) \prod_{i=2}^np(y(\bs_i)\mid \by_{\tNe(\bs_i)}),
\end{equation}
where $\tNe(\bs_i)$ is a subset of $\{\bs_1,\dots,\bs_{i-1}\}$, and
$\by_{\tNe(\bs_i)}$ is the vector formed by stacking the process realization 
%of $Y(\bs)$ 
over $\tNe(\bs_i)$. Traditionally, the elements of $\tNe(\bs_i)$ are selected as the 
nearest neighbors of $\bs_i$ within $\{\bs_1,\dots,\bs_{i-1}\}$, for $i = 2,\dots,n$, 
according to a specified distance in $\D$.
Ordering the elements of $\tNe(\bs_i)$ in  ascending order with respect to 
distance to $\bs_i$, we have $\tNe(\bs_i) = (\bs_{(i1)},\dots,\bs_{(i,i_L)})$,
where $i_L:= (i-1)\wedge L$.  
The joint density in \eqref{eq:approx_dag} 
constructed using  nearest neighbors has been explored for fast 
likelihood approximations \citep{vecchia1988estimation, katzfuss2021general}, 
and extended to nearest-neighbor
Gaussian process models for Gaussian data \citep{datta2016hierarchical},
and to NNMPs for continuous, non-Gaussian data \citep{zheng2021nnmp}. 
We note that the factorization in \eqref{eq:dag} implicitly 
requires a topological ordering on the locations as they are not naturally ordered.
Effects of the ordering on the approximation have been studied in the literature. 
Here, we adopt a random ordering, which is shown to give sharper approximation 
than coordinate-based orderings \citep{guinness2018permutation}.

Here, we introduce NNMPs for discrete-valued spatial processes,
referred to as discrete NNMPs. Such models are derived by the following
two steps. The first step consists of building a valid joint density 
over $\BS$ by modeling the conditional densities in the product of the right hand side of \eqref{eq:approx_dag} with a weighted combination of conditional pmfs:
\begin{equation}\label{eq:nnmp1}
    p(y(\bs_i)\,|\, \y_{\tNe(\bs_i)}) =
    \sum_{l=1}^{i_L} w_l(\bs_i) \, f_{\bs_i,l}(y(\bs_i)\,|\,y(\bs_{(il)})),
\end{equation}
where $w_l(\bs_i) \geq 0$ for every $\bs_i\in\BS$ and for all $l$, and 
$\sum_{l=1}^{i_L}w_l(\bs_i) = 1$.

There are two model elements in \eqref{eq:nnmp1} that describe spatial variability: 
the mixture component pmfs $f_{\bs_i,l}$, and the weights $w_l(\bs_i)$. We defer
the specification of the pmfs $f_{\bs_i,l}$ to the next section. 
Following \cite{zheng2021nnmp}, we define the weights as increments of a logit Gaussian 
cdf $G_{\bs_i}$, i.e., $w_l(\bs_i) = G_{\bs_i}(r_{\bs_i,l})-G_{\bs_i}(r_{\bs_i,l-1})$,
for $l = 1,\dots,i_L$. Here, $0 = r_{\bs_i,0} < r_{\bs_i,1} < \dots < 
r_{\bs_i,i_L-1} < r_{\bs_i,i_L}=1$ are random cutoff points such that 
$r_{\bs_i, l}-r_{\bs_i, l-1} = k'(\bs_i,\bs_{(il)})/\sum_{l=1}^{i_L}k'(\bs_i,\bs_{(il)})$,
for some bounded kernel $k': \D\times\D \rightarrow [0,1]$. 
Convenient choices for $k'$ are kernels that compute the correlation between two points.
The underlying Gaussian distribution for $G_{\bs_i}$ has mean 
$\mu(\bs_i) = \gamma_0 + \gamma_1 s_{i1} + \gamma_2 s_{i2}$, and variance $\kappa^2$,
with $\bs_i =$ $(s_{i1},s_{i2})$ where $s_{i1}$ and $s_{i2}$ correspond 
to the $x-$ and $y-$ coordinates of location $\bs_i$. This formulation allows for 
spatial dependence among the weights through $\mu(\bs_i)$. Also, the random cutoff 
points can flexibly reflect  the neighbor structure of $\bs_i$.

The second step completes the construction of a valid stochastic process over $\D$ by extending \eqref{eq:nnmp1} to an arbitrary finite set of locations outside $\BS$, denoted as $\U =$
$( \su_1,\dots,\su_r )$, where $\U\subset\D\setminus\BS$. In particular, we define the pmf of 
$\y_{\U}$ conditional on $\y_{\BS}$ as
\begin{equation}\label{eq:nnmp2}
    \tilde{p}(\y_{\U}\,|\,\y_{\BS}) = \prod_{i=1}^rp(y(\su_i)\,|\,\by_{\tNe(\su_i)}) =
    \prod_{i=1}^r\sum_{l=1}^{L} w_l(\su_i) \, f_{\su_i,l}(y(\su_i)\,|\,y(\su_{(il)})),
\end{equation}
where the weights and conditional pmfs are defined analogously to Equation \eqref{eq:nnmp1},
and the points $(\su_{(i1)},\dots,\su_{(iL)})$ in $\tNe(\su_i)$ are the 
first $L$ locations in $\BS$ that are closest to $\su_i$.

In fact, given \eqref{eq:nnmp1} and \eqref{eq:nnmp2}, a discrete-valued 
spatial process over $\D$ is well defined, based on the definition of nearest-neighbor 
processes \citep{datta2016hierarchical}. For any finite set $\V\subset\D$ that is 
not a subset of $\BS$, the joint pmf over $\V$ is obtained by marginalizing 
$\tilde{p}(\y_{\U}\,|\,\y_{\BS})\tilde{p}(\y_{\BS})$ over $\y_{\BS\setminus\V}$,
where $\U = \V\setminus\BS$.

We note that the model involves the neighborhood size $L$ in both \eqref{eq:nnmp1} 
and \eqref{eq:nnmp2}. 
Our prior model for the spatially varying weights supports the strategy of using an 
over-specified $L$ that gives a large neighbor set, with important neighbors assigned 
large weights a posteriori. For specific data examples, a sensitivity analysis for $L$ 
can be further carried out to find an optimal $L$ according to standard model comparison 
metrics or scoring rules. This is illustrated with the real data application; see 
Section \ref{sec:bbs} and the supplementary material.

Practically, Equations \eqref{eq:nnmp1} and \eqref{eq:nnmp2} serve different purposes. 
The reference set $\BS$ is often reserved for observed data, so model estimation is 
based on \eqref{eq:nnmp1}, while spatial prediction at new locations outside the 
reference set relies on \eqref{eq:nnmp2}. Henceforth, we use 
\begin{equation}\label{eq:nnmp3}
p(y(\sv)\mid \y_{\tNe(\sv)}) = 
\sum_{l=1}^L w_l(\sv) \, f_{\sv,l}(y(\sv)\,|\,y(\sv_{(l)}))
\end{equation}
to characterize discrete NNMPs, where $\sv$ is a generic location in $\D$. The 
neighbor set $\tNe(\sv)$ contains the first $L$ locations in $\BS$ that are closest 
to $\sv$. We place these locations in ascending order according to distance, denoted 
as $\tNe(\sv)=$ $(\sv_{(1)},\dots, \sv_{(L)})$.

The discrete NNMP formulation implies two distinct features that set it apart 
from SGLMMs. In a SGLMM, responses $y(\sv)$ are conditionally independent with 
distribution $f(y(\sv)\,|\, z(\sv),\bbeta,r) =$ 
$a(y(\sv),r)\exp\left( r\{y(\sv)\eta(\sv) - \psi(\eta(\sv))\} \right)$, where $z(\sv)$ is 
a spatial random effect, $\bbeta$ are regression parameters, $r$ is a dispersion 
parameter, and $h(\eta(\sv)) =$
$\bx(\sv)^\top\bbeta + z(\sv)$ for some link function $h$. 
The joint distribution of observations $(y(\bs_1),\dots,y(\bs_n))$ involves 
integrating out the spatial random effects, i.e., 
$\int\{\prod_{i=1}^nf(y(\bs_i)\,|\,z(\bs_i),\bbeta,r)\}p(\bz_{\BS})d\bz_{\BS}$,
where $\bz_{\BS} = (z(\bs_1),\dots,z(\bs_n))^\top$.
This restricts the choice of $z(\sv)$ to stochastic processes for which the 
corresponding joint densities are easy to work with, limiting the range of 
spatial variability the SGLMM can describe over the domain.
In practice, $z(\sv)$ is commonly assumed to be a Gaussian process.
This limitation, however, does not affect discrete NNMPs, as the spatial 
dependence is introduced at the data level. The joint pmf of a discrete NNMP 
is fully specified through \eqref{eq:nnmp1} and \eqref{eq:nnmp2}, which is a 
finite mixture of generic spatial components that can flexibly capture spatial 
variability. In addition, the mixture model structure of discrete NNMPs allows 
for efficient implementation, using inference approaches for mixtures.

\subsection{Model construction with spatially varying marginals}

The key ingredient in constructing discrete NNMPs lies in the specification 
of the mixture component conditional pmfs $f_{\sv,l}$. There are many avenues 
to specify $f_{\sv,l}$. As each conditional pmf corresponds to a bivariate
random vector, say $(U_{\sv,l},V_{\sv,l})$, our strategy is to model $f_{\sv,l}$ 
through its bivariate pmf, denoted as $f_{U_{\sv,l},V_{\sv,l}}$. 
Let $f_{U_{\sv,l}}$ and $f_{V_{\sv,l}}$ be the marginal pmfs of 
$(U_{\sv,l},V_{\sv,l})$, such that $f_{\sv,l}\equiv f_{U_{\sv,l}|V_{\sv,l}}=$
$f_{U_{\sv,l},V_{\sv,l}}/f_{V_{\sv,l}}$. The benefits of this strategy are twofold.
First, it simplifies the multivariate dependence specification by focusing 
on the bivariate random vectors $(U_{\sv,l},V_{\sv,l})$. The multivariate 
dependence will be induced by bivariate distributions through the model's 
mixture formulation. Second, the strategy allows for the construction of models 
with a pre-specified family of marginal distributions, facilitating the study 
of local variability. For example, it is common in discrete geostatistical data modeling 
to include covariates through the (transformed) mean of the marginal distribution.

The second feature of this strategy relies on an extension of the first-order 
strict stationarity result from \cite{zheng2021nnmp}. Based on that result, an NNMP 
has stationary marginal pmf $f_Y$ if $f_{U_{\sv,l}}=f_{V_{\sv,l}} = f_Y$, for all 
$\sv$ and all $l$. Here, we generalize the result such that discrete NNMPs can be
built from pre-specified spatially varying marginal pmfs $g_{\sv}$, where 
$g_{\sv}$ is the marginal pmf of $Y(\sv)$. The generalization of the stationarity 
proposition applies to all NNMPs. For the interest of this paper, we summarize 
the result in the following proposition for discrete NNMPs.
\begin{prop}\label{prop:sta}
Consider a discrete NNMP model for spatial process $\{Y(\sv): \, \sv\in\D\}$,
and a collection of spatially varying pmfs $\{g_{\sv}: \, \sv\in\D\}$.
If, for each $\sv$, the marginal pmfs of the mixture component 
bivariate distributions are such that $f_{U_{\sv,l}} = g_{\sv}$ and 
$f_{V_{\sv,l}} = g_{\sv_{(l)}}$, the discrete NNMP has marginal pmf
$g_{\sv}$ for $Y(\sv)$, for every $\sv\in\D$.
\end{prop}

A natural example for $\{g_{\sv}: \, \sv\in\D\}$ is a family of distributions with 
(at least) one of its parameters indexed in space, i.e., $g_{\sv}(\cdot) =$
$g(\cdot\,|\,\theta(\sv),\bxi)$, in particular, through spatially varying covariates. 
Using a link function for $\theta(\sv)$, we can include such covariates that provide 
additional spatially referenced information. A more general example involves 
partitioning the domain into several regions, where in each region, $g_{\sv}$ is 
associated with a different family of marginal distributions. A relevant application 
is estimation of the abundance of a species that shows overdispersion in most areas, 
but underdispersion in areas where the species is less prevalent \citep{wu2015bayesian}. 
Overall, Proposition \ref{prop:sta} provides flexibility for construction of 
discrete-valued spatial models with specific marginal pmfs.

We develop next a key component of the methodology, that is, discrete copula NNMP 
model construction and inference. Given a family of marginal pmfs $g_{\sv}$, we 
create spatial copulas for random vectors $(U_{\sv,l},V_{\sv,l})$. We begin with 
copulas for a set of base random vectors $(U_l,V_l)$, and extend them to be spatially 
dependent by modeling the copula parameter that controls the dependence structure as 
spatially varying. Together with Proposition \ref{prop:sta}, this strategy allows 
for construction of discrete NNMPs with marginal pmfs in general families.

\section{Discrete copula NNMPs}

\subsection{Copula functions}

A bivariate copula function $C: [0,1]^2 \rightarrow [0,1]$ is a 
distribution function whose marginals are uniform distributions on $[0,1]$. 
Following \cite{sklar1959fonctions}, 
given a random vector $(Z_1,Z_2)$ with joint probability distribution $F$ 
and marginals $F_1$ and $F_2$, there exists a copula function $C$ such that 
$F(z_1,z_2) = C(F_1(z_1),F_2(z_2))$. 
If $F_1$ and $F_2$ are continuous, $C$ is unique. 
In this case, the copula density is
$c(z_1,z_2) = \partial C(F_1(z_1),F_2(z_2))/(\partial F_1\partial F_2)$,
and the joint density is $f(z_1,z_2) = c(z_1,z_2)f_1(z_1)f_2(z_2)$, 
where $f_1$ and $f_2$ are the densities of $F_1$ and $F_2$, respectively.

If both marginals are discrete, the copula $C$ is only unique on the set $\mathrm{Ran}(F_1)\times\mathrm{Ran}(F_2)$, where $\mathrm{Ran}(F_j)$ consists 
of all possible values of $F_j,\,j=1,2$ \citep{joe2014dependence}.
Nevertheless, if $C$ is a copula and $F_1$ and $F_2$ are discrete 
distribution functions, then $F(z_1,z_2) =$ $C(F_1(z_1),F_2(z_2))$ is a valid 
joint distribution; in practice, we select a parametric family for $C$
\citep{smith2012estimation}. Note that, in contrast with the continuous case, 
when the marginals are discrete, some popular dependence measures, such as 
Kendall's $\tau$, will depend on the marginals 
\citep{denuit2005constraints, genest2007primer}. Consequently, the Kendall's 
$\tau$ of the random vector $(Z_1,Z_2)$ will not be equivalent to the 
Kendall's $\tau$ of the copula. Without loss of generality, hereafter, 
we assume the bivariate copula carries a single parameter.

\subsection{Copula NNMPs for discrete geostatistical data}
\label{sec:prop2}

Here, we introduce copula NNMPs with discrete marginals,
with focus on using copulas to specify the bivariate distributions of 
the mixture components. Dropping the dependence on $l$ for clarity, consider 
a random vector $(U,V)$ with discrete marginal distributions $F_U,F_V$, and 
marginal pmfs $f_U,f_V$. Let $a_u = F_U(u^-)$ and $b_u = F_U(u)$, where 
$F_U(u^-)$ denotes the left limit of 
$F_U$ at $u$. If $U$ is ordinal, $F_U(u^-) = F_U(u - 1)$. 
Analogous definitions of $a_v$ and $b_v$ apply for $V$. The joint pmf
$f_{U,V}$ of $(U,V)$ is obtained by finite differences, 
\begin{equation}\label{eq:finite_diff}
f_{U,V}(u,v) = C(b_u, b_v) - C(b_u,a_v) - C(a_u,b_v) + C(a_u,a_v).
\end{equation}
Let $c(u, v)=$ $f_{U,V}(u,v)/(f_U(u)f_V(v))$, such that
$f_{U,V}(u,v) = c(u,v)f_U(u)f_V(v)$, using a notation that is analogous 
to that of the joint density when $(U,V)$ is continuous. Therefore, 
the conditional pmf, $f_{U|V}(u\,|\,v) = c(u,v)f_U(u)$.

To specify the distribution of base random vector $(U_l,V_l)$, we use
copula $C_l$ with parameter $\eta_l$. For a parsimonious location-dependent 
model, we create spatially varying copulas $C_{\sv,l}$ on
$(U_{\sv,l},V_{\sv,l})$ by extending $\eta_l$ to $\eta_l(\sv)$. In practice, 
we associate $\eta_l(\sv)$ to a spatial kernel that depends on $\sv\in\D$
through a link function. Using Proposition \ref{prop:sta} with a family of 
marginal pmfs $g_{\sv}$, the joint pmf on $(U_{\sv,l},V_{\sv,l})$ is 
$f_{U_{\sv,l},V_{\sv,l}}(u, v) =$ $c_{\sv,l}(u, v) f_{U_{\sv,l}}(u)f_{V_{\sv,l}}(v)$,
where $f_{U_{\sv,l}} = g_{\sv}$ and $f_{V_{\sv,l}} = g_{\sv_{(l)}}$, and 
the conditional pmf is $f_{\sv,l}(u\,|\,v) = c_{\sv,l}(u,v)g_{\sv}(u)$.
Finally, the conditional pmf of the discrete copula NNMP model is given by
\begin{equation}\label{eq:copula_nnmp}
\begin{aligned}
p(y(\sv)\mid \by_{\tNe(\sv)}) =
\sum_{l=1}^L w_l(\sv) \, c_{\sv,l}(y(\sv),y(\sv_{(l)})) \, g_{\sv}(y(\sv)),
\end{aligned}
\end{equation}
where the marginal pmf for $Y(\sv)$ is $g_{\sv}$.

Recall that an NNMP model involves two sets of locations, the reference set $\BS$ 
and nonreference set $\U$. As done in practice, we take the reference set $\BS$ to 
correspond to the observed locations, and consider a generic finite set $\U$ 
such that $\BS\cap\U=\emptyset$. 
Then, the joint pmf $\tilde{p}(\y_{\V})$ over set $\V = \BS\cup\U$ describes
the NNMP distribution over any finite set of locations that includes the 
observed locations. In general, for a discrete NNMP, an explicit expression for 
$\tilde{p}(\y_{\V})$ is not available, since working with a bivariate discrete 
distribution and its conditional pmf is difficult. However, using copulas to 
specify the bivariate mixture component yields a structured conditional pmf 
and allows for the study of the joint pmf. The following proposition provides 
an explicit expression for $\tilde{p}(\y_{\V})$ under a discrete copula NNMP. 
The proof of the proposition can be found in the supplementary material.

\begin{prop}\label{prop:copnnmp_joint}
Consider a discrete copula NNMP model for spatial process $\{Y(\sv): \, \sv\in\D\}$,
with $\BS = \{\bs_1,\dots,\bs_n\}$ and $\U = \{\su_1,\dots,\su_m\}$,
where $n\geq 2$, $m\geq1$, and $\BS\cap\U=\emptyset$. 
Take $\V = \BS\cup\U$, and 
let $\y_{\V} = (y(\bs_1),\dots,y(\bs_n),y(\su_1),\dots,y(\su_m))^\top$.
Then the joint pmf of $\y_{\V}$ is $\tilde{p}(\y_{\V}) = \tilde{p}(\by_{\U}\,|\,\by_{\BS})\tilde{p}(\by_{\BS})$, where
\begin{equation}\label{eq:joint_pmf}
\begin{aligned}
\tilde{p}(\y_{\BS}) & = \prod_{i=1}^ng_{\bs_i}(y(\bs_i))
\sum_{l_n=1}^{n_L}\dots\sum_{l_2=1}^{2_L}w_{\bs_n,l_n}\dots w_{\bs_2,l_2}
c_{\bs_n,l_n}\dots c_{\bs_2,l_2},\\
\tilde{p}(\y_{\U}\,|\,\y_{\BS}) & = \prod_{i=1}^mg_{\su_i}(y(\su_i))
\sum_{l_m=1}^L\dots\sum_{l_1=1}^Lw_{\su_m,l_m}\dots w_{\su_1,l_1}
c_{\su_m,l_m}\dots c_{\su_1,l_1}.
\end{aligned}
\end{equation}
where $w_{\bs_i,l_i}\equiv w_{l_i}(\bs_i)$ and
$c_{\bs_i,l_i}\equiv c_{\bs_i,l_i}(y(\bs_i),y(\bs_{(i,l_i)}))$, for
$l_i = 1,\dots,i_L$, $i = 2,\dots, n$, and
$w_{\su_i,l_i}\equiv w_{l_i}(\su_i)$ and $c_{\su_i,l_i}\equiv c_{\su_i,l_i}(y(\su_i),y(\su_{(i,l_i)}))$, 
for $l_i = 1,\dots,L$, $i = 1,\dots, m$.
\end{prop}

We note that Proposition \ref{prop:copnnmp_joint} also applies 
when $\y_{\V}$ is continuous. It indicates that, given the sequence of pmfs $g_{\sv}$, 
the joint pmf of $\by_{\V}$ is determined by the collection of bivariate copulas,
motivating the use of different copula families to construct discrete NNMPs.
To balance flexibility and scalability, our strategy is to take all copulas $C_l$ 
in one family with the same link function for the copula parameters. 
Table \ref{tbl:copula} presents three examples with copula parameters modeled 
via a link function $k: \D\times\D\rightarrow[0,1]$. 
In particular, the Gumbel and Clayton copulas are asymmetric.
They exhibit greater dependence in the positive and negative tails, respectively.
In the first simulation example, we demonstrate that 
when the underlying spatial dependence is non-Gaussian,
it may be appropriate to choose asymmetric copulas.
We present next an example of a discrete copula NNMP construction.

\begin{example}\label{ex:nb}
\textit{Gaussian copula NNMP with negative binomial marginals}.
For the family of marginal pmfs $g_{\sv}$, consider the negative binomial distribution 
with mean $\alpha(\sv)$ and dispersion parameter $r$, denoted as $\mathrm{NB}(\alpha(\sv),r)$. Therefore, $g_{\sv}(y) = \binom{y+r-1}{y}(p(\sv))^r(1-p(\sv))^y$,
with $p(\sv) = r/(\alpha(\sv)+r)$. To include a vector of covariates $\x(\sv)$, 
we take a log-link function for $\alpha(\sv)$ such that
$\log(\alpha(\sv)) = \x(\sv)^\top\bbeta$, where $\bbeta$ is a vector of regression parameters.
We first specify Gaussian copulas $C_l$ with correlation parameters $\rho_l$ for the base random vectors $(U_l,V_l)$.
We then modify the correlation parameters $\rho_l$ using a correlation
function $k$ for all $l$ such that $\rho_l(\sv) := k(\sv,\sv_{(l)})$, creating 
a sequence of spatially varying copulas $C_{\sv,l}$. 
The resulting model is given by \eqref{eq:copula_nnmp} with $g_{\sv} =$ $\mathrm{NB}(\alpha(\sv),r)$.
\end{example}

\begin{table}[t!]
\caption{Examples of spatial copulas $C_{\sv,l}$ and corresponding 
link functions, $k: \D\times\D\rightarrow[0,1]$.}
    \centering
    \begin{threeparttable}
    \begin{tabular*}{\hsize}{@{\extracolsep{\fill}}lcc}
\hline
  & $C_{\sv,l}(z_1,z_2)$ & link function \\
\hline
Gaussian & $\Phi_2(\Phi^{-1}(z_1),\,\Phi^{-1}(z_2))$ 
& $\rho_l(\sv) = k(\sv, \sv_{(l)})$ \\
\hline
Gumbel & $\exp(-((-\log z_1)^{\eta_l(\sv)}+(-\log z_2)^{\eta_l(\sv)})^{1/\eta_l(\sv)})$ 
& $\eta_l(\sv) = (1 - k(\sv,\sv_{(l)}))^{-1}$ \\
\hline
Clayton & $(z_1^{-\delta_l(\sv)} + z_2^{-\delta_l(\sv)} - 1)^{-1/\delta_l(\sv)}$
&  $\delta_l(\sv) = 2k(\sv,\sv_{(l)}) / (1 - k(\sv,\sv_{(l)}))$\\
\hline
    \end{tabular*}
    \begin{tablenotes}[para,flushleft]
        \small 
        \vspace{20pt}
        Note: the bivariate cdf $\Phi_2$ corresponds to the
        standard bivariate Gaussian distribution with 
        correlation $\rho \in (0,1)$, and the cdf $\Phi$
        corresponds to the standard univariate Gaussian distribution.
    \end{tablenotes}
    \end{threeparttable}
    \label{tbl:copula}
\end{table}

\subsection{Inference for discrete copula NNMPs}
\label{sec:ce}

A traditional copula model for an $n$-variate discrete-valued 
vector involves evaluating $2^n$ terms of $n$-dimensional copulas.
Unless $n$ is very small, the computation is infeasible. Notable exceptions are
discrete vine copula models \citep{panagiotelis2012pair} that decompose a 
multivariate pmf into bivariate copulas and marginals under a set of trees. 
The computations for likelihood evaluations grow quadratically in $n$.
Discrete copula NNMPs compare favorably with discrete vine models, 
as the structured mixture formulation results in only $4nL$ bivariate 
copula function evaluations for the likelihood, providing linear growth in $n$.

Here, we develop a framework for discrete copula NNMP inference, based on 
transforming the discrete random variables to continuous ones by adding 
auxiliary variables, using the continuous extension (CE) approach in \cite{denuit2005constraints}. Working with continuous marginals improves 
computational efficiency and stability: the likelihood requires only $nL$ 
bivariate copula density evaluations; and, computing the conditional pmf 
using the finite differences in \eqref{eq:finite_diff} is bypassed, thus 
avoiding numerical instability especially for copulas that are not 
analytically available, such as the Gaussian copula.
Moreover, this framework makes more efficient the key task of spatial 
prediction over unobserved sites by avoiding computation that involves 
inverting the conditional cdf based on \eqref{eq:finite_diff}.

We associate each $Y(\sv)$ with a continuous random variable 
$Y^*(\sv)$, such that $Y^*(\sv)= Y(\sv) - O(\sv)$, where $O(\sv)$ is a continuous 
uniform random variable on $(0,1)$, independent of $Y(\sv)$ and of 
$O(\sv')$, for $\sv'\neq\sv$. We refer to $Y^*(\sv)$ as the continued $Y(\sv)$ 
by $O(\sv)$. Let $Q_{\sv}$ and  $g_{\sv}$ be the marginal cdf and pmf of $Y(\sv)$,
respectively. Then, the marginal cdf and density of $Y^*(\sv)$ are 
$Q^*_{\sv}(y^*(\sv)) = Q_{\sv}([y^*(\sv)]) + (y^*(\sv)-[y^*(\sv)])g_{\sv}([y^*(\sv)+1])$,
and $g^*_{\sv}(y^*(\sv)) = g_{\sv}([y^*(\sv)+1])$, respectively, 
where $[x]$ denotes the integer part of $x$.

Based on marginal densities $g^*_{\sv}$, we take spatial copulas
$C^*_{\sv,l} = C_{\sv,l}$ for continuous random vectors $(U^*_{\sv,l},V^*_{\sv,l})$, 
with marginals
$f_{U^*_{\sv,l}} = g^*_{\sv}$ and $f_{V^*_{\sv,l}} = g^*_{\sv_{(l)}}$, using
copulas $C_{\sv,l}$ from the original NNMP model. The joint density on $(U^*_{\sv,l},V^*_{\sv,l})$
is $f_{U^*_{\sv,l},V^*_{\sv,l}}(u,v) = c^*_{\sv,l}(u,v)g^*_{\sv}(u)g^*_{\sv_{(l)}}(v)$,
and the conditional density is $f^*_{\sv,l}(u\,|\,v) = c^*_{\sv,l}(u,v)g^*_{\sv}(u)$,
where $c^*_{\sv,l}$ is the copula density.
Denote by $\y^*_{\tNe(\sv)}$ the vector that contains the continued elements
of $\y_{\tNe(\sv)}$, and $\bm o_{\tNe(\sv)}$ the vector of auxiliary variables for
elements of $\y_{\tNe(\sv)}$. Then, the implied model on $y^*(\sv)$ is
\begin{equation}\label{eq:ce-nnmp}
\begin{aligned}
p(y^*(\sv)\,|\,D^*(\sv)) 
& = \sum_{l=1}^L w_l(\sv) \, c^*_{\sv,l}(y^*(\sv),y^*(\sv_{(l)})) \, g^*_{\sv}(y^*(\sv))
\end{aligned}
\end{equation}
where $y^*(\sv) = y(\sv) - o(\sv)$, and 
$D^*(\sv) = \{\by^*_{\tNe(\sv)}, o(\sv), \bm o_{\tNe(\sv)}\}$. Based on 
Proposition \ref{prop:sta}, model \eqref{eq:ce-nnmp} has marginal density 
$g^*_{\sv}$ for $Y^*(\sv)$. To recover $y(\sv)$, we first generate $y^*(\sv)$ 
from the extended model, and then set $y(\sv) = [y^*(\sv)+1]$.

Regarding the existing literature, 
statistical inference for spatial copula models based on the CE approach is typically 
conducted by maximizing the expected likelihood with respect to the auxiliary variables 
\citep{madsen2009maximum, hughes2015copcar}. We develop inferential methods under the 
Bayesian framework. Posterior simulation based on \eqref{eq:ce-nnmp} takes advantage 
of copula properties for continuous random variables, thus providing efficient computation 
for both model estimation and prediction.

\section{Bayesian implementation}
\label{sec:inference}

\subsection{Hierarchical model formulation}

Assume that $\y_{\BS}=(y(\bs_1),\dots,y(\bs_n))^\top$ is a realization of a 
discrete copula NNMP with spatially varying marginal pmfs through spatially 
dependent covariates, $g_{\bs_i}(y(\bs_i))\equiv$ $g(y(\bs_i)\,|\,\bbeta, \bxi)$.
Here, $\bbeta = (\beta_0, \beta_1,\dots,\beta_p)^\top$, where
$\beta_0$ is an intercept and $(\beta_1,\dots,\beta_p)^\top$ is the regression 
parameter vector for covariates $\x(\bs_i)$, and $\bxi$ collects all other 
parameters of $g$. The copula parameter is modeled through
a link function $k$ with parameter(s) $\bphi$. We use the CE approach 
associating each $y(\bs_i)$ with $y^*(\bs_i)$, 
such that $y^*(\bs_i) = y(\bs_i) - o_i$, where $o_i\equiv o(\bs_i)$ 
is uniformly distributed on $(0,1)$, independent of $y(\bs_i)$
and of $o_j$, for $j\neq i$. Moreover, denote by $\bzeta$ the parameter 
of the cutoff point kernel for the mixture weights, 
defined in Section \ref{sec:framework}.

The formulation of the mixture weights allows us to augment the model with a sequence of 
auxiliary variables, $\{ t_i: i = 3,\dots, n \}$, where $t_i$ is a Gaussian random variable
with mean $\mu(\bs_i)$ and variance $\kappa^2$. The augmented model for the data 
can be expressed as 
$$
\begin{aligned}
&y(\bs_i) = y^*(\bs_i) + o_i,\;\;o_i\stackrel{i.i.d.}{\sim}\mathrm{Unif}(0,1),\;i=1,\dots,n,\\
&y^*(\bs_1)\mid\bbeta,\bxi\,  \sim g^*_{\bs_1}(y^*(\bs_1)),\;\;
y^*(\bs_2)\mid y^*(\bs_1),\bphi,\bbeta,\bxi \sim f^*_{\bs_2,1}(y^*(\bs_2)\,|\,y^*(\bs_1)),\\ 
&y^*(\bs_i)\mid\{y^*(\bs_{(il)})\}_{l=1}^{i_L},t_i,\bphi,\bbeta,\bxi,\bzeta\, 
\stackrel{ind.}{\sim}\sum_{l=1}^{i_L}f^*_{\bs_i,l}(y^*(\bs_i)\,|\,y^*(\bs_{(il)}))
\, \mathbbm{1}_{(r^*_{\bs_i,l-1},r^*_{\bs_i,l})}(t_i),\;i=3,\dots,n,\\
& t_i\mid\bga,\kappa^2\, \stackrel{ind.}{\sim}
N(t_i\mid\gamma_0+\gamma_1s_{i1}+\gamma_2s_{i2},\kappa^2),\;i = 3,\dots, n,
\end{aligned}
$$
where $f^*_{\bs_i,l}(y^*(\bs_i)\,|\,y^*(\bs_{(il)})) =
c^*_{\bs_i,l}(y^*(\bs_i),y^*(\bs_{(il)}))g^*_{\bs_i}(y^*(\bs_i))$,
and $r^*_{\bs_i,l}=$ $\log\{ r_{\bs_i,l}/(1-r_{\bs_i,l}) \}$, 
for $l=1,\dots,i_L$. The full Bayesian model is completed with prior 
specification for parameters $\bbeta, \bxi, \bphi, \bzeta, 
\bga = (\gamma_0,\gamma_1,\gamma_2)^\top$ 
and $\kappa^2$.
The priors for $\bxi$, $\bphi$, and $\bzeta$ depend on the choices of the 
pmf $g_{\bs_i}$, the copula $C^*_{\bs_i,l}$, 
and the kernel $k'$, respectively.
For parameters $\bbeta$, $\bga$, and $\kappa^2$, we consider
$N(\bbeta\,|\, \mu_{\bbeta},\bV_{\bbeta})$, $N(\bga\,|\,\mu_{\bga},\bV_{\bga})$,
and $\mathrm{IG}(\kappa^2\,|\, u_{\kappa^2},v_{\kappa^2})$ priors,
where $\mathrm{IG}$ denotes the inverse gamma distribution.

\subsection{Model estimation, validation and prediction}

We outline the MCMC sampler for parameters
$(\bbeta,\bxi,\bphi,\bzeta,\bga,\kappa^2)$, and latent variables
$\{t_i\}_{i=3}^n$ and $\{o_i\}_{i=1}^n$. 
We note that there is a set of configuration variables $\{\ell_i\}_{i=3}^n$
in one-to-one correspondence with $t_i$, i.e., $\ell_i = l$ if and only if $t_i\in(r_{\bs_i,l-1}^*,r_{\bs_i,l}^*)$, for $l = 1,\dots,i_L$.

The updates for parameters $\bbeta$, $\bxi$ and $\bphi$ require
Metropolis steps, since they enter in copula densities $c^*_{\bs_i,l}$.
We use a Metropolis step also for kernel $k'$ parameter $\bzeta$, 
which is involved in the definition of the mixture weights. Let $\bD$ be 
the $(n-2)\times 3$ matrix with $i$th row $(1,s_{i+2,1},s_{i+2,2})$.
The posterior full conditional of $\bga$ is 
$N(\bga\,|\,\bmu_{\bga}^*, \bV_{\bga}^*)$, where 
$\bV_{\bga}^* =$ $(\bV_{\bga}^{-1} + \kappa^{-2}\bD^\top\bD)^{-1}$
and $\bmu_{\bga}^* =$ $\bV_{\bga}^*(\bV_{\bga}^{-1}\bmu_{\bga} + 
\kappa^{-2}\bD^\top\bm{t})$, with the vector $\bm{t} = (t_3,\dots,t_n)^\top$.
The posterior full conditional distribution of $\kappa^2$ is 
$\mathrm{IG}(\kappa^2\,|\, u_{\kappa^2}+(n-2)/2, v_{\kappa^2}+
\sum_{i=3}^n(t_i-\mu(\bs_i))^2/2)$.

The posterior full conditional distribution for each latent variable 
$t_i$, $i =3,\dots,n$, can be expressed as 
$\sum_{l=1}^{i_L} q_l(\bs_i) \,
\mathrm{TN}(t_i \,|\, \mu(\bs_i),\kappa^2; r^*_{\bs_i,l-1} < t_i \leq r^*_{\bs_i,l})$,
where $\mathrm{TN}$ denotes the truncated normal distribution over the indicated interval, 
and $q_l(\bs_i) \propto$ $w_l(\bs_i) \, c^*_{\bs_i,l}(y^*(\bs_i),y^*(\bs_{(il)}))$,
for $l=1,...,i_L$. Hence, each $t_i$ can be readily updated by sampling from the 
$l$-th truncated normal with probability proportional to $q_l(\bs_i)$.
For auxiliary variables $o_i$, the posterior full conditional of $o_1$ 
is proportional to 
$\prod_{\{ j:\bs_{(j,\ell_j)}=\bs_1 \}} c^*_{\bs_j,\ell_j}(y(\bs_j)-o_j, y(\bs_1)-o_1)$,
and that of $o_i$, $i \geq 2$, is proportional to
$c^*_{\bs_i,\ell_i}(y(\bs_i)-o_i, y(\bs_{(i,\ell_i)})-o_{(i,\ell_i)})
\prod_{\{ j:\bs_{(j,\ell_j)}=\bs_i \}}c^*_{\bs_j,\ell_j}(y(\bs_j)-o_j, y(\bs_i)-o_i)$,
where $\ell_2=1$ and $o_{(i,\ell_i)}\equiv o(\bs_{(i,\ell_i)})$.
We update each latent variable $o_i$ with an independent Metropolis step with 
a $\mathrm{Unif}(0,1)$ proposal distribution.

The likelihood of the continued model admits the form 
$g_{\bs_1}(y^*(\bs_1))\prod_{i=2}^np(y^*(\bs_i)\,|\,D^*(\bs_i))$.
The product formulation allows for model validation, using a generalization of 
the randomized quantile residuals  proposed by \cite{dunn1996randomized} for 
independent data. Specifically, we define the marginal quantile residual, 
$r_1 =$ $\Phi^{-1}(Q^*_{\bs_1}(y^*(\bs_1)))$, and the $i$th conditional quantile 
residual, $r_i =$ $\Phi^{-1}(F(y^*(\bs_i)\,|\,D^*(\bs_i)))$,
$i = 2,\dots, n$, where $F$ is the conditional cdf of $y^*(\bs_i)$.
If the model is correctly specified, the residuals $r_i$, $i = 1,\dots, n$, 
would be independent and identically distributed as a standard Gaussian distribution.

Finally, we turn to posterior predictive inference at a new location $\sv_0$.
If $\sv_0\notin\BS$, for each posterior sample, we first compute the cutoff points 
$r_{\sv_0,l}$, such that $r_{\sv_0,l} - r_{\sv_0,l-1} = k'(\sv_0,\sv_{(0l)})/\sum_{l=1}^Lk'(\sv_0,\sv_{(0l)})$,
and the weights $w_l(\sv_0) = G_{\sv_0}(r_{\sv_0,l})-G_{\sv_0}(r_{\sv_0,l-1})$,
for $l = 1,\dots, L$. We then generate $y^*(\sv_0)$ based on \eqref{eq:ce-nnmp},
and set $y(\sv_0) = [y^*(\sv_0) + 1]$.
If $\sv_0\equiv\bs_i\in\BS$, we generate $y(\sv_0)$ similarly, the difference
being that we now use the posterior samples for the mixture weights obtained 
from the MCMC algorithm.

\section{Data illustrations}

To illustrate the proposed methodology, we present two synthetic data examples 
and a real data analysis. The goal of the first simulation experiment is to 
investigate the flexibility of discrete copula NNMPs, using different copula 
functions to define the NNMP mixture components. In the second experiment, 
we demonstrate the inferential and computational advantages of our approach 
for count data modeling, compared to SGLMMs. Implementation details for the models 
are provided in the supplementary material. Since our purpose is primarily 
demonstrative, we took $L = 10$ for the simulation experiments. A comprehensive
sensitivity analysis for $L$ was conducted for the real data application of 
Section \ref{sec:bbs}, with details provided in the supplementary material.

In both simulated data examples, we ran the MCMC algorithm for each copula NNMP model 
for 20000 iterations, discarding the first 4000 iterations, and collecting posterior 
samples every four iterations. The SGLMM models were implemented using the spBayes 
package in R \citep{r-spBayes}; 
we ran the algorithm for 40000 iterations and collected posterior samples every 
five iterations, with the first 20000 as burn-in. 

We compare models based on parameter estimates, root mean squared prediction error (RMSPE),
$95\%$ credible interval width ($95\%$ CI width), $95\%$ credible interval coverage rate ($95\%$ CI cover), 
continuous ranked probability score (CRPS; \citealt{gneiting2007strictly}), energy score (ES; \citealt{gneiting2007strictly}), and variogram score of order one (VS; \citealt{scheuerer2015variogram}). The energy score is a multivariate extension of 
the CRPS, while the variogram score examines pairwise differences of the components 
of the multivariate quantity. Both the ES and VS allow for comparison of model 
predictive performance with respect to dependence structure.

\subsection{First simulation experiment}
\label{sec: sim1}

We first generated sites over a regular grid of $120\times 120$ resolution 
on  a unit square domain, and then simulated data from $y(\sv) =$
$F_Y^{-1}\big(F_Z(z(\sv))\big)$, where $F_Y$ corresponds to the Poisson 
distribution with rate parameter $\lambda_0 = 5$, and
$z(\sv)$ is the skew-Gaussian random field from \cite{zhang2010spatial} with stationary 
marginal distribution $F_Z$. More specifically, $z(\sv) =$ 
$\sigma_1 \, |\omega_1(\sv)| + \sigma_2 \, \omega_2(\sv)$, where 
both $\omega_1(\sv)$ and $\omega_2(\sv)$ are standard Gaussian processes with 
exponential correlation function based on range parameter $0.1$. 
The density of $F_Z$ is $f_Z(z) =$ $2 \, N(z\,|\, 0,\sigma_1^2+\sigma_2^2) \, \Phi(\sigma_1z/(\sigma_2\sqrt{\sigma_1^2+\sigma_2^2}))$, 
where $\sigma_1\in\R$ controls the skewness, and $\sigma_2>0$ is a scale parameter.
We took $\sigma_2 = 1$, and $\sigma_1 = 1, 3, 10$, which corresponds to positive weak, 
moderate, and strong skewness.

\begin{table}[t!]
\caption{Simulation example 1: posterior mean and 95\% CI estimates for the rate
parameter $\lambda$ of the Poisson NNMP marginal distribution, and scores for 
comparison of Gaussian, Gumbel and Clayton copula NNMP models, under each 
of the three simulation scenarios for $\sigma_1$.}
    \centering
    \begin{tabular*}{\hsize}{@{\extracolsep{\fill}}lccccccccccc}
    \hline
\multicolumn{1}{l}{} & 
\multicolumn{3}{c}{$\sigma_1 = 1$} & \multicolumn{1}{c}{} & 
\multicolumn{3}{c}{$\sigma_1 = 3$} & \multicolumn{1}{c}{} & 
\multicolumn{3}{c}{$\sigma_1 = 10$}\\
\cline{2-4} \cline{6-8} \cline{10-12}
\multicolumn{1}{l}{} & 
\multicolumn{3}{c}{$\lambda$} & \multicolumn{1}{c}{} & 
\multicolumn{3}{c}{$\lambda$} & \multicolumn{1}{c}{} & 
\multicolumn{3}{c}{$\lambda$}\\
\hline
Gaussian & 
\multicolumn{3}{c}{4.55 (4.16, 4.94)} & \multicolumn{1}{c}{} & 
\multicolumn{3}{c}{4.71 (4.37, 5.07)} & \multicolumn{1}{c}{} & 
\multicolumn{3}{c}{4.88 (4.55, 5.22)}\\
Gumbel & 
\multicolumn{3}{c}{4.78 (4.39, 5.21)} & \multicolumn{1}{c}{} & 
\multicolumn{3}{c}{4.88 (4.56, 5.24)} & \multicolumn{1}{c}{} & 
\multicolumn{3}{c}{4.94 (4.66, 5.23)}\\
Clayton & 
\multicolumn{3}{c}{5.33 (4.99, 5.68)} & \multicolumn{1}{c}{} & 
\multicolumn{3}{c}{5.25 (4.96, 5.56)} & \multicolumn{1}{c}{} & 
\multicolumn{3}{c}{5.36 (5.08, 5.65)}\\
\hline
\multicolumn{1}{l}{} & 
\multicolumn{3}{c}{$\sigma_1 = 1$} & \multicolumn{1}{c}{} & 
\multicolumn{3}{c}{$\sigma_1 = 3$} & \multicolumn{1}{c}{} & 
\multicolumn{3}{c}{$\sigma_1 = 10$}\\
\cline{2-4} \cline{6-8} \cline{10-12}
\multicolumn{1}{l}{} & \multicolumn{1}{c}{CRPS} & \multicolumn{1}{c}{ES} & \multicolumn{1}{c}{VS} & \multicolumn{1}{c}{} &
\multicolumn{1}{c}{CRPS} & \multicolumn{1}{c}{ES} & \multicolumn{1}{c}{VS} & \multicolumn{1}{c}{} &
\multicolumn{1}{c}{CRPS} & \multicolumn{1}{c}{ES} & \multicolumn{1}{c}{VS} \\
\hline
Gaussian &$0.69$ &$12.77$ & $94855$ & 
         &$0.85$ &$15.54$ & $124893$ & 
         &$0.93$ &$16.98$ & $138592$ \\
Gumbel &$0.69$ &$12.58$ & $92278$ & 
         &$0.85$ &$15.32$ & $120932$ & 
         &$0.92$ &$16.71$ & $134774$ \\
Clayton &$0.75$ &$14.34$ & $125800$ & 
         &$0.90$ &$17.36$ & $164148$ & 
         &$1.00$ &$18.70$ & $174123$ \\
\hline
    \end{tabular*}
    \label{tbl:sim1}
\end{table}

We considered three discrete copula NNMPs with stationary Poisson marginals, i.e., 
$g_{\sv}=f_Y$, for all $\sv$, where $f_Y$ is the Poisson pmf with rate $\lambda$.
The three models correspond to the copulas in Table \ref{tbl:copula},
with the link function $k$ given by an exponential correlation
function with range parameter denoted by $\phi_1$, $\phi_2$, and $\phi_3$ for the 
Gaussian, Gumbel, and Clayton copula models, respectively.
We specified the cutoff point kernel through an exponential
correlation function with range parameter $\zeta_1$, $\zeta_2$, and $\zeta_3$
for the Gaussian, Gumbel, and Clayton copula models, respectively. The Bayesian models 
are fully specified with an $\mathrm{IG}(3, 1)$ prior for the $\phi$ and $\zeta$ 
parameters, and with $N(\bga\,|\,(-1.5, 0, 0)^\top,\,2\mathbf{I}_3)$ and 
$\mathrm{IG}(\kappa^2\,|\,3,1)$ priors. Finally, the prior for the rate parameter 
$\lambda$ was taken as $\mathrm{Ga}(1,1)$, where $\mathrm{Ga}(a,b)$ denotes the
gamma distribution with mean $a/b$.
We simulated 1000 responses and used 800 of them to fit the three NNMP models. 
The remaining 200 observations were used for model comparison.

Table \ref{tbl:sim1} provides estimates for the rate parameter $\lambda$ of
the Poisson marginal distribution, and predictive performance metrics. 
For all three cases for $\sigma_1 = 1, 3, 10$, the Gumbel model yields the 
more accurate estimates for $\lambda$. In particular, the Gumbel model's
$95\%$ CIs include the true parameter value, whereas those of the Gaussian and 
Clayton models failed to cover it when $\sigma_1 = 1$ and $\sigma_1 = 10$, respectively.
Regarding predictive performance, the Gumbel model outperforms to a smaller or 
larger extent the other two models across different scenarios. Predictive random 
fields under the three models are provided in the supplementary material. 
We found that prediction by the Clayton model was not able to recover large values.
Compared to the Gaussian model, the Gumbel model recovered large values slightly better.
Overall, this example demonstrates that, when the underlying spatial dependence is 
driven by non-Gaussian processes, it is practically useful to consider copulas
from asymmetric families, including use of appropriate model comparison tools.

\subsection{Second simulation experiment}
\label{sec: sim2}

\begin{figure}[t!]
    \centering
    \begin{subfigure}[b]{0.4\textwidth}
         \centering
         \includegraphics[width=\textwidth]{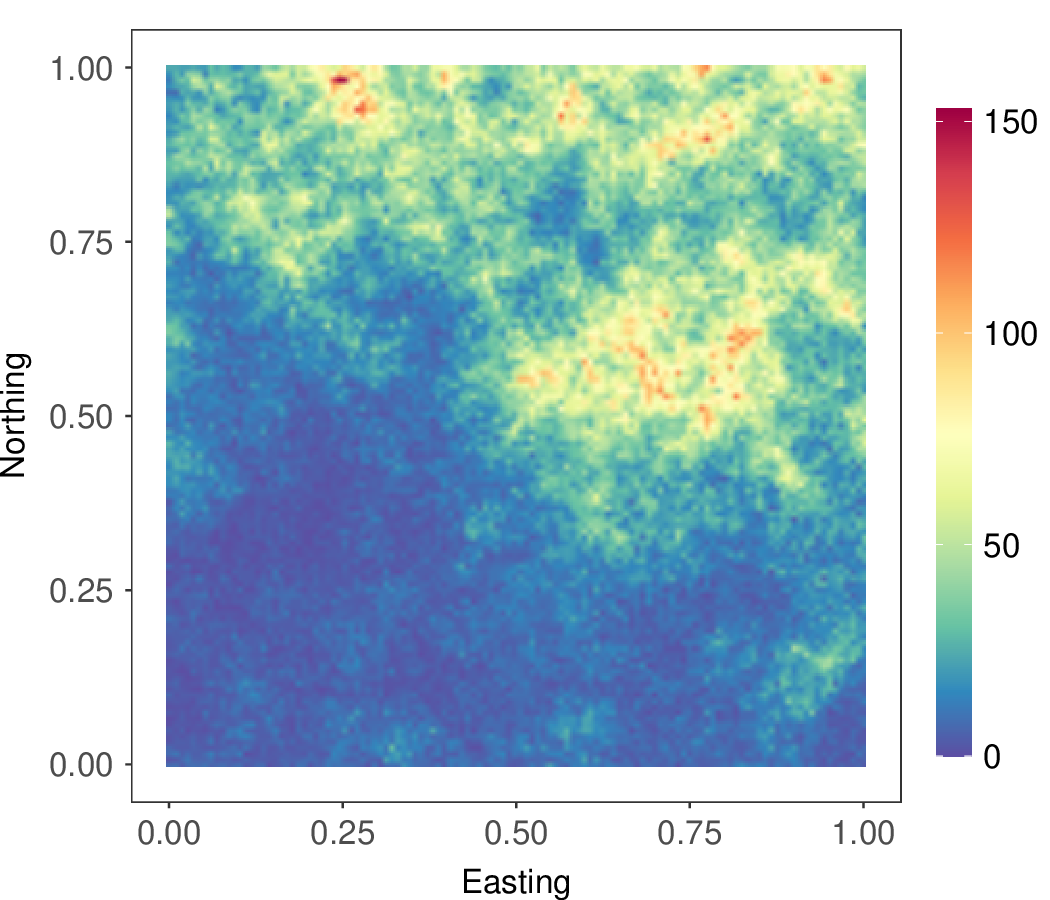}
         \caption{True $y(\sv)$}
     \end{subfigure}
    %  \hfill
    \begin{subfigure}[b]{0.4\textwidth}
         \centering
         \includegraphics[width=\textwidth]{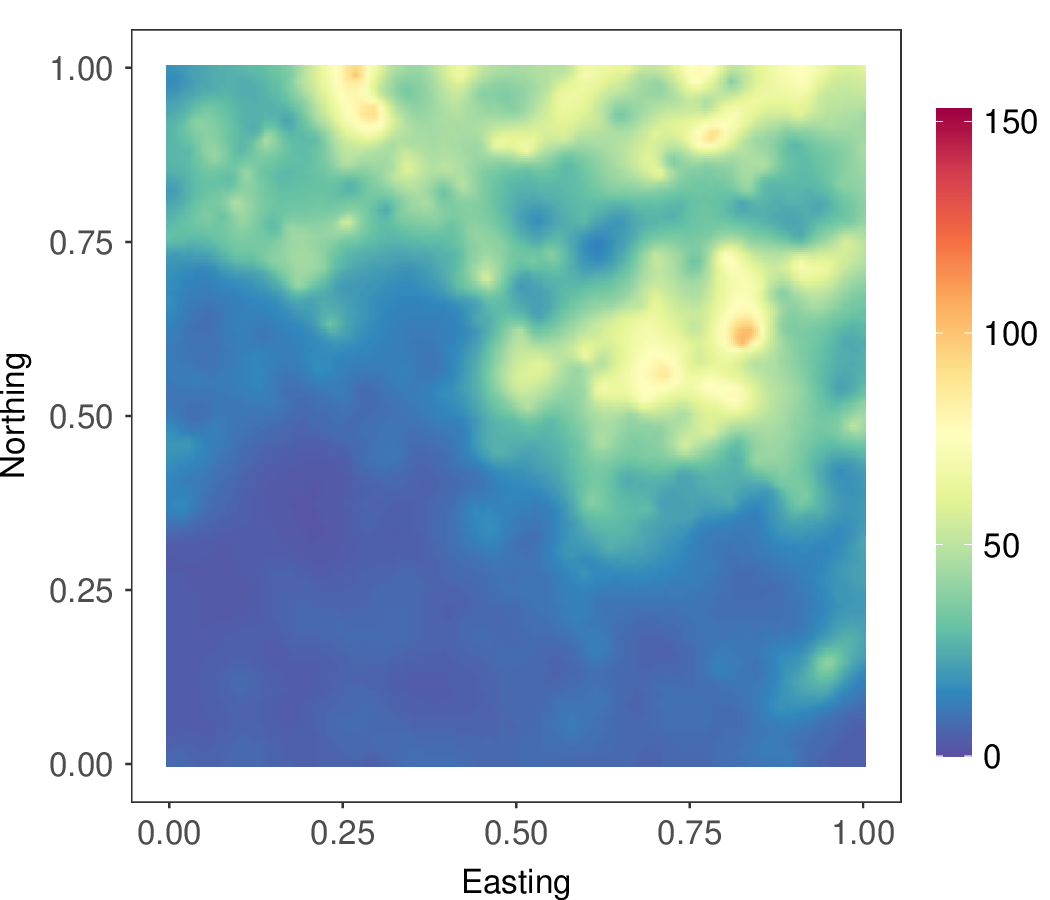}
         \caption{SGLMM-GP}
     \end{subfigure}     \\
     \medskip
     \begin{subfigure}[b]{0.4\textwidth}
         \centering
         \includegraphics[width=\textwidth]{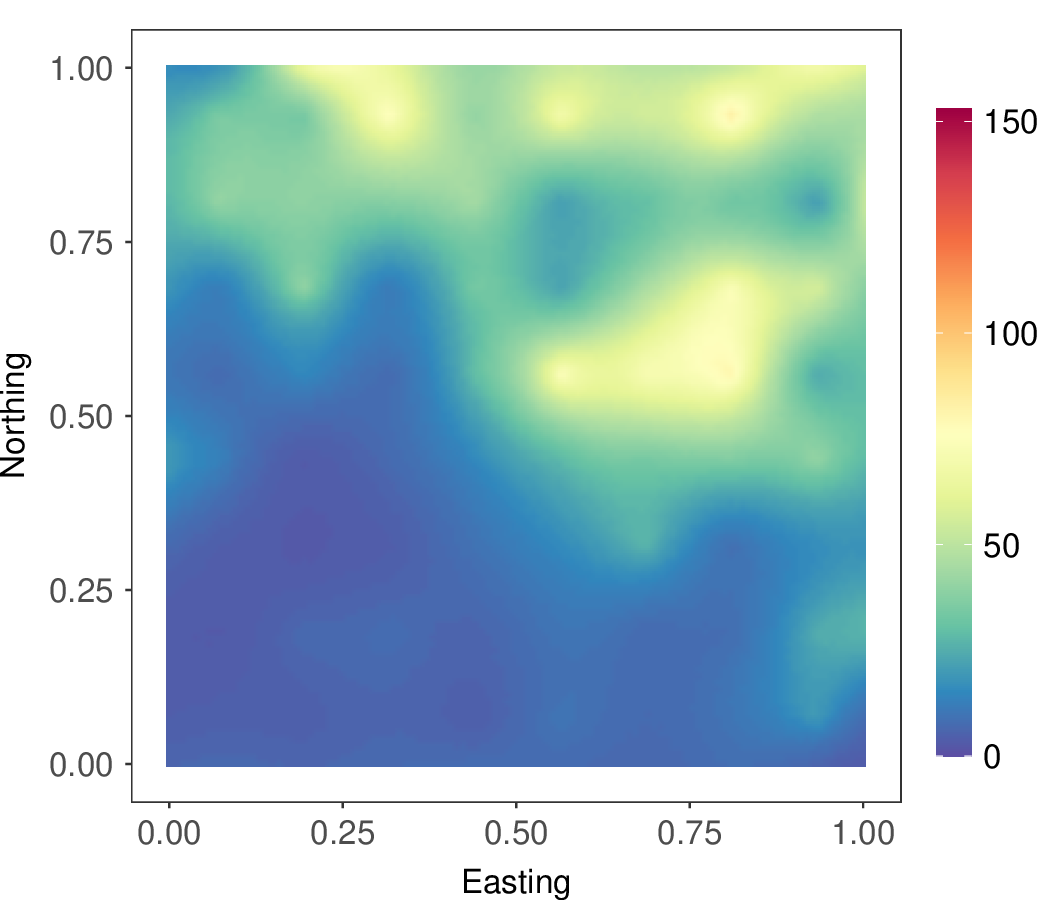}
         \caption{SGLMM-GPP}
     \end{subfigure}
     \begin{subfigure}[b]{0.4\textwidth}
         \centering
         \includegraphics[width=\textwidth]{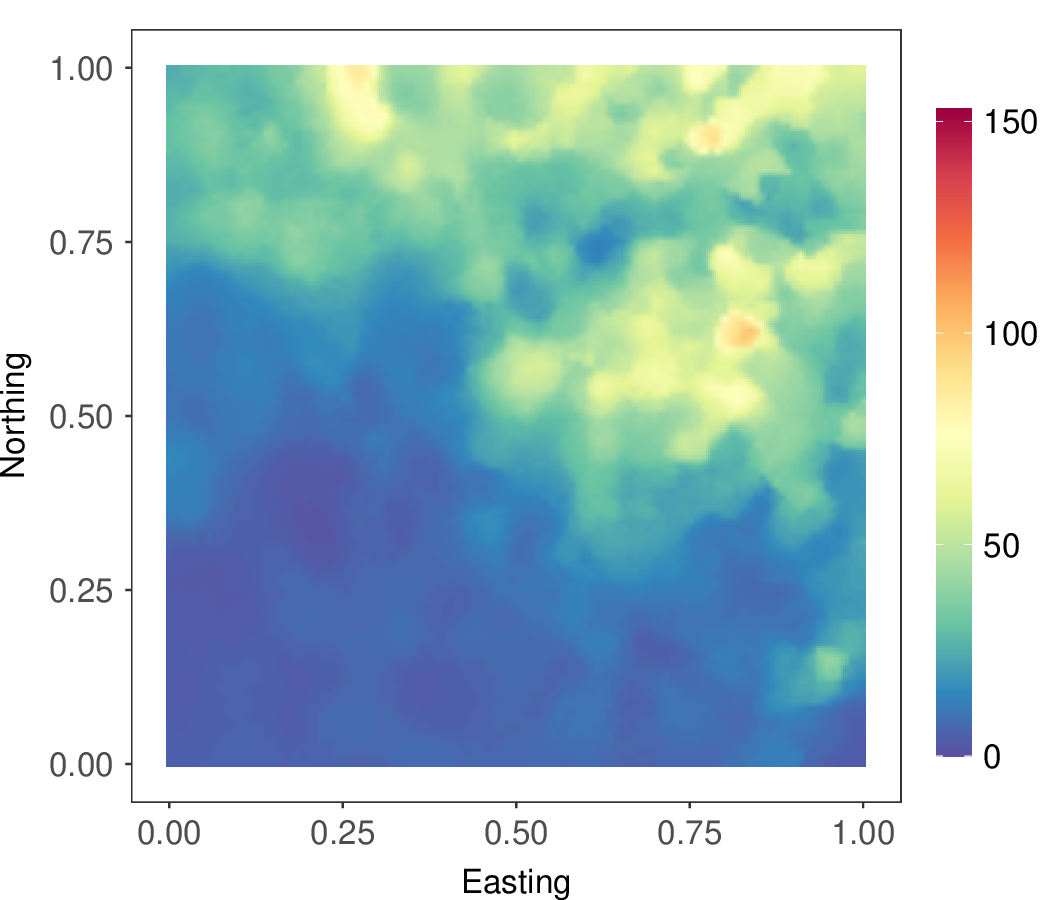}
         \caption{NBNNMP}
     \end{subfigure}     
    \caption{
    Second simulation example. Interpolated surfaces of the true model and 
    posterior median estimates of the SGLMM-GP, SGLMM-GPP and NBNNMP models.
    }
    \label{fig:sim}
\end{figure}

We generated data over a grid of sites with
$120\times 120$ resolution, uniformly on the square $[0,1]\times [0,1]$,
using a Poisson SGLMM with 
$y(\sv)\,|\,\eta(\sv)  \sim \mathrm{Pois}(\eta(\sv))$, and 
$\log(\eta(\sv))  = \beta_0 + v_{1}\beta_1 + v_{2}\beta_2 + z(\sv)$,
where $\sv = (v_1,v_2)$, and $z(\sv)$ is a zero-centered Gaussian process (GP) 
with variance parameter $\sigma^2 = 0.2$ and an exponential correlation function 
with range parameter $\phi_0 = 1/12$. We set the regression
coefficients $\bbeta = (\beta_0,\beta_1,\beta_2)^\top = (1.5, 1, 2)^\top$, 
resulting in a random field with a trend, as shown in Figure \ref{fig:sim}(a).

We considered three models. The first is the negative binomial NNMP model (NBNNMP) 
with a Gaussian copula, as discussed in Example \ref{ex:nb}. The second model (SGLMM-GP) 
is a Poisson SGLMM with a GP prior assigned to $z(\sv)$. For the last model (SGLMM-GPP), 
we considered a Poisson SGLMM with spatial random effects $z(\sv)$ corresponding to a 
Gaussian predictive process (GPP, \citealt{banerjee2008gaussian}), with $10 \times 10$ 
knots placed on a grid over the domain. We chose the number of knots such that the 
computing times for the SGLMM-GPP and NBNNMP models are similar. As in the first 
simulation example, all models were fit to 800 observations and compared on the basis 
of 200 additional observations.

The regression coefficients for all models were assigned %flat priors. 
mean-zero, dispersed normal priors.
We worked with an exponential correlation function for all models,
used for $\rho_l(\sv)$ of the Gaussian copula in the NBNNMP model, 
and as the correlation function for the GP and GPP in the SGLMMs. 
The range parameter was assigned an inverse gamma prior $\mathrm{IG}(3, 1)$
for the NBNNMP model, and a uniform prior $\mathrm{Unif}(1/30,1/3)$ for the 
other two models. 
The cutoff point kernel of the NBNNMP was also specified an exponential correlation function, with
an $\mathrm{IG}(3,1)$ prior for the range parameter.
The variance parameter for the SGLMM models was assigned an 
inverse gamma prior $\mathrm{IG}(2,1)$. For the logit Gaussian distribution 
parameters $\bga$ and $\kappa^2$ of the NBNNMP, we used 
$N((-1.5,0,0)^\top,\,2\mathbf{I}_3)$ and $\mathrm{IG}(3, 1)$ priors, 
respectively. Finally, we placed a $\mathrm{Ga}(1,1)$ prior on the NBNNMP
dispersion parameter $r$.

Estimates of the regression parameters
and performance metrics for out-of-sample prediction are provided in Table \ref{tbl:sim2}.
We observe that, overall, the NBNNMP model provided the more accurate estimation for 
$\bbeta$. Regarding predictive performance, the NBNNMP model outperformed the SGLMM-GPP 
model by a large margin, and was comparable to the SGLMM-GP model, which corresponds 
to the data generating process for this simulation experiment. Moreover, the last row of the 
table highlights the NBNNMP model's huge gains in computing time compared to the SGLMM-GP
model.

\begin{table}[t!]
\caption{Simulation example 2: posterior mean and 95\% CI estimates for the
regression parameters, performance metrics, and computing time, under the 
NBNNMP model and the two SGLMM models.}
    \centering
    \begin{tabular*}{\hsize}{@{\extracolsep{\fill}}lcccc}
    % \\[-5pt]
\hline
    & True & NBNNMP & SGLMM-GP & SGLMM-GPP \\
\hline
$\beta_0$ & 1.5 & $1.61\,(1.29, 1.97)$ & $1.53\,(1.22, 1.81)$ & $1.41\,(1.02, 1.73)$ \\
$\beta_1$ & 1   & $0.90\,(0.51, 1.31)$ & $0.70\,(0.25, 1.15)$ & $0.91\,(0.43, 1.34)$ \\
$\beta_2$ & 2   & $1.94\,(1.51, 2.32)$ & $2.18\,(1.91, 2.53)$ & $2.25\,(1.81, 2.84)$ \\
RMSPE     & -    & 9.06 & 8.88 & 10.00\\
$95\%$ CI cover & -       & 0.98 & 0.97 & 0.78\\
$95\%$ CI width & - & 37.02 & 32.24 & 19.02\\
CRPS    & -        & 4.58 & 4.52 & 5.37\\
ES      & -        & 92.07 & 91.41 & 107.46 \\
VS      & -        & 5175591 & 5199629 & 6378263\\
Time (mins)  & -   & 11.18 & 935.02 & 11.68 \\
\hline
    \end{tabular*}
    \label{tbl:sim2}
\end{table}

Figure \ref{fig:sim}(b)-\ref{fig:sim}(d) plots the posterior median estimates of the 
random field for the three models. The SGLMM-GPP yields an overly smooth 
estimate, whereas the SGLMM-GP and NBNNMP models provide similar estimates that approximate 
well the true surface. Overall, this example illustrates the inferential and computational
advantages of discrete copula NNMPs for modeling count data.

\subsection{North American Breeding Bird Survey data analysis}
\label{sec:bbs}

The primary source of information on population evolution for birds
is count data surveys. Since 1966, the North American Breeding Bird Survey (BBS)
has been conducted to monitor bird population change. There are over 4000 sampling units
in the survey, each with a 24.5-mile roadside route. Along each route, volunteer observers
count the number of birds by sight or sound, in a 3-min period at each of 50 stops \citep{pardieck2020north}. 
The BBS data are often used to determine temporal or geographical patterns of relative abundance.
Spatial maps of relative abundance are crucial for ecological studies.

\begin{figure}[t!]
    \centering
    \begin{subfigure}[b]{0.4\textwidth}
         \centering
         \includegraphics[width=\textwidth]{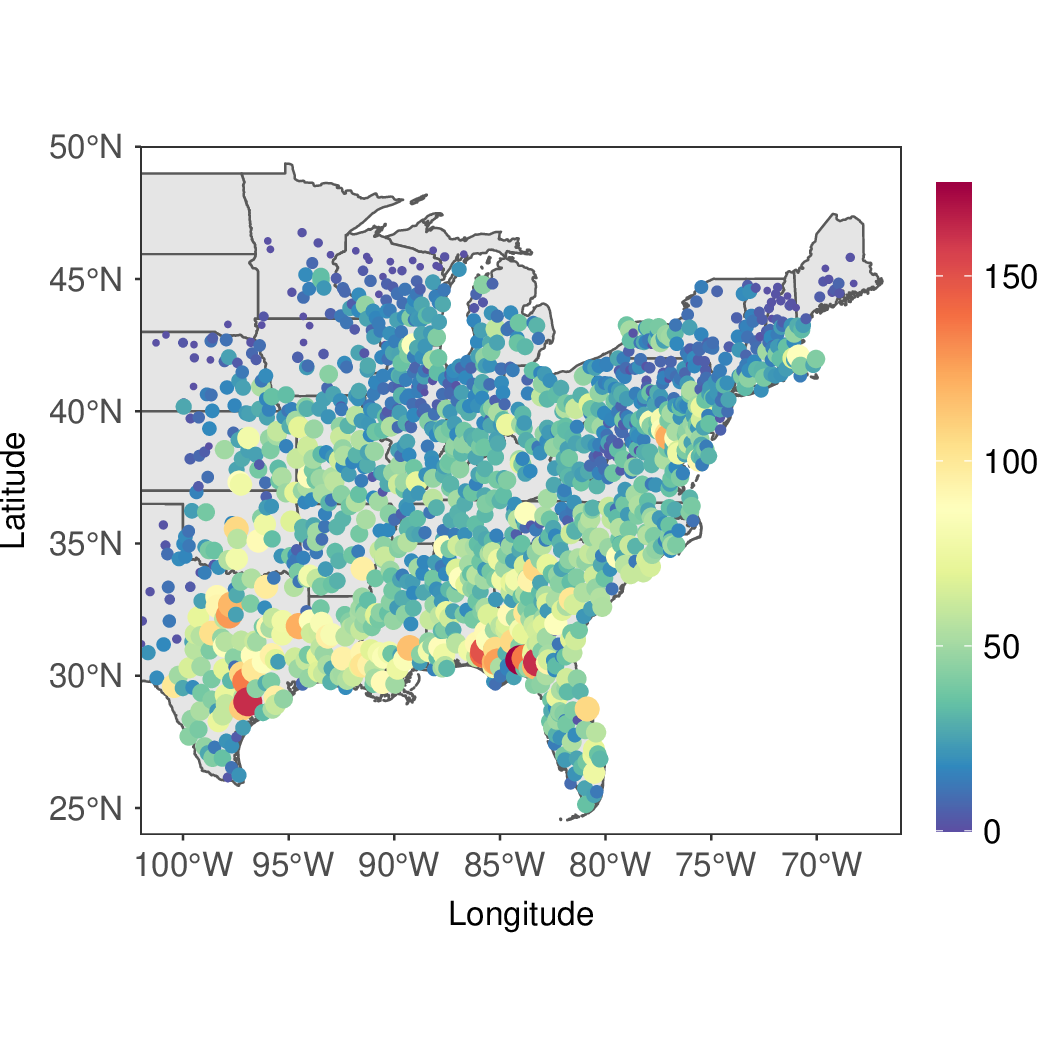}
         \caption{Observed counts}
     \end{subfigure}
    %  \hfill
    \begin{subfigure}[b]{0.4\textwidth}
         \centering
         \includegraphics[width=\textwidth]{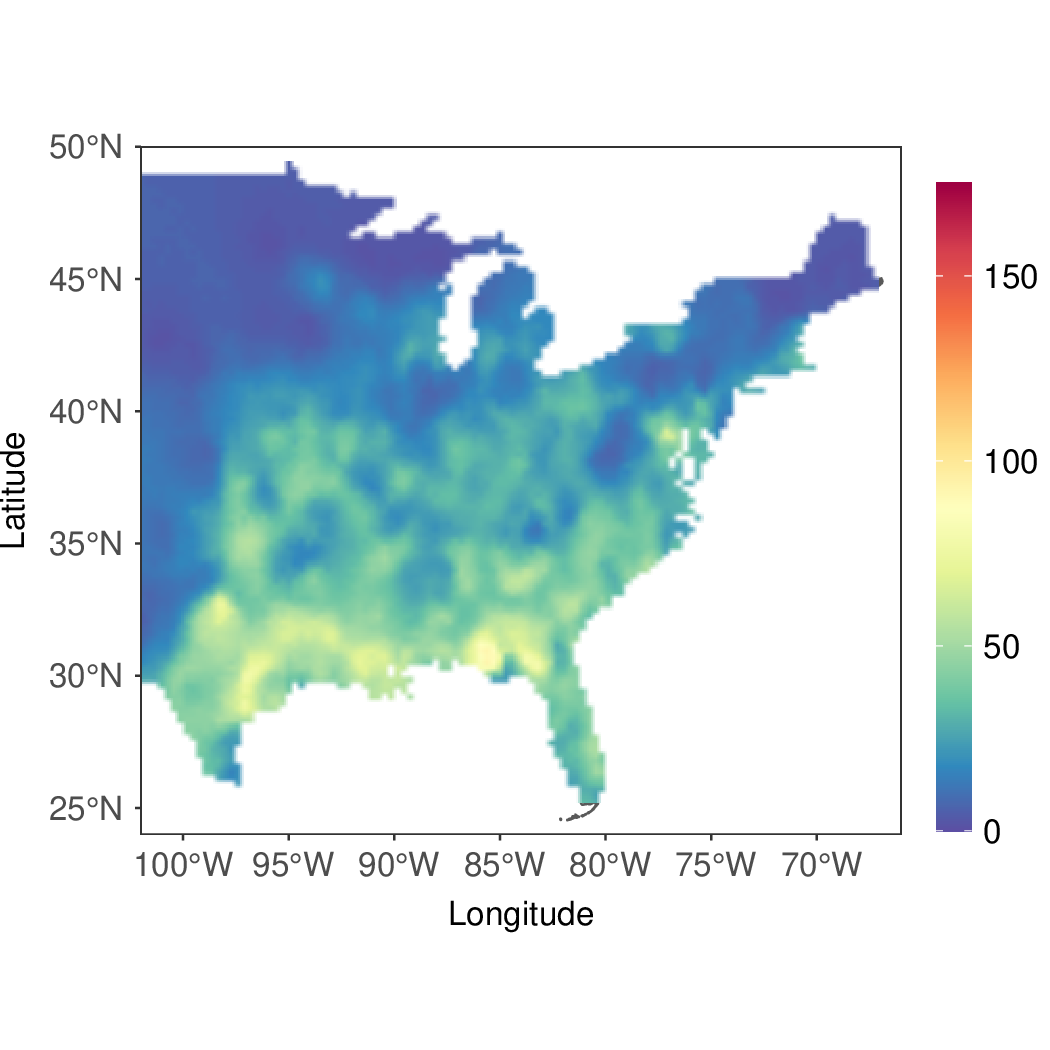}
         \caption{Predicted counts}
     \end{subfigure}     
     \medskip
     \begin{subfigure}[b]{0.4\textwidth}
         \centering
         \includegraphics[width=\textwidth]{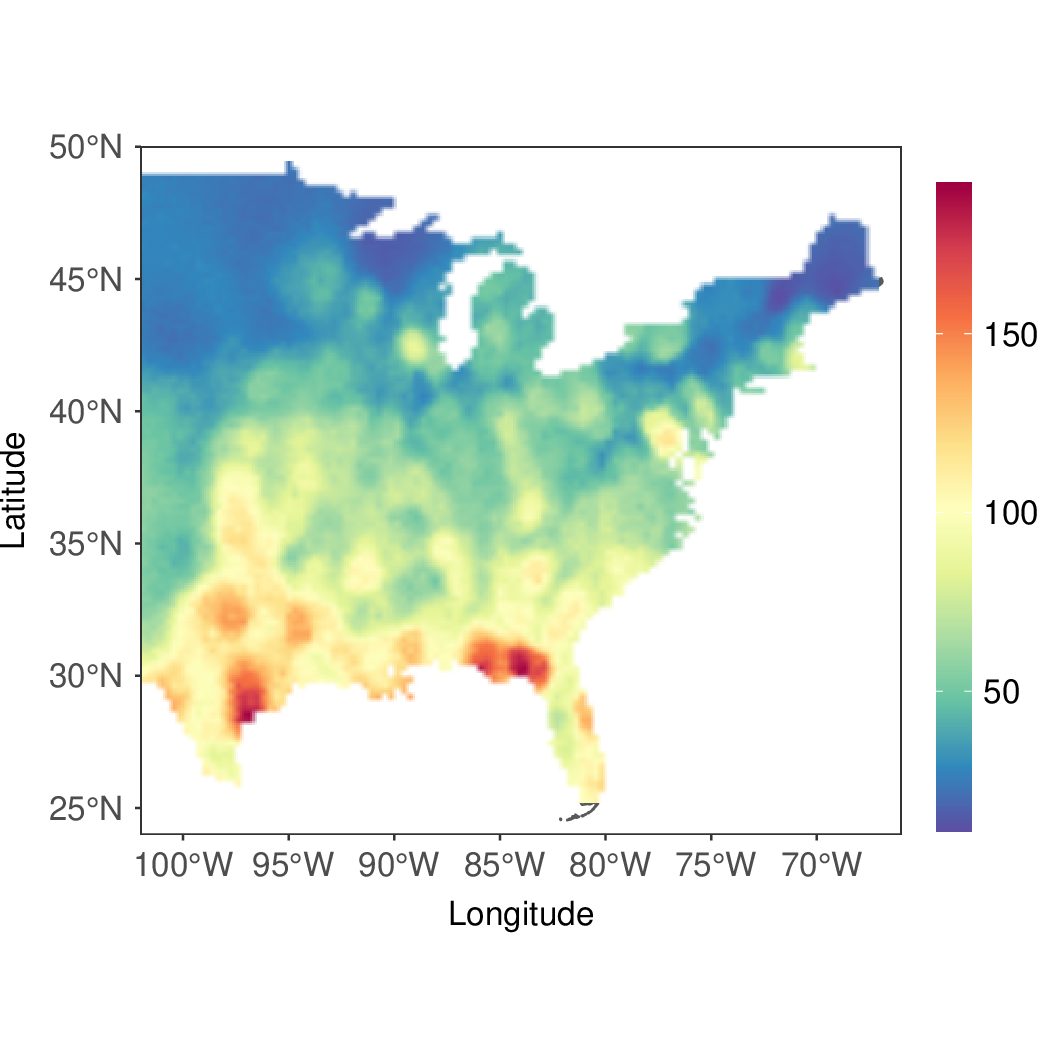}
         \caption{95\% CI widths}
     \end{subfigure}
     \begin{subfigure}[b]{0.4\textwidth}
         \centering
         \includegraphics[width=\textwidth]{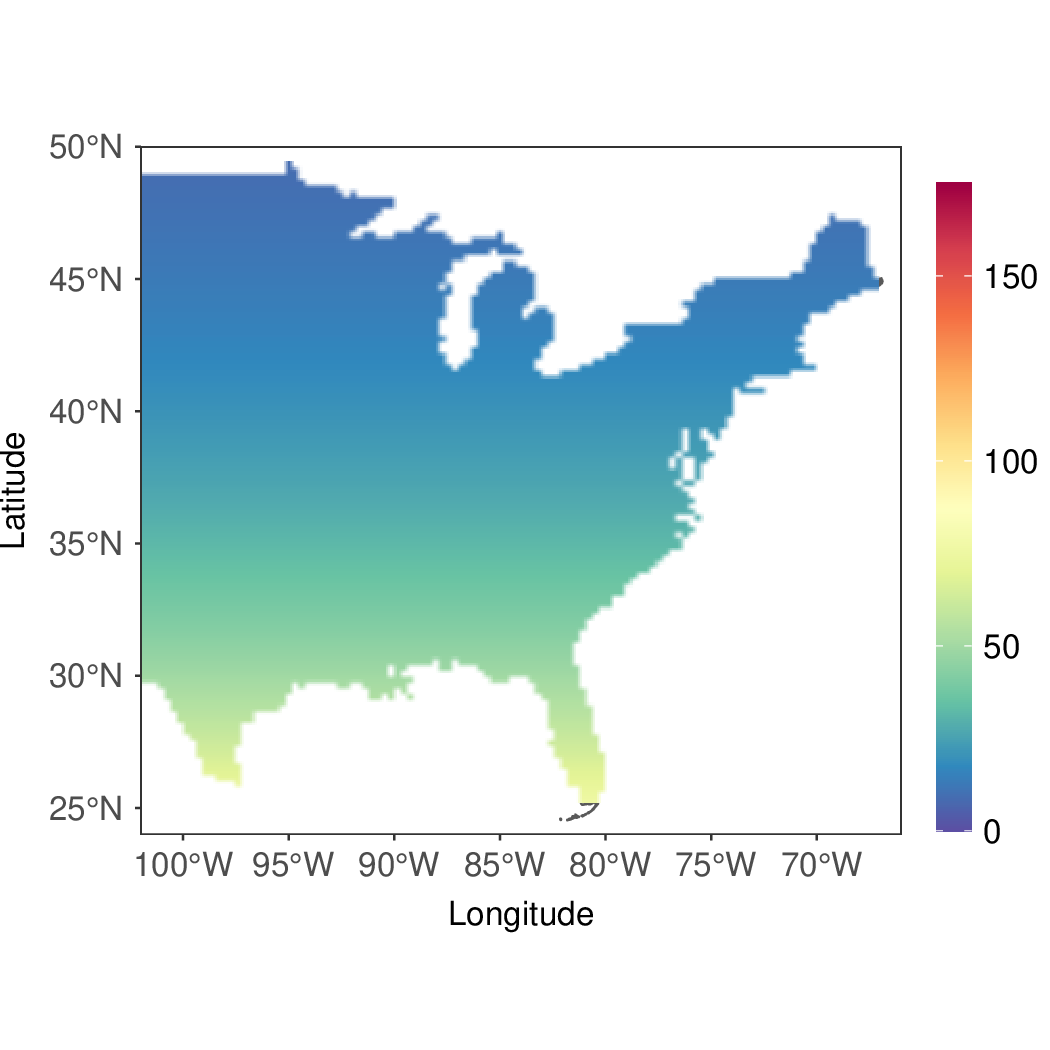}
         \caption{Posterior mean of $\exp(\x(\sv)^\top\bbeta)$}
     \end{subfigure}    
\caption{
North American Breeding Bird Survey data analysis:
(a) observed counts for 2019 BBS of Northern Cardinal, with circle radius proportional 
to the observed counts; 
(b) median of the posterior predictive distribution for Northern Cardinal count;
    (c) widths of the 95\% CI of the posterior predictive distribution for Northern Cardinal count;
    (d) posterior mean of $\exp(\x(\sv)^\top\bbeta)$.
    }
    \label{fig:bird}
\end{figure}

We are interested in the relative abundance of the Northern Cardinal, a bird 
species that is prevalent in Eastern United States. Figure \ref{fig:bird}(a) shows
the number of birds observed in 2019, with the sizes of the circle radii proportional
to the number of birds at each sampling location. The dataset was extracted with the help of the 
R package \textit{bbsAssistant} \citep{burnett2019bbsassistant}; it contains 1515 irregular
sampling locations. From Figure \ref{fig:bird}(a) we observe that the counts tend 
to increase as latitude decreases, and we thus take latitude as a covariate to account 
for the long range variability in the population.

We considered the Gaussian copula NBNNMP model defined in Example \ref{ex:nb},
with spatially varying marginal
$\mathrm{NB}(\exp(\x(\sv)^\top\bbeta),r)$, where $\bbeta = (\beta_0,\beta_1)^\top$.
We used the same link functions and prior specifications as in Section \ref{sec: sim2}. 
We first examined model performance under different values of $L$. Overall, parameter 
estimates were quite robust. The estimates of mixture weights suggested
that the effective number of neighbors for each location 
was quite consistent for $L$ between $10$ and $20$. Also, there was no discernible
differences for out-of-sample predictive performance. Therefore, we 
took $L = 20$ as a reasonable upper bound.
We also compared NBNNMP models with the three copulas listed in Table \ref{tbl:copula},
using the same link functions for copulas as in Section \ref{sec: sim1}. 
The three models were evaluated based on their predictive performance. 
Overall, the Gaussian copula outperformed the other two. Details of these 
analyses are provided in the supplementary material.

\begin{figure}[t!]
    \centering
    \begin{subfigure}[b]{0.32\textwidth}
         \centering
         \includegraphics[width=\textwidth]{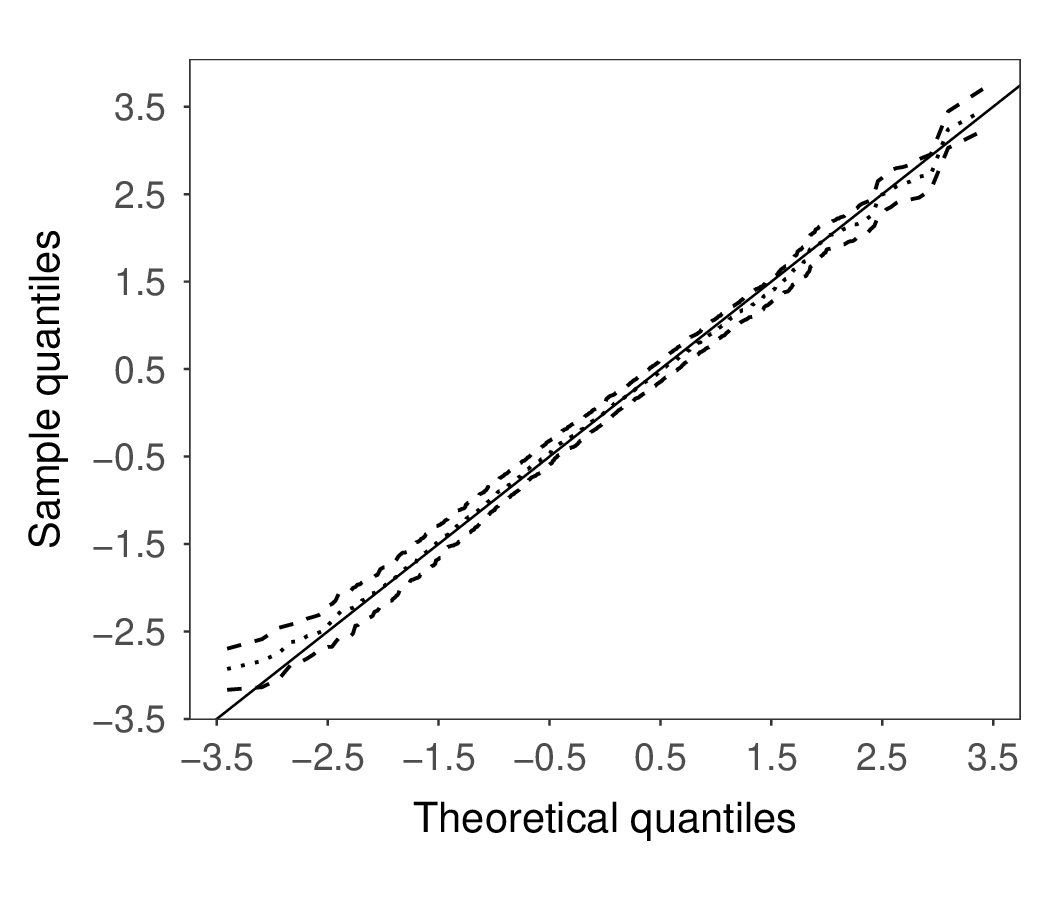}
         \caption{Quantile-quantile plot}
     \end{subfigure}
    %  \hfill
    \begin{subfigure}[b]{0.32\textwidth}
         \centering
         \includegraphics[width=\textwidth]{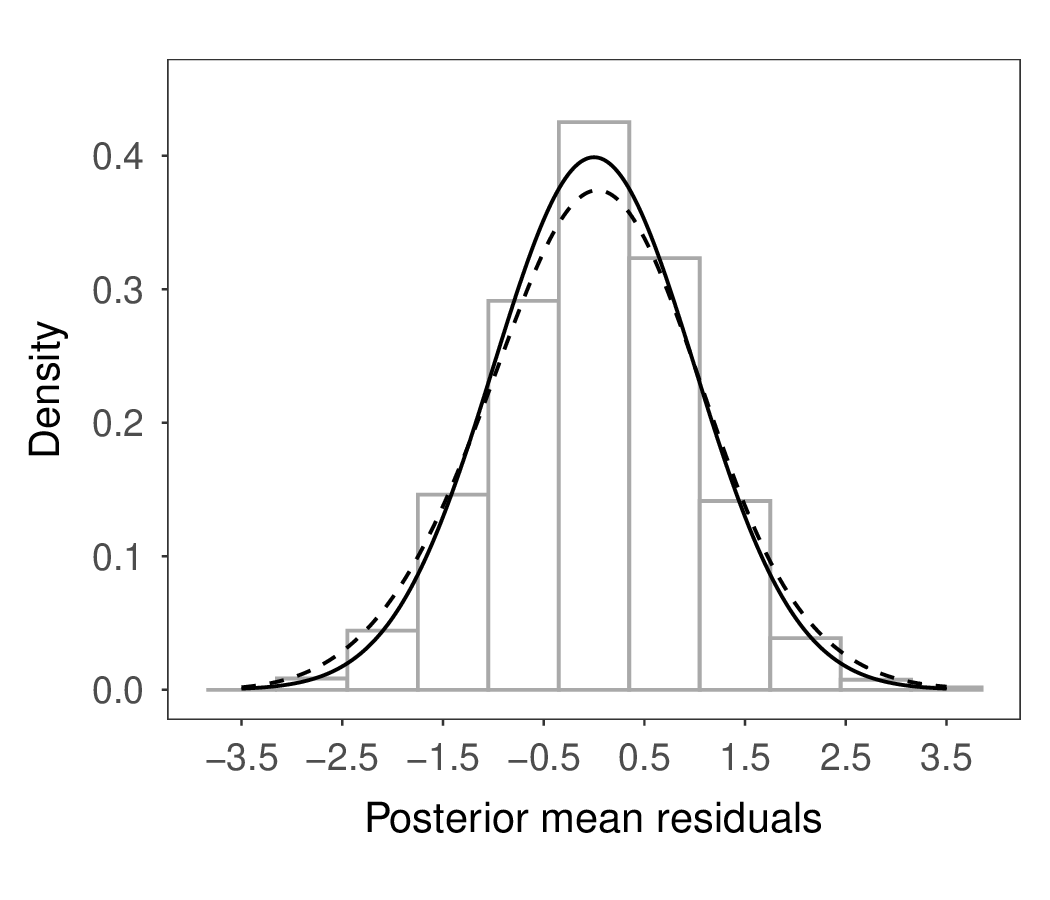}
         \caption{Histogram}
     \end{subfigure}    
     \begin{subfigure}[b]{0.32\textwidth}
         \centering
         \includegraphics[width=\textwidth]{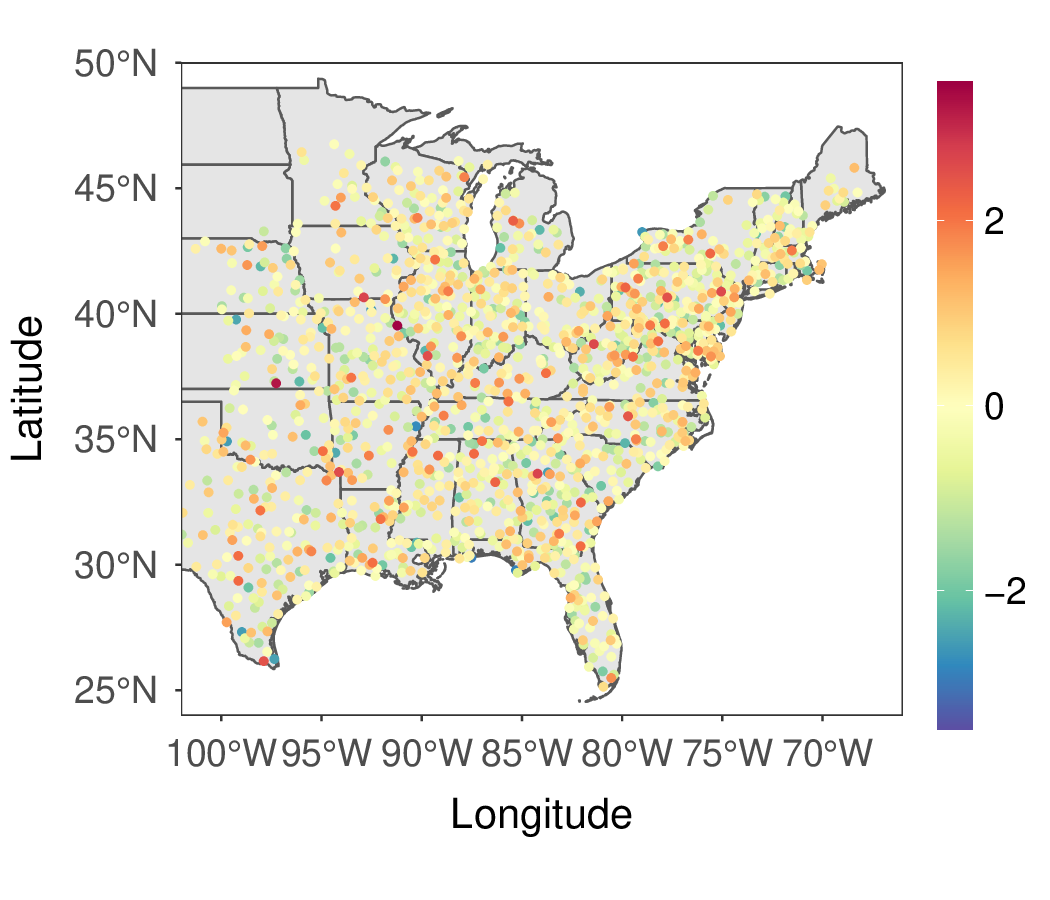}
         \caption{Posterior mean residuals}
     \end{subfigure}
\caption{
North American Breeding Bird Survey data analysis.
Randomized quantile residual analysis:
(a) dotted and dashed lines correspond to the posterior mean and 95\% interval 
bands, respectively; (b) solid and dashed lines are the standard Gaussian 
density and the kernel density estimate of the posterior means of the residuals, 
respectively; (c) spatial plot of the posterior means of the residuals.
}
    \label{fig:bird2}
\end{figure}

We proceeded to analyze the BBS data with the Gaussian copula NBNNMP model with $L = 20$.
The posterior mean and 95\% CI estimates of the regression parameters $\beta_0$ and $\beta_1$
are $6.53\,(5.61, 7.38)$ and $-0.09\,(-0.11, -0.06)$, respectively, 
suggesting an increasing trend in the Northern Cardinal counts as the latitude decreases.
The corresponding estimates of the dispersion parameter $r$ are $1.88\,(1.55, 2.22)$,
indicating that there is overdispersion over the domain. Figure \ref{fig:bird}(b) and 
\ref{fig:bird}(c) show the posterior predictive median of the counts and the $95\%$ 
posterior predictive CI width, respectively. 
Figure \ref{fig:bird}(b) displays the domain's spatial variability.
The estimated uncertainty, as shown in Figure \ref{fig:bird}(c), is 
meaningful, as areas with high uncertainty correspond to those where the observed 
counts are quite heterogeneous. Figure \ref{fig:bird}(d) provides a spatial map of 
the mean of the negative binomial marginals, which depicts a 
North--South trend. Model checking results are shown in Figure \ref{fig:bird2},
including a posterior summary of the Gaussian quantile-quantile plot, and the histogram and 
spatial plot of the posterior means of the residuals. The results suggest good model fit.

Finally, we compared the NBNNMP with the SGLMM-GP model (details are given in the 
supplementary material). 
The parameter estimates of $\bbeta$ were quite close under the two models.
On the other hand, the NBNNMP model resulted in better out-of-sample predictive 
performance, and, notably, it was substantially more efficient to implement, 
with computing time 110 times faster than that for the SGLMM-GP model.

\section{Discussion}

We have introduced a new class of models for discrete geostatistical data, with
particular focus on using different families of bivariate copulas to build modeling
and inference. Compared to traditional SGLMM methods, the proposed class of models 
is scalable, and is able to accommodate complex dependence structures.

In general, multivariate discrete distributions are not as tractable
as certain families of multivariate continuous distributions, in particular, the Gaussian 
family. This is the fundamental difficulty of process-based modeling for discrete 
geostatistical data. Our methodology overcomes this difficulty 
through a structured mixture model formulation, reducing the specification of a 
multivariate pmf to that of bivariate copulas that define the 
mixture components. This formulation yields models for spatial processes that provide 
flexibility and deliver computational scalability.

In the present work, we explored the strategy of using 
a single copula family for all bivariate distributions.
Exploring the alternative which builds from different copula families 
for the bivariate distributions remains an interesting question to
investigate. We can cast this as a model selection problem
and develop algorithms to select models;
see examples in \cite{panagiotelis2017model} and \cite{gruber2018bayesian}
in the context of regular vine copula models.
Different copula families for bivariate distributions yield more flexibility 
for the model to capture complex dependence, albeit at the cost of computational scalability.
If the main purpose of the application is prediction, rather than model selection, 
one could explore calibrating the prediction using all candidate copula families. 
This could be done, for example, with the pseudo Bayesian model averaging approach,
where the weight for each model is estimated based on stacking \citep{yao2018using}.

We conduct inference for the discrete copula NNMPs based on the continuous extension approach.
Apart from the aforementioned benefits, this approach may allow discrete copula NNMPs 
to make use of alternative algorithms for faster computation, 
which are currently being developed for continuous NNMP models.
Moreover, with the CE approach, it is possible to develop a class of 
NNMPs for a multivariate response that consists of both continuous and discrete components, 
while at the same time retaining computational efficiency.

\section*{Supplementary Material}

The supplementary material includes proofs and other technical details, sampling algorithm details, and additional 
results on the data examples of Section \ref{sec: sim1} and \ref{sec:bbs}.

\bibliographystyle{jasa3} 
\bibliography{ref}

% \newpage
\clearpage\pagebreak\newpage
%%%%%%%%%% Merge with supplemental materials %%%%%%%%%%

\spacingset{1.5} 

\begin{center}
\LARGE\bf Supplementary Material for ``Bayesian Geostatistical Modeling for Discrete-Value Processes"
\end{center}
% \section*{\hfil \LARGE\bf Supplementary Material\hfil}
%%%%%%%%%% Prefix a "S" to all equations, figures, tables and reset the counter %%%%%%%%%%
\setcounter{section}{0}
\setcounter{equation}{0}
\setcounter{figure}{0}
\setcounter{table}{0}
% \setcounter{page}{1}
% \makeatletter
% \renewcommand{\theequation}{S\arabic{equation}}
% \renewcommand{\thefigure}{S\arabic{figure}}
% \renewcommand{\bibnumfmt}[1]{[S#1]}
% \renewcommand{\citenumfont}[1]{S#1}
%%%%%%%%%% Prefix a "S" to all equations, figures, tables and reset the counter %%%%%%%%%%

\renewcommand{\thesection}{\Alph{section}}  

\section{Proof of Proposition 2}

\begin{proof}
Consider a discrete copula NNMP characterized by
$$
p(y(\sv)\,|\,\y_{\tNe(\sv)})=\sum_{l=1}^Lw_l(\sv)\,c_{\sv,l}(y(\sv),y(\sv_{(l)}))g_{\sv}(y(\sv)),
$$
where $g_{\sv}$ is the marginal pmf of $Y(\sv)$.

Let $\y_{\V} = (y(\bs_1),\dots,y(\bs_n), y(\su_1),\dots,y(\su_m))^\top$ for $n\geq 2$ and $m\geq 1$,
where $\V = \BS\cup\U$, $\BS = \{\bs_1,\dots,\bs_n\}$, 
$\U = \{\su_1,\dots,\su_m\}$, and $\BS\cap\U=\emptyset$.
The joint pmf of $\y_{\V}$ can be written as $\tilde{p}(\y_{\V}) = \tilde{p}(\by_{\U}\,|\,\by_{\BS})\tilde{p}(\by_{\BS})$. We will first derive
the joint pmf $\tilde{p}(\by_{\BS}) = \tilde{p}(y(\bs_1),\dots,y(\bs_n))$ and then $\tilde{p}(\by_{\U}\,|\,\by_{\BS})$, where $\by_{\U} = (y(\su_1),\dots,y(\su_m))^\top$.

Let $c_{\bs_i,l_i}\equiv c_{\bs_i,l_i}(y(\bs_i),y(\bs_{(i,l_i)}))$ and
$w_{\bs_i,l_i}\equiv w_{l_i}(\bs_i)$ with $l_i = 1,\dots,i_L$ and $i_L = (i-1)\wedge L$, 
for all $i$. Then
$$
\tilde{p}(y(\bs_1),y(\bs_2)) = p(y(\bs_2)\,|\,y(\bs_1))g_{\bs_1}(y(\bs_1)) = 
c_{\bs_2,1}g_{\bs_2}(y(\bs_2))g_{\bs_1}(y(\bs_1)). 
$$
Note that by definition of the discrete NNMP, $w_{\bs_2,1} = 1$. Then
$$
\begin{aligned}
\tilde{p}(y(\bs_1),y(\bs_2),y(\bs_3)) 
& = p(y(\bs_3)\,|\,y(\bs_1),y(\bs_2))\tilde{p}(y(\bs_1),y(\bs_2))\\
& = \left(\sum_{l_3=1}^2w_{\bs_3,l_3}c_{\bs_3,l_3}g_{\bs_3}(y(\bs_3))\right)
c_{\bs_2,1}g_{\bs_2}(y(\bs_2))g_{\bs_1}(y(\bs_1))\\
& = \prod_{i=1}^3g_{\bs_i}(y(\bs_i))
\sum_{l_3=1}^2w_{\bs_3,l_3}c_{\bs_3,l_3}c_{\bs_2,1}\\
& = \prod_{i=1}^3g_{\bs_i}(y(\bs_i))
\sum_{l_3=1}^{2}\sum_{l_2=1}^{1}w_{\bs_3,l_3}w_{\bs_2,l_2}c_{\bs_3,l_3}c_{\bs_2,l_2}.
\end{aligned}
$$
Similarly, for $4\leq n\leq L$, the joint pmf is
$$
\begin{aligned}
& \tilde{p}(y(\bs_1),\dots,y(\bs_n))\\
& = p(y(\bs_n)\,|\,\y_{\tNe(\bs_n)})\tilde{p}(y(\bs_1),\dots,y(\bs_{n-1}))\\
& = \left(\sum_{l_n=1}^{n-1}w_{\bs_n,l_n}c_{\bs_n,l_n}g_{\bs_n}(y(\bs_n))\right)
\left(\prod_{i=1}^{n-1}g_{\bs_i}(y(\bs_i))
\sum_{l_{n-1}=1}^{n-2}\dots\sum_{l_2=1}^1w_{\bs_{n-1},l_{n-1}}\dots w_{\bs_2,l_2}c_{\bs_{n-1},l_{n-1}}\dots c_{\bs_2,l_2}\right)\\
& = \prod_{i=1}^{n}g_{\bs_i}(y(\bs_i))
\sum_{l_{n}=1}^{n-1}\dots\sum_{l_2=1}^1w_{\bs_n,l_n}\dots w_{\bs_2,l_2}c_{\bs_{n},l_{n}}\dots c_{\bs_2,l_2}.
\end{aligned}
$$
Finally, for $n > L$, it is easy to show that the joint pmf is
$$
\begin{aligned}
& \tilde{p}(y(\bs_1),\dots,y(\bs_n))\\ 
& = p(y(\bs_n)\,|\,\y_{\tNe(\bs_n)})\tilde{p}(y(\bs_1),\dots, y(\bs_{n-1}))\\
& = \left(\sum_{l_n=1}^Lw_{\bs_n,l_n}c_{\bs_n,l_n}g_{\bs_n}(y(\bs_n))\right)\\
& \;\;\;\;\;\;\prod_{i=1}^{n-1}g_{\bs_i}(y(\bs_i))
\sum_{l_{n-1}=1}^{L}\dots\sum_{l_{L+1}=1}^{L}\sum_{l_L=1}^{L-1}\dots\sum_{l_2=1}^1
w_{\bs_{n-1},l_{n-1}}\dots w_{\bs_2,l_2}c_{\bs_{n-1},l_{n-1}}\dots c_{\bs_2,l_2}\\
& = \prod_{i=1}^{n}g_{\bs_i}(y(\bs_i))
\sum_{l_{n}=1}^{L}\dots\sum_{l_{L+1}=1}^{L}\sum_{l_L=1}^{L-1}\dots\sum_{l_2=1}^1
w_{\bs_n,l_n}\dots w_{\bs_2,l_2}c_{\bs_{n},l_{n}}\dots c_{\bs_2,l_2}.
\end{aligned}
$$
Therefore, we have that, for $n\geq 2$, the joint pmf
$$
\tilde{p}(\by_{\BS}) = \tilde{p}(y(\bs_1),\dots,y(\bs_n)) = \prod_{i=1}^{n}g_{\bs_i}(y(\bs_i))
\sum_{l_{n}=1}^{n_L}\dots\sum_{l_2=1}^{2_L}w_{\bs_n,l_n}\dots w_{\bs_2,l_2}c_{\bs_{n},l_{n}}\dots c_{\bs_2,l_2}.
$$

Turning to the non-reference set $\U$.
Let $c_{\su_i,l_i}\equiv c_{\su_i,l_i}(y(\su_i),y(\su_{(i,l_i)}))$ and
$w_{\su_i,l_i}\equiv w_{l_i}(\su_i)$ with $l_i = 1,\dots,L$, for all $i$.
When $m = 1$, $\tilde{p}(\by_{\U}\,|\,\by_{\BS}) = p(y(\su_1)\,|\,\by_{\tNe(\su_1)})$.

When $m\geq2$, without loss of generality, we consider the case of $m = 2$, i.e., 
we take $\U = \{\su_1,\su_2\}$. Then we have that
$$
\begin{aligned}
p(\y_{\U}\,|\,\y_{\BS}) & = p(y(\su_1)\,|\,\y_{\tNe(\su_1)})p(y(\su_2)\,|\,\y_{\tNe(\su_2)})\\
& = \left(\sum_{l_1=1}^{L}w_{\su_1,l_1}c_{\su_1,l_1}g_{\su_1}(y(\su_1))\right)
\left(\sum_{l_2=1}^{L}w_{\su_2,l_2}c_{\su_2,l_2}g_{\su_2}(y(\su_2))\right)\\
& = \prod_{i=1}^2g_{\su_i}(y(\su_i))\sum_{l_2=1}^L\sum_{l_1=1}^Lw_{\su_2,l_2}w_{\su_1,l_1}
c_{\su_2,l_2}c_{\su_1,l_1}.
\end{aligned}
$$
Obviously, the result is easily generalized for $\U = \{\su_1,\dots,\su_m\}$ for any
$m > 2$.
\end{proof}

\section{Gaussian, Gumbel, and Clayton copulas}

We introduce properties of the Gaussian, Gumbel and Clayton copulas that are useful 
for the discrete copula NNMP's model estimation and prediction.
For more details we refer to Joe (2014).
Consider a bivariate vector $(X_1,X_2)$
with marginal cumulative distribution functions (cdfs) 
such that $F_1(x_1) = t_1$ and $F_2(x_2) = t_2$.

\paragraph{\textbf{Gaussian copula}}
A Gaussian copula with correlation $\rho\in(0,1)$ for $(X_1,X_2)$ is 
$$
C(F_1(x_1), F_2(x_2)\,|\,\rho) = C(t_1,t_2\,|\,\rho) = \Phi_2(\Phi^{-1}(t_1),\,\Phi^{-1}(t_2)\,|\,\rho).
$$
If both $X_1$ and $X_2$ are continuous random variables, the copula has density
$$
\frac{1}{\sqrt{1-\rho^2}}\exp\left(\frac{2\rho\Phi^{-1}(t_1)\Phi^{-1}(t_2) - 
\rho^2\{(\Phi^{-1}(t_1))^2 + (\Phi^{-1}(t_2))^2)\}}{2(1-\rho^2)}\right).
$$
The conditional cdf of $T_1$ given $T_2 = t_2$, denoted as $C_{1|2}(t_1\,|\,t_2)$, is given by
$$
C_{1|2}(t_1\,|\,t_2) = \frac{\partial C(t_1,t_2)}{\partial t_2} = 
\Phi\left(\frac{\Phi^{-1}(t_1) - \rho\Phi^{-1}(t_2)}{\sqrt{1-\rho^2}}\right).
$$
To simulate $X_1$ given $X_2 = x_2$, we first compute $t_2 = F_2(x_2)$.
We then generate a random number $z$ from a uniform distribution on $[0,1]$, and compute $t_1 = C_{1|2}^{-1}(z\,|\,t_2)$ where
$C_{1|2}^{-1}(z\,|\,t_2) = \Phi\left(\sqrt{(1-\rho^2)}\Phi^{-1}(z) + \rho\Phi^{-1}(t_2)\right)$ is the inverse of $C_{1|2}(t_1\,|\,t_2)$.
Finally, we obtain $x_1$ from the inverse cdf $F_1^{-1}(t_1)$.

\paragraph{\textbf{Gumbel copula}}
A Gumbel copula with parameter $\eta\in[1,\infty)$ for $(X_1,X_2)$ is 
$$
C(F_1(x_1), F_2(x_2)\,|\,\eta) = C(t_1,t_2\,|\, \eta) = \exp(-((-\log(t_1))^{\eta} + (-\log(t_2))^{\eta})^{1/\eta}).
$$
Let $u_1 = -\log(t_1)$ and $u_2 = -\log(t_2)$. 
If both $X_1$ and $X_2$ are continuous random
variables, the copula has density
$$
\exp(-(u_1^{\eta}+u_2^{\eta})^{1/\eta})((u_1^{\eta}+u_2^{\eta})^{1/\eta} + \eta - 1)
(u_1^{\eta}+u_2^{\eta})^{1/\eta-2}(u_1u_2)^{\eta-1}(t_1t_2)^{-1}.
$$
The conditional cdf of $T_1$ given $T_2 = t_2$ is
$$
C_{1|2}(t_1\,|\,t_2) = \overline{C}_{1|2}(u_1\,|\,u_2) = t_2^{-1}\exp(-(u_1^{\eta}+u_2^{\eta})^{1/\eta})
(1 + (u_1/u_2)^{\eta})^{1/\eta-1},
$$
where the conditional cdf $\overline{C}_{1|2}(u_1\,|\,u_2)$ corresponds to the copula
$\overline{C}(u_1,u_2\,|\,\eta) = \exp(-(u_1^{\eta} + u_2^{\eta})^{1/\eta})$ 
which is a bivariate exponential survival function, with marginals corresponding
to a unit rate exponential distribution. 
The inverse conditional cdf $C_{1|2}^{-1}(\cdot\,|\,t_2)$ does not have a closed form. 
To generate $X_1$ given $X_2 = x_2$, following Joe (2014),
we first define $y = (u_1^{\eta}+u_2^{\eta})^{1/\eta}$.
Then we have a realization of $X_1$, say 
$x_1 = (y_0^{\eta}-u_2^{\eta})^{1/\eta}$, where $y_0$ is the root of 
$h(y) = y + (\eta-1)\log(y) - (u_2 + (\eta-1)\log(u_2) - \log z) = 0$,
where $y\geq u_2$, and $z$ is a random number generated from a uniform distribution on $[0,1]$.

\paragraph{\textbf{Clayton copula}}
A Clayton copula with parameter $\delta\in[0,\infty)$ for $(X_1,X_2)$ is 
$$
C(F_1(x_1), F_2(x_2)\,|\,\delta) = C(t_1,t_2\,|\,\delta) = (t_1^{-\delta} + t_2^{-\delta} 
- 1)^{-1/\delta}.
$$
If both $X_1$ and $X_2$ are continuous random
variables, the copula has density
$$
(1+\delta)(t_1t_2)^{-\delta-1}(t_1^{-\delta}+t_2^{-\delta}-1)^{-2-1/\delta}.
$$
The conditional cdf of $T_1$ given $T_2 = t_2$ is
$$
C_{1|2}(t_1\,|\,t_2) = (1 + t_2^{\delta}(t_1^{-\delta}-1))^{-1-1/\delta}.
$$
To simulate $X_1$ given $X_2$, we first compute $t_2 = F_2(x_2)$, and
generate a uniform random number $z$ on $[0,1]$. Then we compute 
$t_1 = C_{1|2}^{-1}(z\,|\,t_2)$ where 
$C_{1|2}^{-1}(z\,|\,t_2) = ((z^{-\delta/(1+\delta)}-1)t_2^{-\delta}+1)^{-1/\delta}$.
Finally, we obtain $x_1$ from the inverse cdf $F_1^{-1}(t_1)$.

\section{Implementation details}

In this section, we introduce necessary posterior simulation steps for the Poisson NNMP (PONNMP)
and negative binomial NNMP (NBNNMP) models illustrated in the data examples. For both models, 
we use an exponential correlation function with range parameter $\phi$ to create spatial copulas.
More specifically, given two different sites $\sv\neq\sv'$,
the link functions for parameters of the Gaussian, Gumbel and Clayton copulas, respectively, are
$$
\begin{aligned}
\rho(||\sv-\sv'||) & = \exp(-||\sv-\sv'||/\phi),\\
\eta(||\sv-\sv'||) & = \min\{(1 - \exp(-||\sv-\sv'||/\phi))^{-1}, 50\},\\
\delta(||\sv-\sv'||) & = \min\{2\exp(-||\sv-\sv'||/\phi)/(1-\exp(-||\sv-\sv'||/\phi)), 98\},
\end{aligned}
$$
where the upper bounds 50 and 98 for Gumbel and Clayton copulas are chosen for numerical stability.
When $\eta(d_0) = 50$, $\exp(-d_0/\phi) = 0.98$. 
Similarly, when $\delta(d_0) = 98$, $\exp(-d_0/\phi) = 0.98$.
Both link functions imply that given $\phi$, the dependence implied by the copulas stays the same 
for any distance between $\sv$ and $\sv'$ smaller than $d_0$. 

We assume that $\y_{\BS} = (y(\bs_1),\dots,y(\bs_n))^\top$ is a vector of observations,
where $\BS = \{\bs_1,\dots,\bs_n\}$ is the reference set. Each $y(\bs_i)$ is associated
with $y^*(\bs_i)$ such that $y^*(\bs_i) = y(\bs_i)- o_i$, where $o_i\equiv o(\bs_i)$,
$o(\bs_i)\stackrel{i.i.d.}{\sim}\mathrm{Unif}(0,1)$, for $i = 1,\dots,n$.
The auxiliary variables $o_i$ is independent of $y(\bs_i)$ and of $o_j$ for $j\neq i$.
Let $\y^*_{\tNe(\bs_i)} = (y^*(\bs_{(i1)}),\dots, y^*(\bs_{(i,i_L)}))^\top$ and
$\bm o_{\tNe(\bs_i)} = (o(\bs_{(i1)}),\dots, o(\bs_{(i,i_L)}))^\top$, for $i = 2,\dots,n$.

\subsection{Poisson NNMP models and inference}

The conditional density of the continued Poisson NNMP (PONNMP) over the reference set is given by
$$
p(y^*(\bs_i)\,|\,\by^*_{\tNe(\bs_i)}, o(\bs_i), \bm o_{\tNe(\bs_i)}) 
= \sum_{l=1}^{i_L}w_l(\bs_i)\,c^*_{\bs_i,l}(y^*(\bs_i),y^*(\bs_{(il)}))f^*_Y(y^*(\bs_i)),
$$
for $i = 2,\dots,n$, where $f^*_Y(y^*(\bs_i)) = f_Y([y^*(\bs_i)+1])$,
and $f_Y$ is a Poisson distribution with rate parameter $\lambda$.
The component $c^*_{\bs_i,l}$ is the copula density of a spatial copula. 
We will illustrate the posterior inference using the Gaussian case as an example.
The copula density of the spatial Gaussian copula is given by
$$
\begin{aligned}
& c^*_{\bs_i,l}(y^*(\bs_i),y^*(\bs_{(il)})) = \\ 
&\;\;\;\;\;\;\;\;\;\frac{1}{\sqrt{1-(\rho_l(\bs_i))^2}}\exp\left(\frac{2\rho(\bs_i)
\Phi^{-1}(q_i)\Phi^{-1}(q_{il}) - 
(\rho_l(\bs_i))^2\{(\Phi^{-1}(q_i))^2 + (\Phi^{-1}(q_{il}))^2\}}{2(1-(\rho_l(\bs_i))^2)}\right),
\end{aligned}
$$
where $\rho_l(\bs_i) \equiv \rho(||\bs_i-\bs_{(il)}'||) = \exp(-||\bs_i-\bs_{(il)}||/\phi)$,
$q_i = F^*_Y(y^*(\bs_i))$, $q_{il} = F^*_Y(y^*(\bs_{(il)}))$, 
and $F_Y^*$ is the cdf of $f^*_Y$.

The formulation of the mixture weights allows us to
augment the model with a sequence of auxiliary variables $t_i$, $i = 3,\dots, n$, 
where $t_i$ is a Gaussian random variable with mean 
$\mu(\bs_i) = \gamma_0 + s_{i1}\gamma_1 + s_{i2}\gamma_2$ and variance $\kappa^2$. 
The conditional density of the augmented model on $y^*(\bs_i)$ is
$$
p(y^*(\bs_i)\,|\,\by^*_{\tNe(\bs_i)}, o(\bs_i), \bm o_{\tNe(\bs_i)}) 
= \sum_{l=1}^{i_L}c^*_{\bs_i,l}(y^*(\bs_i),y^*(\bs_{(il)}))f^*_Y(y^*(\bs_i))
\mathbbm{1}_{(r^*_{\bs_i,l-1},r^*_{\bs_i,l})}(t_i),
$$
for $i = 3,\dots,n$, where $r^*_{\bs_i,l} = \log(r_{\bs_i,l}/(1-r_{\bs_i,l}))$ for $l = 1,\dots, i_L$.
The random cutoff points $r_{\bs_i,l}$ is defined such that
$r_{\bs_i, l}-r_{\bs_i, l-1} = k'(\bs_i,\bs_{(il)})/\sum_{l=1}^{i_L}k'(\bs_i,\bs_{(il)})$,
where $k'(\bs_i,\bs_{(il)}) = \exp(||\bs_i - \bs_{(il)}||/\zeta)$.

Let $\bga = (\gamma_0,\gamma_1,\gamma_2)$. 
The Bayesian model is completed with prior specifications for 
parameters $(\lambda, \phi, \zeta, \bga,\kappa^2)$.
Let $f^*_{\bs_i,l}(y^*(\bs_i)\,|\,y^*(\bs_{(il)})) = c^*_{\bs_i,l}(y^*(\bs_i),y^*(\bs_{(il)}))f^*_Y(y^*(\bs_i))$.
With customary prior specifications, the posterior distribution of the parameters 
and latent variables can be written as
$$
\begin{aligned}
& \mathrm{Ga}(\lambda\,|\,u_{\lambda},v_{\lambda})
\times \mathrm{IG}(\phi\,|\,u_{\phi},v_{\phi})\times
\mathrm{IG}(\zeta\,|\,u_{\zeta},v_{\zeta})
\times N(\bga\,|\,\bmu_{\gamma},\bV_{\gamma}) \times \mathrm{IG}(\kappa^2\,|\, u_{\kappa^2},v_{\kappa^2})\\
&\times N(\bm t\,|\,\bD\bga,\,\kappa^2\mathbf{I}_{n-2}))
\times f^*_Y(y(\bs_1) - o_1\,|\,\lambda) \times 
f^*_{\bs_2,1}(y(\bs_2) - o_2\,|\,y(\bs_1) - o_1, \lambda,\phi)\\
&\times \prod_{i=1}^n\mathrm{Unif}(o_i\,|\,0,1)
\times \prod_{i=3}^n\sum_{l=1}^{i_L}
f^*_{\bs_i,l}(y(\bs_i)-o_i\,|\,y(\bs_{(il)})-o_{(il)}, \lambda,\phi)
\mathbbm{1}_{(r^*_{\bs_i,l-1},r^*_{\bs_i,l})}(t_i),
\end{aligned}
$$
where $o_{(il)}\equiv o(\bs_{(il)})$,
the vector $\bm{t} = (t_3,\dots,t_n)^\top$, and the matrix $\bD$ is $(n-2)\times 3$ such that
the $i$th row is $(1,s_{2+i,1},s_{2+i,2})$.

The Monte Carlo Markov chain (MCMC) algorithm to obtain posterior samples 
consists of updates from the posterior full conditional distribution of each 
of $(\lambda, \phi, \zeta, \bga, \kappa^2, \{t_i\}_{i=3}^n, \{o_i\}_{i=1}^n)$.
The posterior full conditional distributions of each of $(\bga, \kappa^2, \{t_i\}_{i=3}^n, \{o_i\}_{i=1}^n)$ are described in the main paper.
We focus on the posterior updates for $(\lambda,\phi,\zeta)$.
Note that there is a set of configuration variables $\{\ell_i\}_{i=3}^n$
in one-to-one correspondence with $t_i$, i.e., 
$\ell_i = l$ if and only if $t_i\in(r_{\bs_i,l-1}^*,r_{\bs_i,l}^*)$, for $l = 1,\dots,i_L$.
We take $\ell_2 = 1$.
The posterior full conditional distributions of $\lambda$ and $\phi$
are proportional to
$\mathrm{Ga}(\lambda\,|\,u_{\lambda},v_{\lambda})
f^*_Y(y(\bs_1)-o_1)\prod_{i=2}^nf^*_{\bs_i,\ell_i}(y(\bs_i)-o_i\,|\,y(\bs_{(i,\ell_i)})-o_{(i,\ell_i)})$
and $\mathrm{IG}(\phi\,|\,u_{\phi},v_{\phi})
\prod_{i=2}^nc^*_{\bs_i,\ell_i}(y(\bs_i)-o_i,y(\bs_{(i,\ell_i)})-o_{(i,\ell_i)})$, respectively. 
For each parameter, we update it on its log scale with a random walk Metropolis step.
To update $\zeta$, we first marginalize out
the latent variables $t_i$ from the joint posterior distribution.
The posterior full conditional distribution of $\zeta$ is proportional to
$\mathrm{IG}(\zeta\,|\,u_{\zeta},v_{\zeta})
\prod_{i=3}^n\{G_{\bs_i}(r_{\bs_i,\ell_i}\,|\,\mu(\bs_i),\kappa^2)
-G_{\bs_i}(r_{\bs_i,\ell_i-1}\,|\,\mu(\bs_i),\kappa^2)\}$.
We update $\zeta$ on its log scale with a random walk Metropolis step.

\subsection{Negative binomial NNMP models and inference}

The conditional density of the continued negative binomial NNMP (NBNNMP) over the reference set is
given by
$$
p(y^*(\bs_i)\,|\,\by^*_{\tNe(\bs_i)}, o(\bs_i), \bm o_{\tNe(\bs_i)}) 
= \sum_{l=1}^{i_L}w_l(\bs_i)c^*_{\bs_i,l}(y^*(\bs_i),y^*(\bs_{(il)}))g_{\bs_i}(y^*(\bs_i)),
$$
for $i = 2,\dots, n$, where $g^*_{\bs_i}(y^*(\bs_i)) = g_{\bs_i}([y^*(\bs_i)+1])$, and 
$g_{\bs_i}$ is a negative binomial distribution with 
mean $\alpha(\bs_i) = \exp(\x(\bs_i)^\top\bbeta)$ and dispersion parameter $r$. 
Similar to the Poisson case, we illustrate the posterior inference using a 
spatial Gaussian copula with copula density given by
$$
\begin{aligned}
& c^*_{\bs_i,l}(y^*(\bs_i),y^*(\bs_{(il)})) = \\ 
&\;\;\;\;\;\;\;\;\;\frac{1}{\sqrt{1-(\rho_l(\bs_i))^2}}
\exp\left(\frac{2\rho(\bs_i)\Phi^{-1}(q_i)\Phi^{-1}(q_{il}) - 
(\rho_l(\bs_i))^2\{(\Phi^{-1}(q_i))^2 + (\Phi^{-1}(q_{(il)}))^2\}}{2(1-(\rho_l(\bs_i))^2)}\right),
\end{aligned}
$$
where $\rho_l(\bs_i) \equiv \rho(||\bs_i-\bs_{(il)}'||) = \exp(-||\bs_i-\bs_{(il)}||/\phi)$,
$q_i = Q_{\bs_i}^*(y^*(\bs_i))$, 
$q_{il} = Q_{\bs_{(il)}}^*(y^*(\bs_{(il)}))$,
and $Q_{\bs_i}^*$ is the cdf of $g_{\bs_i}^*$ for all $\bs_i$.

Similarly, we use an exponential correlation function for 
the cutoff point kernel $k'$, and augment the model with
a set of Gaussian random variables $t_i$ with mean $\mu(\bs_i)$ and $\kappa^2$.
Let $f^*_{\bs_i,l}(y^*(\bs_i)\,|\,y^*(\bs_{(il)})) = c^*_{\bs_i,l}(y^*(\bs_i),y^*(\bs_{(il)}))g^*_{\bs_i}(y^*(\bs_i))$.
With customary prior specifications, the joint posterior distribution is given by
$$
\begin{aligned}
& N(\bbeta\,|\,\bmu_{\beta},\bV_{\beta})\times \mathrm{IG}(r\,|\,u_{r},v_{r})
\times \mathrm{IG}(\phi\,|\,u_{\phi},v_{\phi})\times
\mathrm{IG}(\zeta\,|\,u_{\zeta},v_{\zeta})
\times N(\bga\,|\,\bmu_{\gamma},\bV_{\gamma}) \times \mathrm{IG}(\kappa^2\,|\, u_{\kappa^2},v_{\kappa^2})\\
&\times N(\bm t\,|\,\bD\bga,\,\kappa^2\mathbf{I}_{n-2}))
\times g^*_{\bs_1}(y(\bs_1) - o_1\,|\,\bbeta,r) \times 
f^*_{\bs_2,1}(y(\bs_2) - o_2\,|\,y(\bs_1) - o_1, \bbeta,r,\phi)\\
&\times \prod_{i=1}^n\mathrm{Unif}(o_i\,|\,0,1)
\times \prod_{i=3}^n\sum_{l=1}^{i_L}
f^*_{\bs_i,l}(y(\bs_i) - o_i\,|\,y(\bs_{(il)}) - o_{(il)}, \bbeta,r,\phi)
\mathbbm{1}_{(r^*_{\bs_i,l-1},r^*_{\bs_i,l})}(t_i),
\end{aligned}
$$
where $o_{(il)}\equiv o(\bs_{(il)})$, the vector $\bm{t} = (t_3,\dots,t_n)^\top$, and the matrix $\bD$ is $(n-2)\times 3$ such that
the $i$th row is $(1,s_{2+i,1},s_{2+i,2})$.

The MCMC algorithm to obtain posterior samples 
consists of updates from the posterior full conditional distribution of each 
of $(\bbeta, r, \phi, \zeta, \bga, \kappa^2, \{t_i\}_{i=3}^n, \{o_i\}_{i=1}^n)$.
The posterior full conditional distributions of each of $(\bga, \kappa^2, \{t_i\}_{i=3}^n, \{o_i\}_{i=1}^n)$ are described in the main paper.
We focus on the posterior updates for $(\bbeta,r,\phi,\zeta)$.
Note that there is a set of configuration variables $\{\ell_i\}_{i=3}^n$
in one-to-one correspondence with $t_i$, i.e., 
$\ell_i = l$ if and only if $t_i\in(r_{\bs_i,l-1}^*,r_{\bs_i,l}^*)$, for $l = 1,\dots,i_L$.
We take $\ell_2 = 1$.
The posterior full conditional distributions of $\bbeta$ and $r$
are proportional to
$N(\bbeta\,|\,\bmu_{\beta},\bV_{\beta})
g^*_{\bs_1}(y(\bs_1)-o_1)\prod_{i=2}^nf^*_{\bs_i,\ell_i}(y(\bs_i)-
o_i\,|\,y(\bs_{(i,\ell_i)})-o_{(i,\ell_i)})$ 
and $\mathrm{IG}(r\,|\,u_{r},v_{r})g^*_{\bs_1}(y(\bs_1)-o_1)
\prod_{i=2}^nf^*_{\bs_i,\ell_i}(y(\bs_i)-o_i\,|\,y(\bs_{(i,\ell_i)})-o_{(i,\ell_i)})$, respectively. 
We use a random walk Metropolis step to update $\bbeta$ and $r$ on its log scale, respectively.
The posterior full conditional distribution of $\phi$ is proportional to
$\mathrm{IG}(\phi\,|\,u_{\phi},v_{\phi})
\prod_{i=2}^nc^*_{\bs_i,\ell_i}(y(\bs_i)-o_i,y(\bs_{(i,\ell_i)})-o_{(i,\ell_i)})$.
We update $\phi$ on its log scale with a random walk Metropolis step.
To update $\zeta$, we first marginalize out
the latent variables $t_i$ from the joint posterior distribution.
The posterior full conditional distribution of $\zeta$ is proportional to
$\mathrm{IG}(\zeta\,|\,u_{\zeta},v_{\zeta})
\prod_{i=3}^n\{G_{\bs_i}(r_{\bs_i,\ell_i}\,|\,\mu(\bs_i),\kappa^2)
-G_{\bs_i}(r_{\bs_i,\ell_i-1}\,|\,\mu(\bs_i),\kappa^2)\}$.
We update $\zeta$ on its log scale with a random walk Metropolis step.

\section{Additional simulation and model checking results}

This section presents additional results of the data examples in the main paper.
In particular, Section D.1 corresponds to the first simulation experiment.
Section D.2 investigates the mixture weights and neighborhood sizes of the 
Gaussian copula NBNNMP, compares three discrete copula NBNNMPs, and compares
the Gaussian copula NBNNMP with the SGLMM-GP, for the real data example. 
Section D.3 shows the model checking results.

\subsection{First simulation experiment}

\begin{figure}[t!]
    \centering
    \begin{subfigure}[b]{0.32\textwidth}
         \centering
         \includegraphics[width=\textwidth]{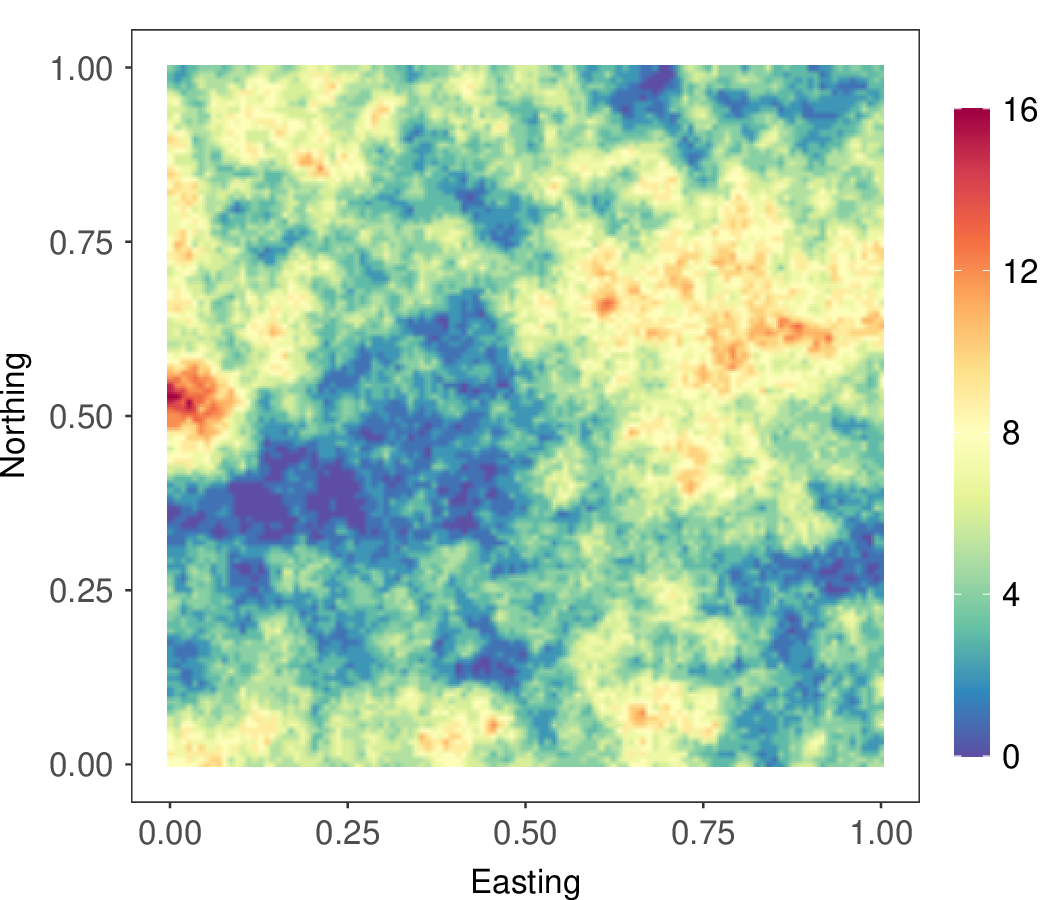}
         \caption{True $y(\sv)$ ($\sigma_1 = 1$)}
     \end{subfigure}
    %  \hfill
    \begin{subfigure}[b]{0.32\textwidth}
         \centering
         \includegraphics[width=\textwidth]{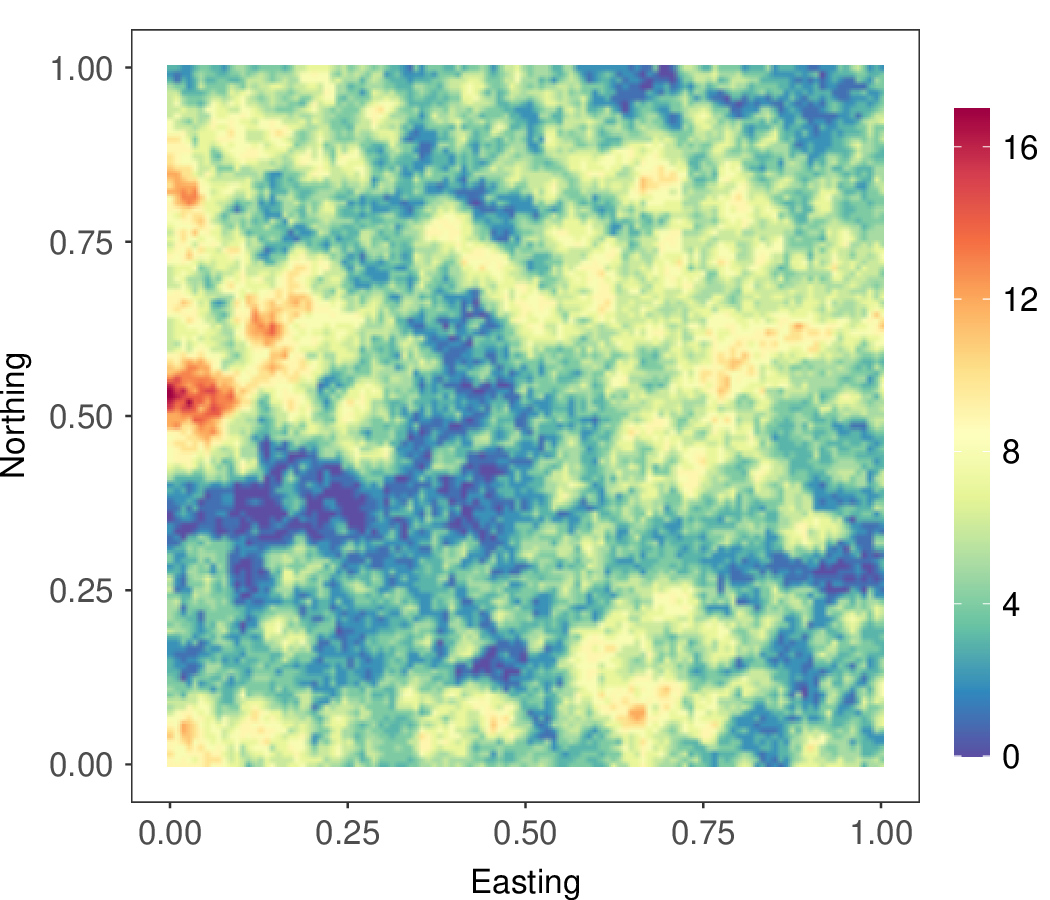}
         \caption{True $y(\sv)$ ($\sigma_1 = 3$)}
     \end{subfigure}     
     \begin{subfigure}[b]{0.32\textwidth}
         \centering
         \includegraphics[width=\textwidth]{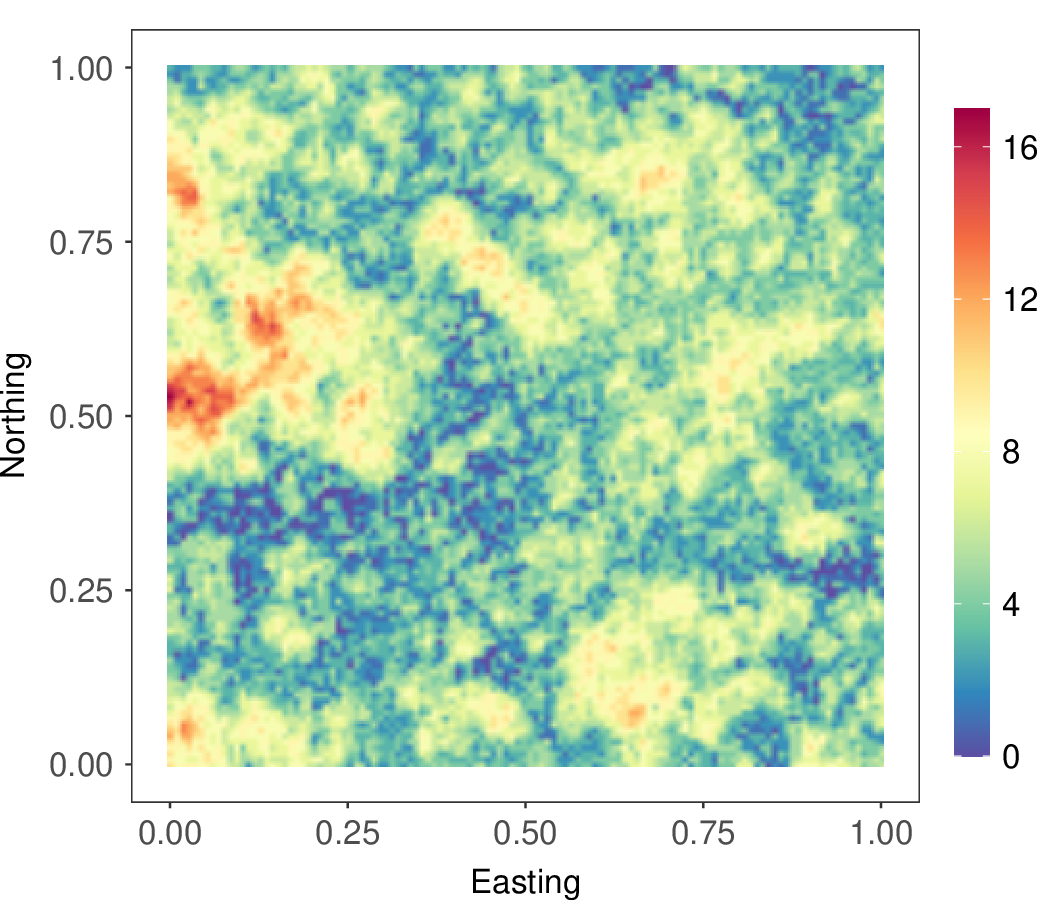}
         \caption{True $y(\sv)$ ($\sigma_1 = 10$)}
     \end{subfigure}
    \begin{subfigure}[b]{0.32\textwidth}
         \centering
         \includegraphics[width=\textwidth]{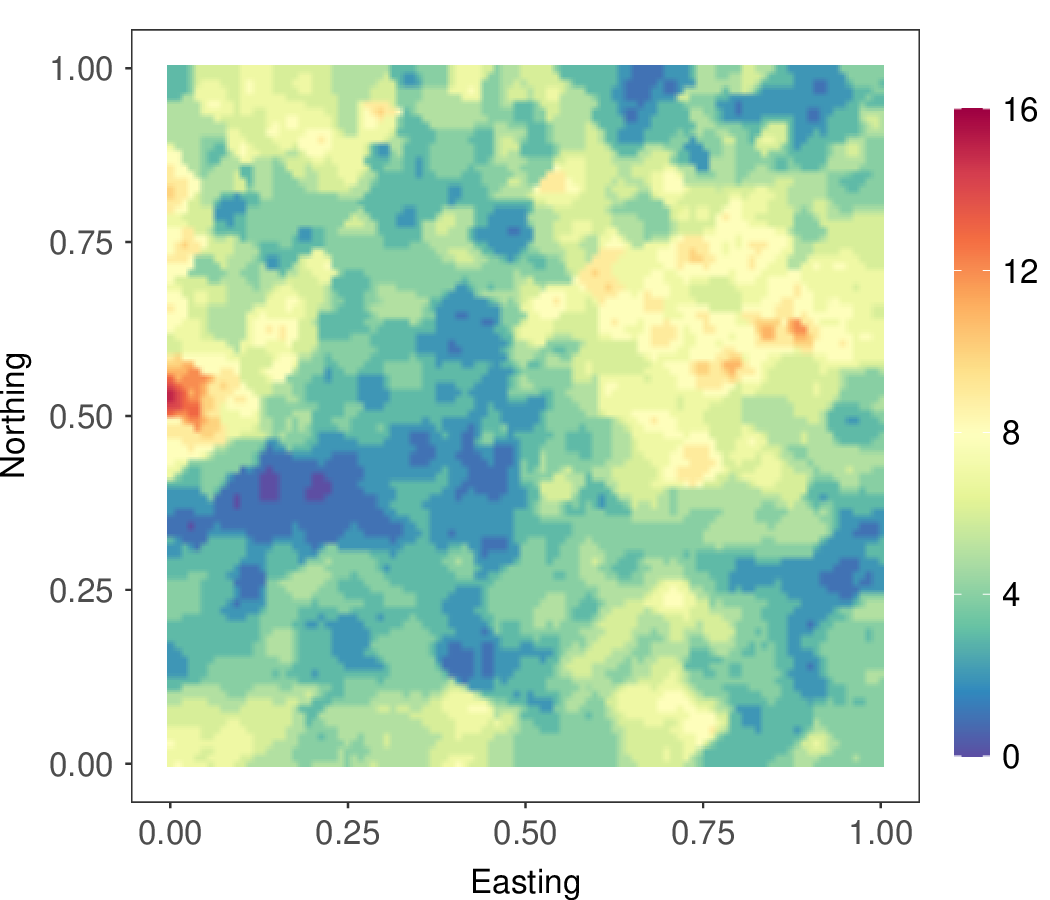}
         \caption{PONNMP (Gaussian)}
     \end{subfigure}
    \begin{subfigure}[b]{0.32\textwidth}
         \centering
         \includegraphics[width=\textwidth]{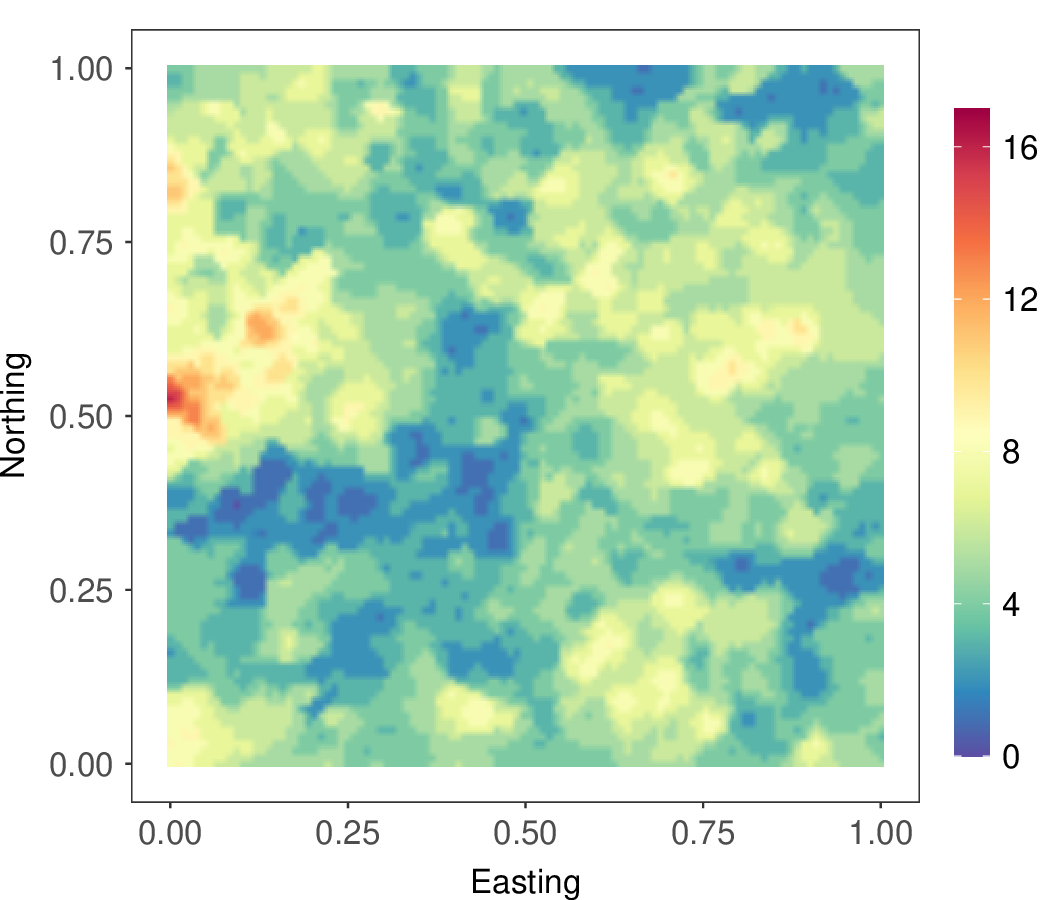}
         \caption{PONNMP (Gaussian)}
     \end{subfigure}
    \begin{subfigure}[b]{0.32\textwidth}
         \centering
         \includegraphics[width=\textwidth]{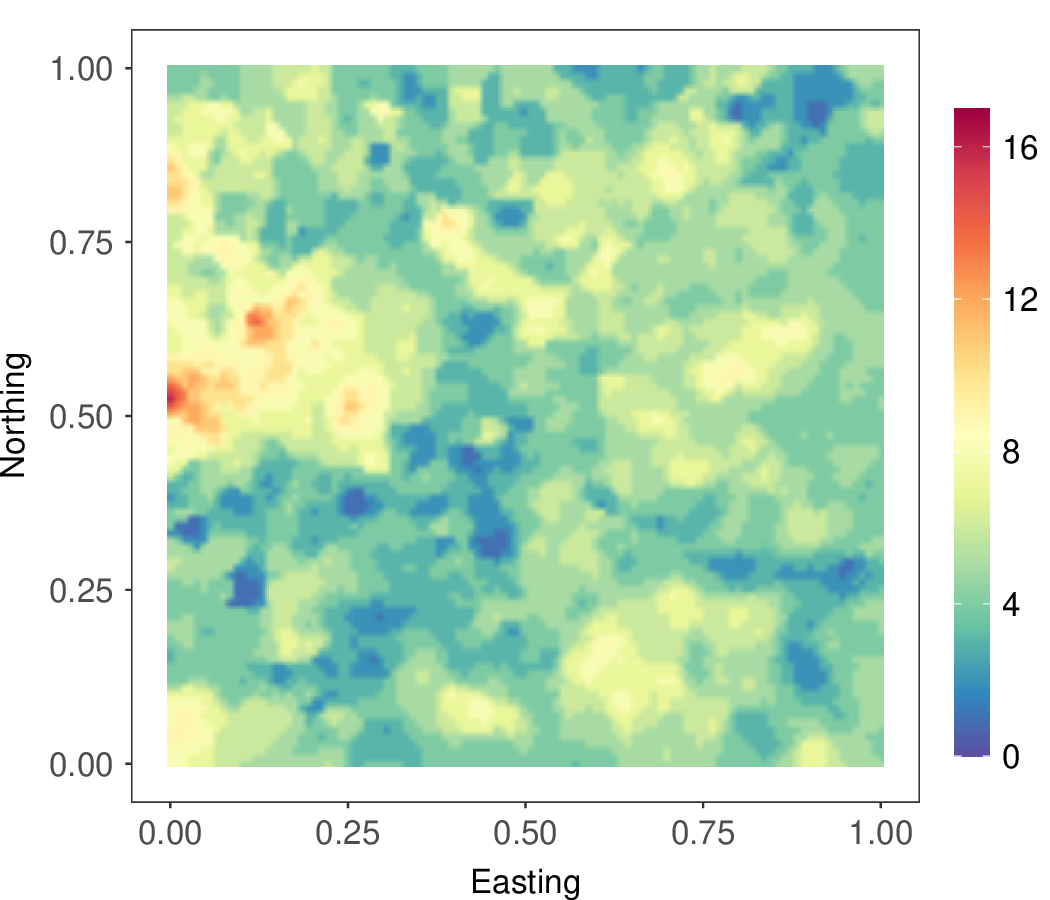}
         \caption{PONNMP (Gaussian)}
     \end{subfigure}     
    %  \hfill
    \begin{subfigure}[b]{0.32\textwidth}
         \centering
         \includegraphics[width=\textwidth]{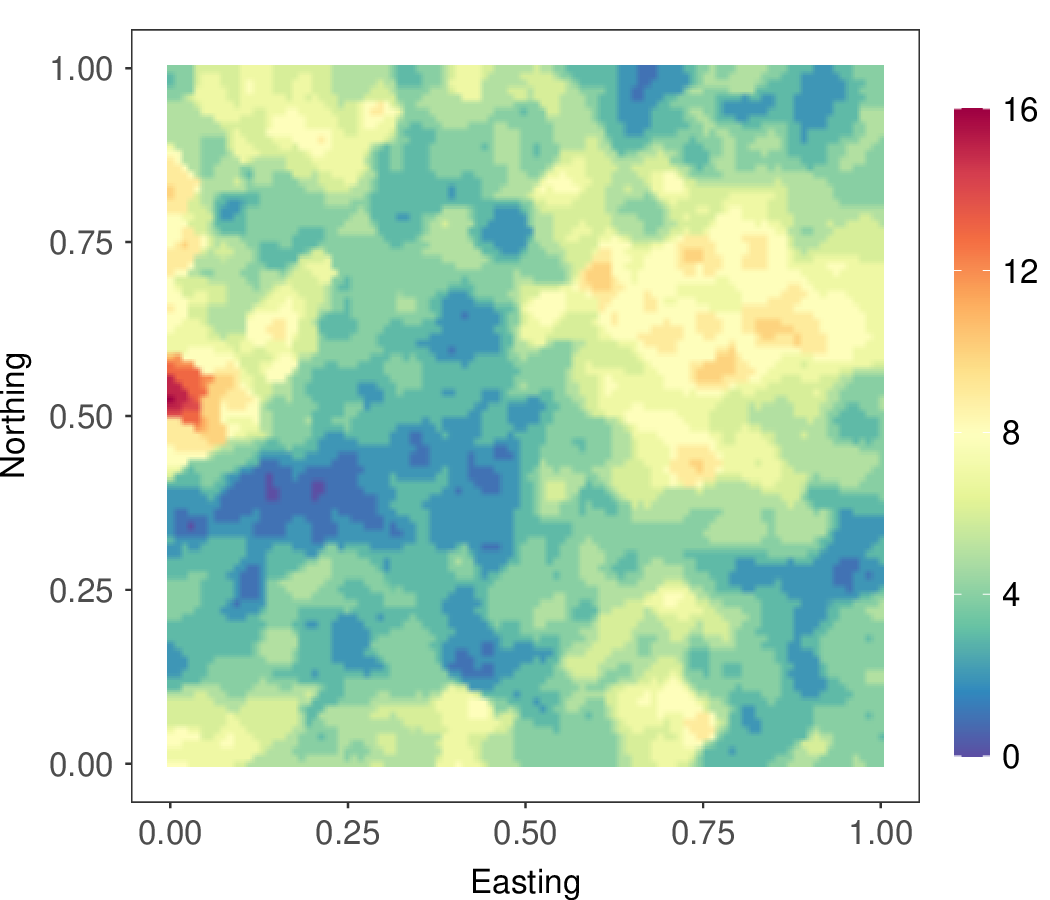}
         \caption{PONNMP (Gumbel)}
     \end{subfigure}     
    \begin{subfigure}[b]{0.32\textwidth}
         \centering
         \includegraphics[width=\textwidth]{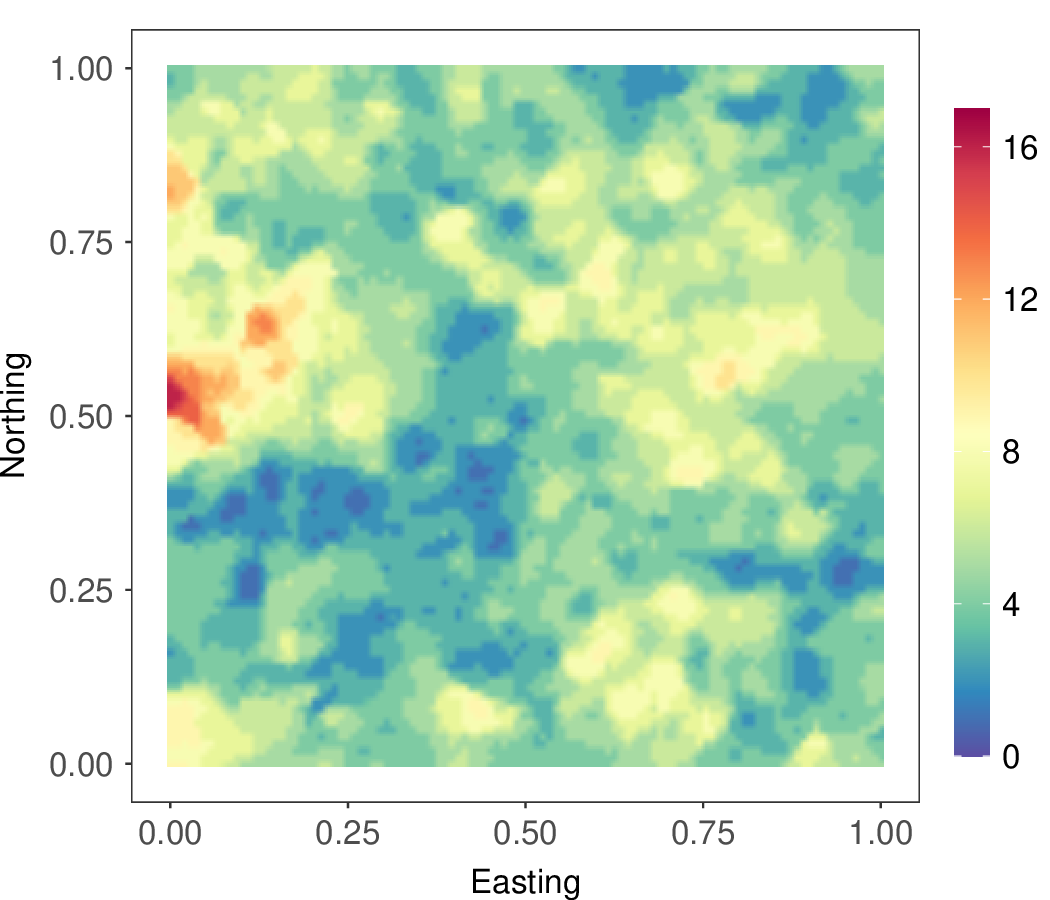}
         \caption{PONNMP (Gumbel)}
     \end{subfigure}     
     \begin{subfigure}[b]{0.32\textwidth}
         \centering
         \includegraphics[width=\textwidth]{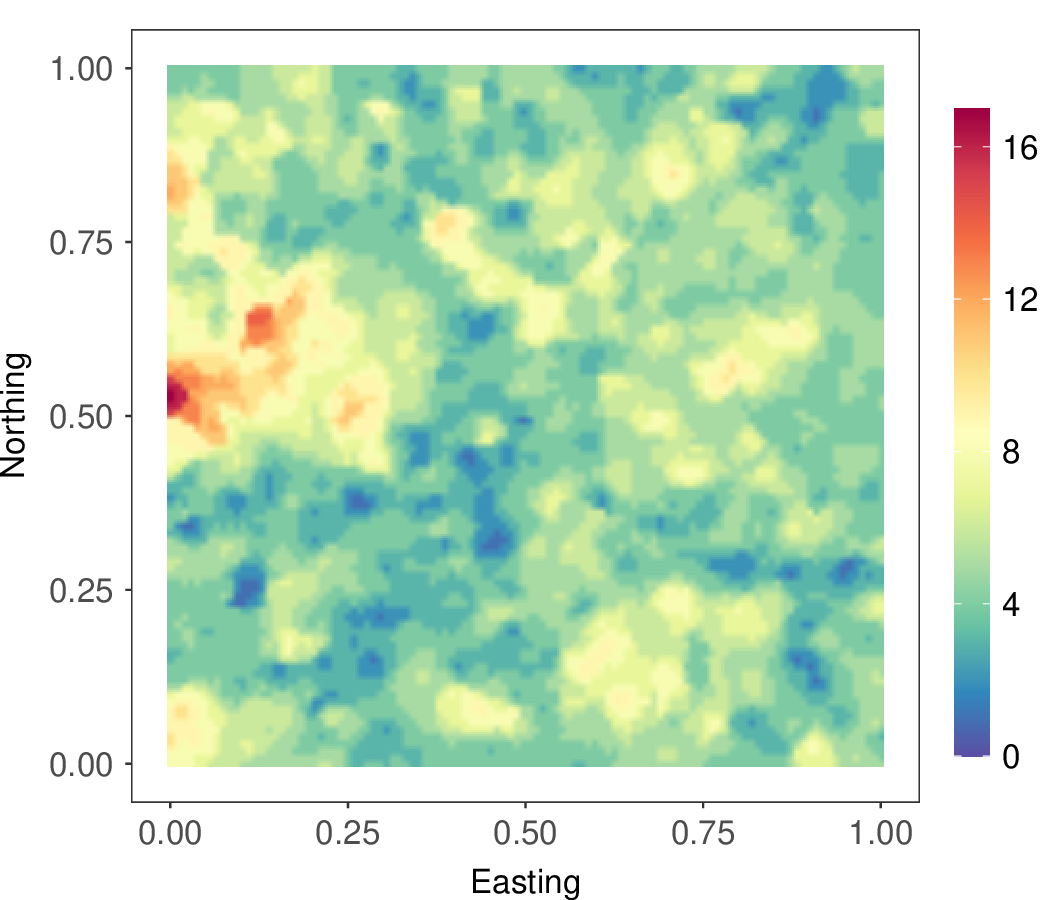}
         \caption{PONNMP (Gumbel)}
     \end{subfigure}
    \begin{subfigure}[b]{0.32\textwidth}
         \centering
         \includegraphics[width=\textwidth]{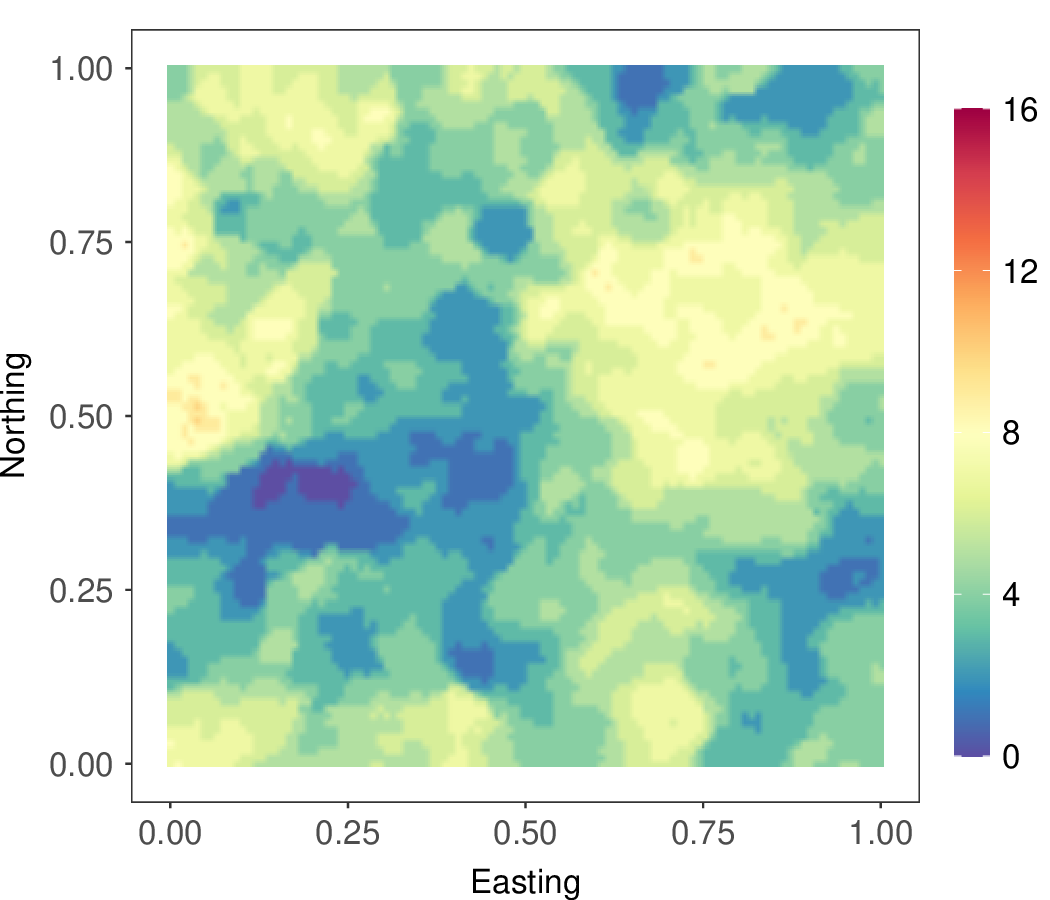}
         \caption{PONNMP (Clayton)}
     \end{subfigure}
    \begin{subfigure}[b]{0.32\textwidth}
         \centering
         \includegraphics[width=\textwidth]{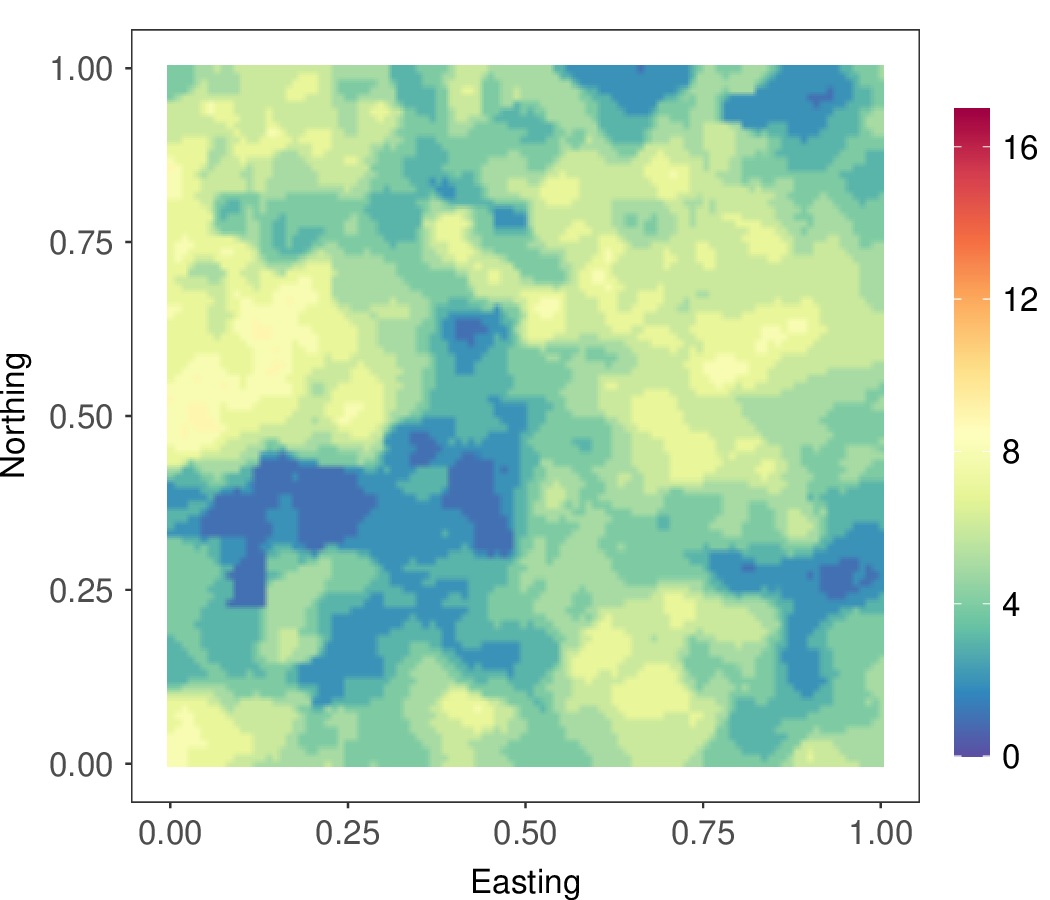}
         \caption{PONNMP (Clayton)}
     \end{subfigure}     
    %  \hfill
    \begin{subfigure}[b]{0.32\textwidth}
         \centering
         \includegraphics[width=\textwidth]{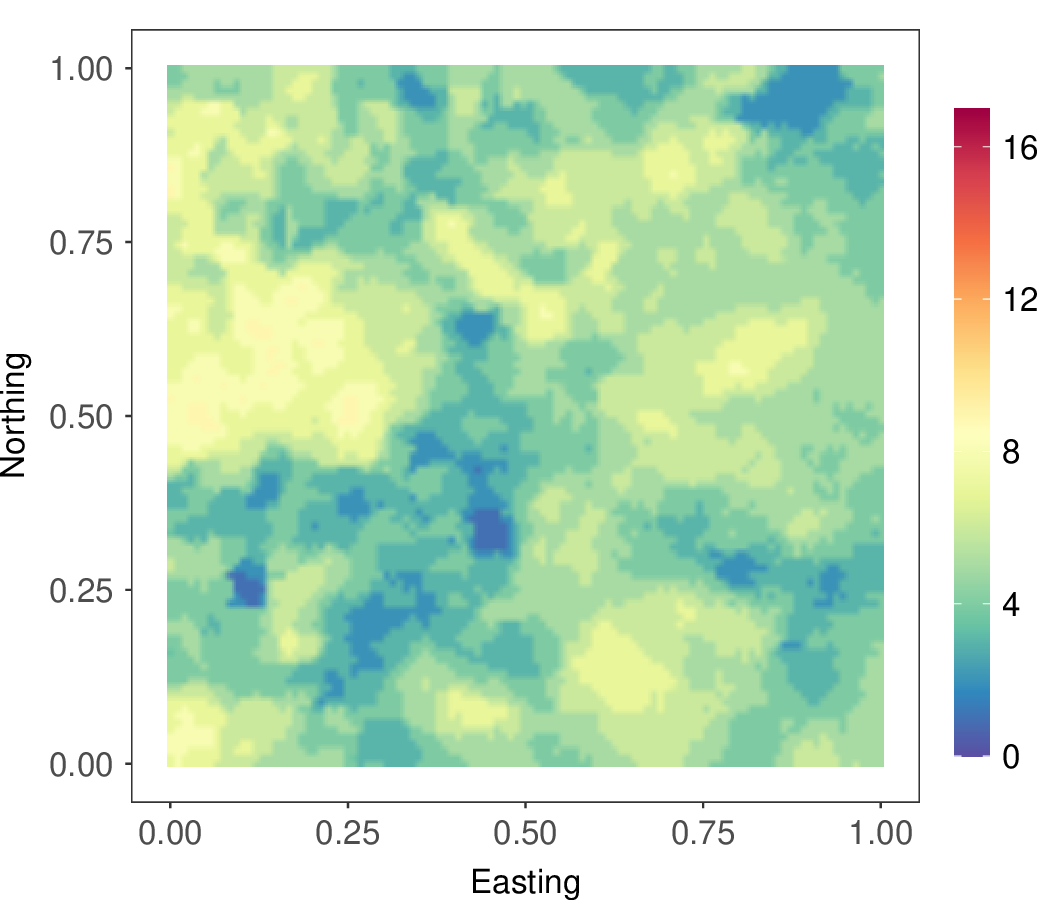}
         \caption{PONNMP (Clayton)}
     \end{subfigure}     
    \caption{
    Simulated data example 1. Interpolated surfaces of the true model (first row),
    and posterior median estimates of the Poisson NNMP (PONNMP) models 
    using Gaussian (second row), Gumbel (third row), and Clayton (fourth row) copulas.
    Columns from left to right correspond to scenarios with $\sigma_1 = 1, 3, 10$, respectively.
    }
    \label{fig:sim1-pred}
\end{figure}

Figure \ref{fig:sim1-pred} shows the predicted random fields, given by the three discrete 
copula NNMP models with Poisson stationary marginals, under different scenarios.

As discussed in the main paper, the Clayton model was not able to recover large values.
The Gumbel model seems to recover large values slightly better than the Gaussian model.

\subsection{North American Breeding Bird Survey data analysis}

\subsubsection{Analysis of $L$}

We applied the Gaussian copula NBNNMP model to the whole data set with $L = 5, 10, 15, 20$. 
For each $L$, we ran the MCMC algorithm for 30000 iterations, discarding the first 
10000 iterations, and collecting posterior samples every 5th iteration.

Table \ref{tbl:real2} provides the posterior means and 95\% CI estimates of the 
model parameters. They were quite robust across different values of $L$, except for 
those of $\phi$ and $\zeta$, even though the different credible intervals have
substantial overlap. Note that $\phi$ and $\zeta$ are 
the range parameters of the exponential correlation functions for 
the Gaussian copula correlation
and for the cutoff point kernel, respectively. 
Since a model with a large value of $L$ includes more distant neighbors, 
$\phi$ and $\zeta$ should be larger as they indicate effective ranges.

\begin{figure}[ht]
    \centering
    \includegraphics[width=.24\textwidth]{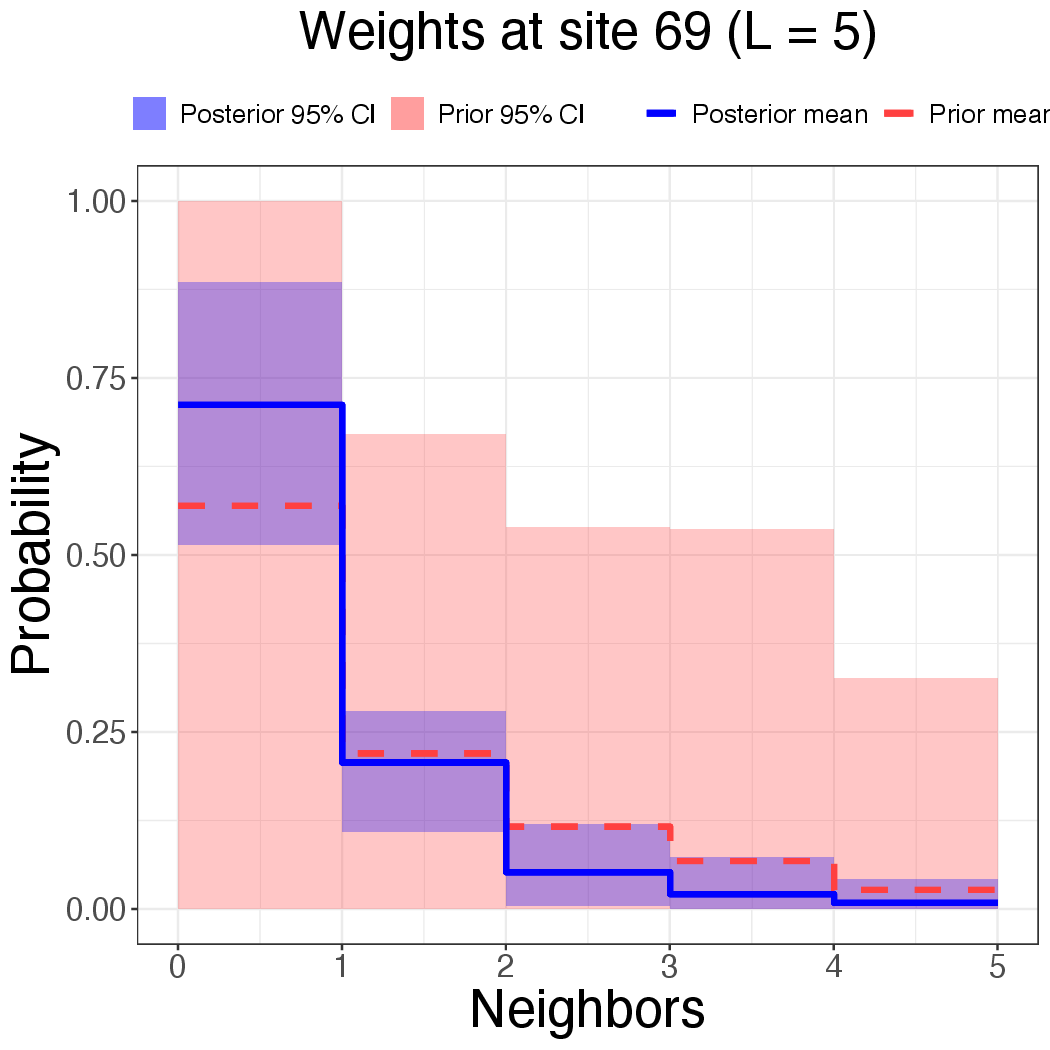}
    \includegraphics[width=.24\textwidth]{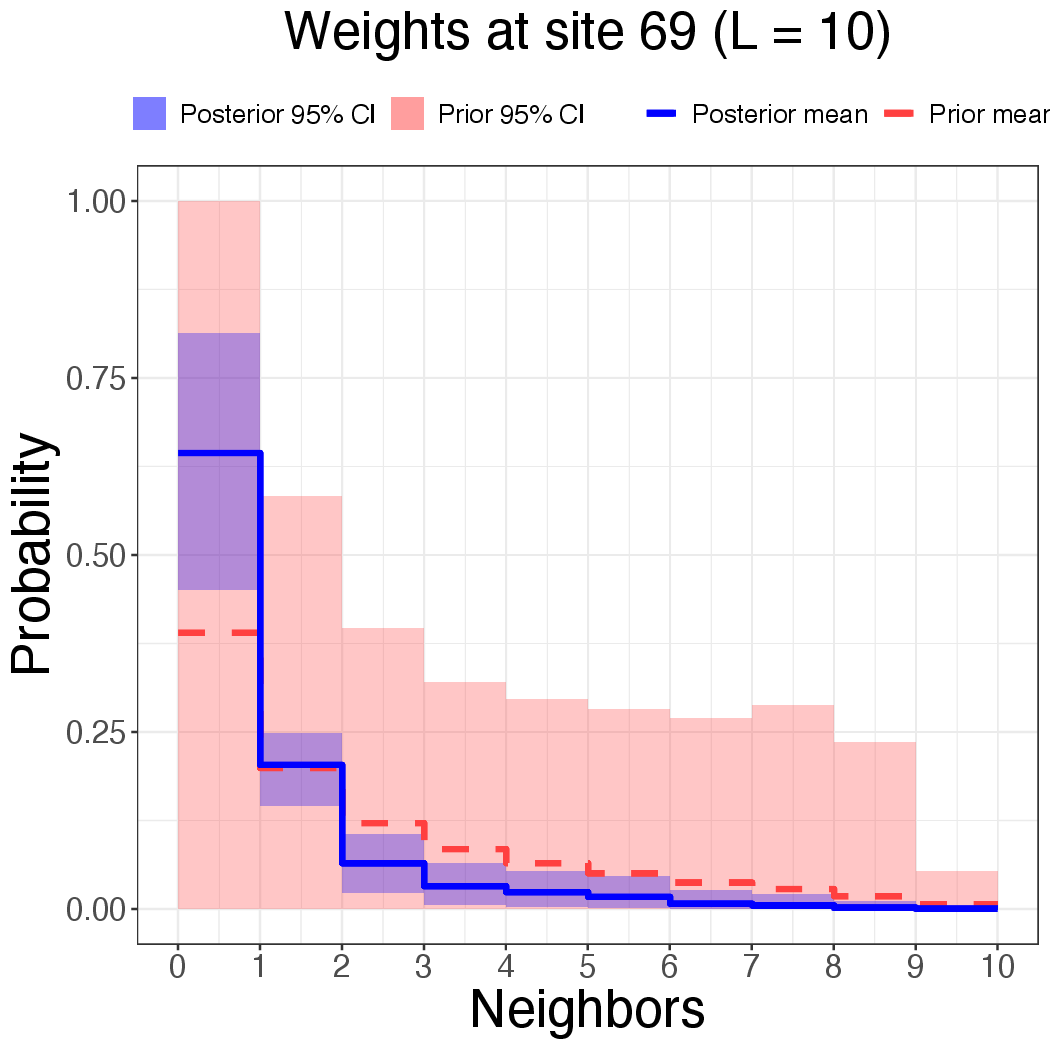}
    \includegraphics[width=.24\textwidth]{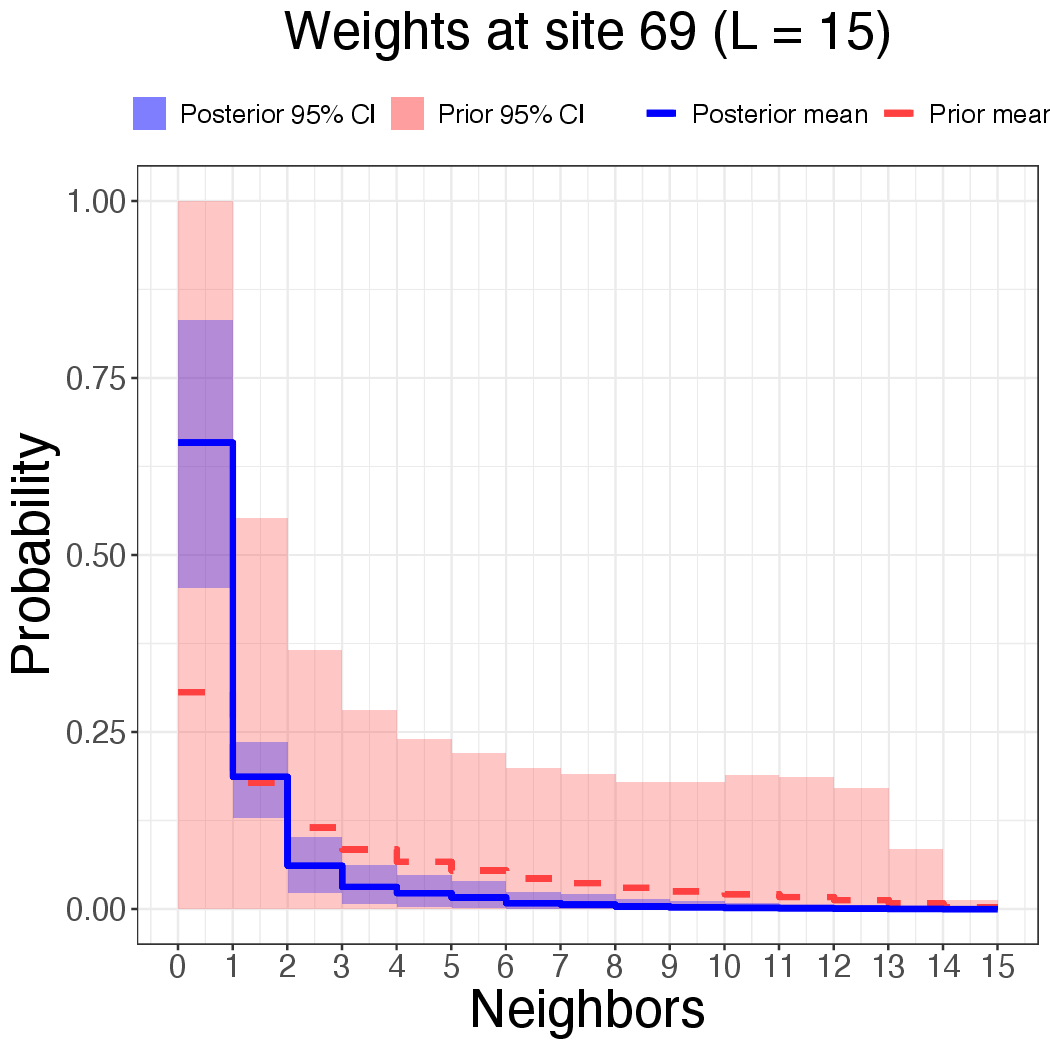}
    \includegraphics[width=.24\textwidth]{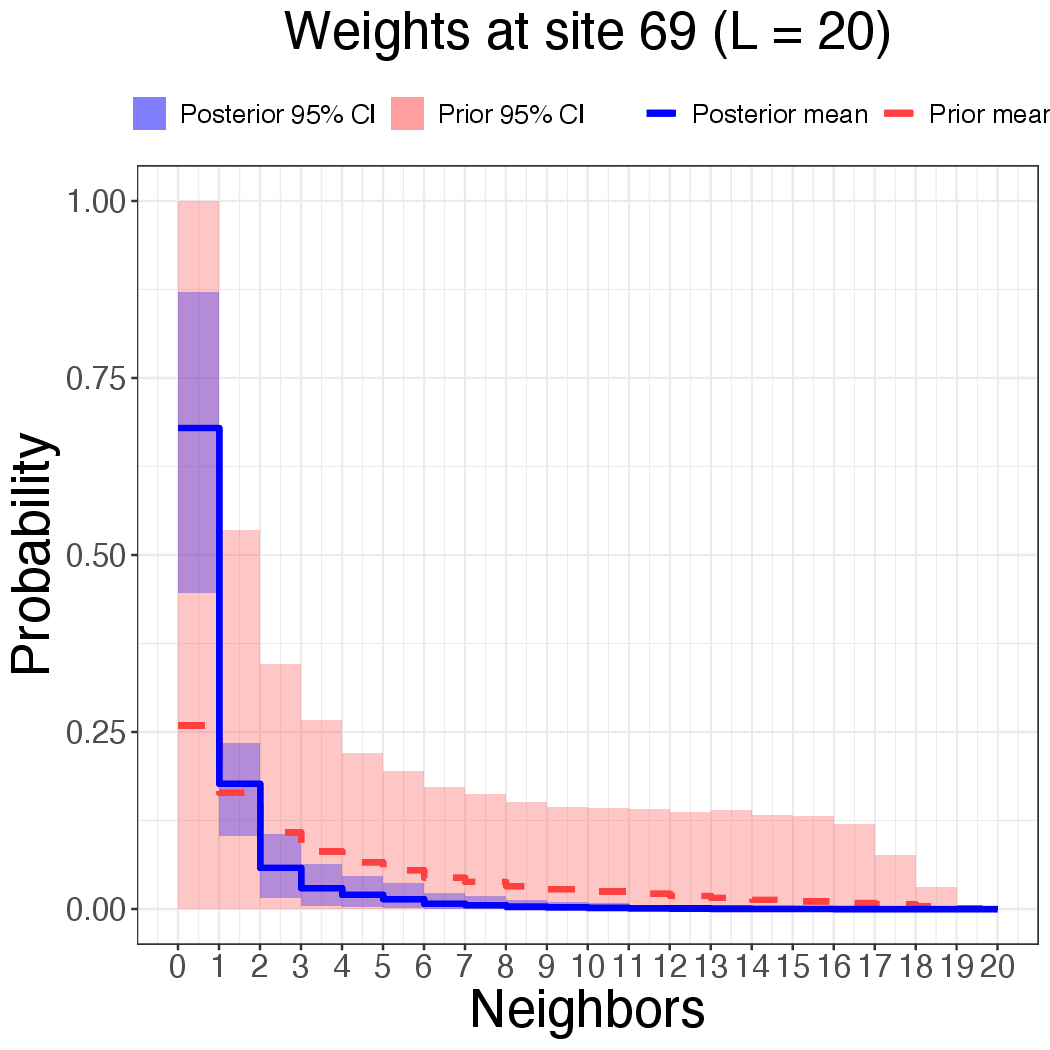}\\
    \bigskip
    \includegraphics[width=.24\textwidth]{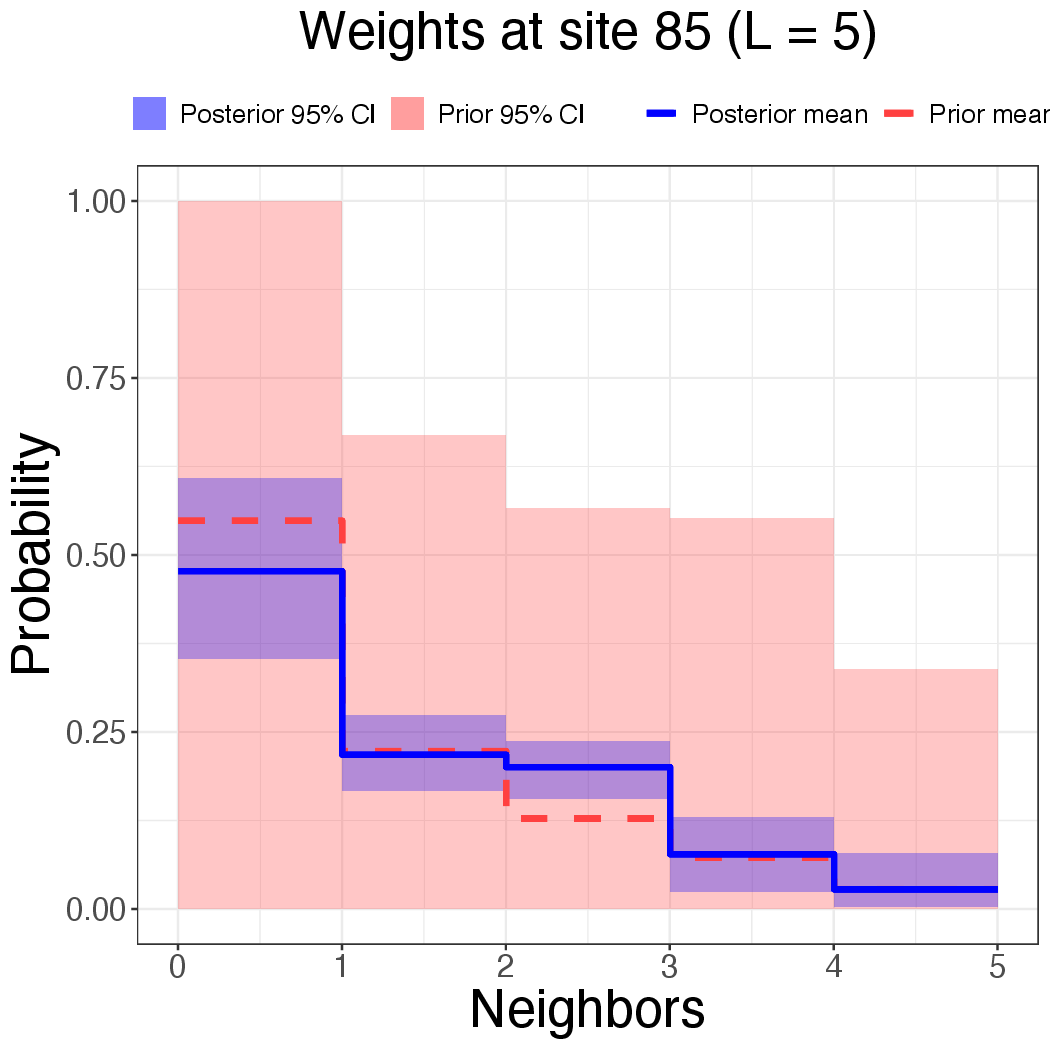}
    \includegraphics[width=.24\textwidth]{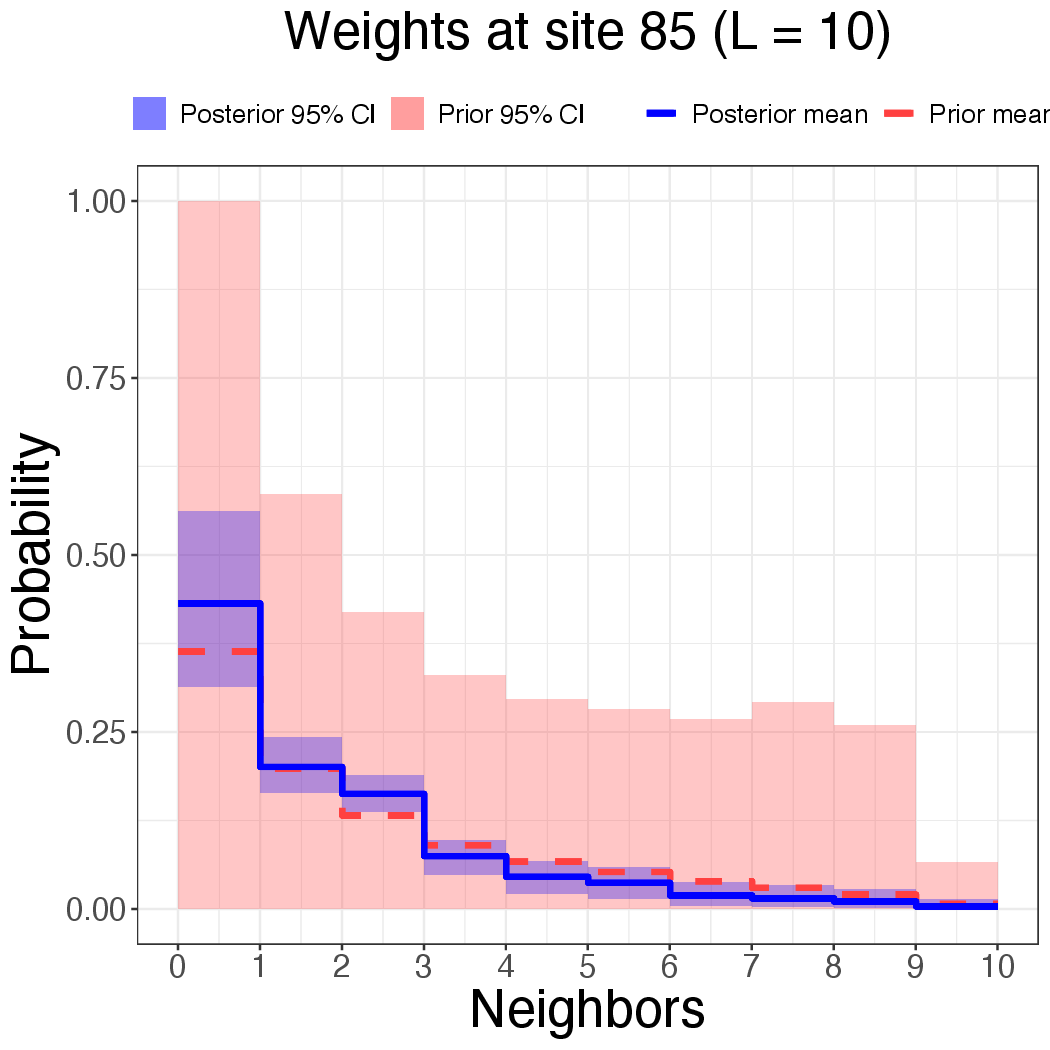}
    \includegraphics[width=.24\textwidth]{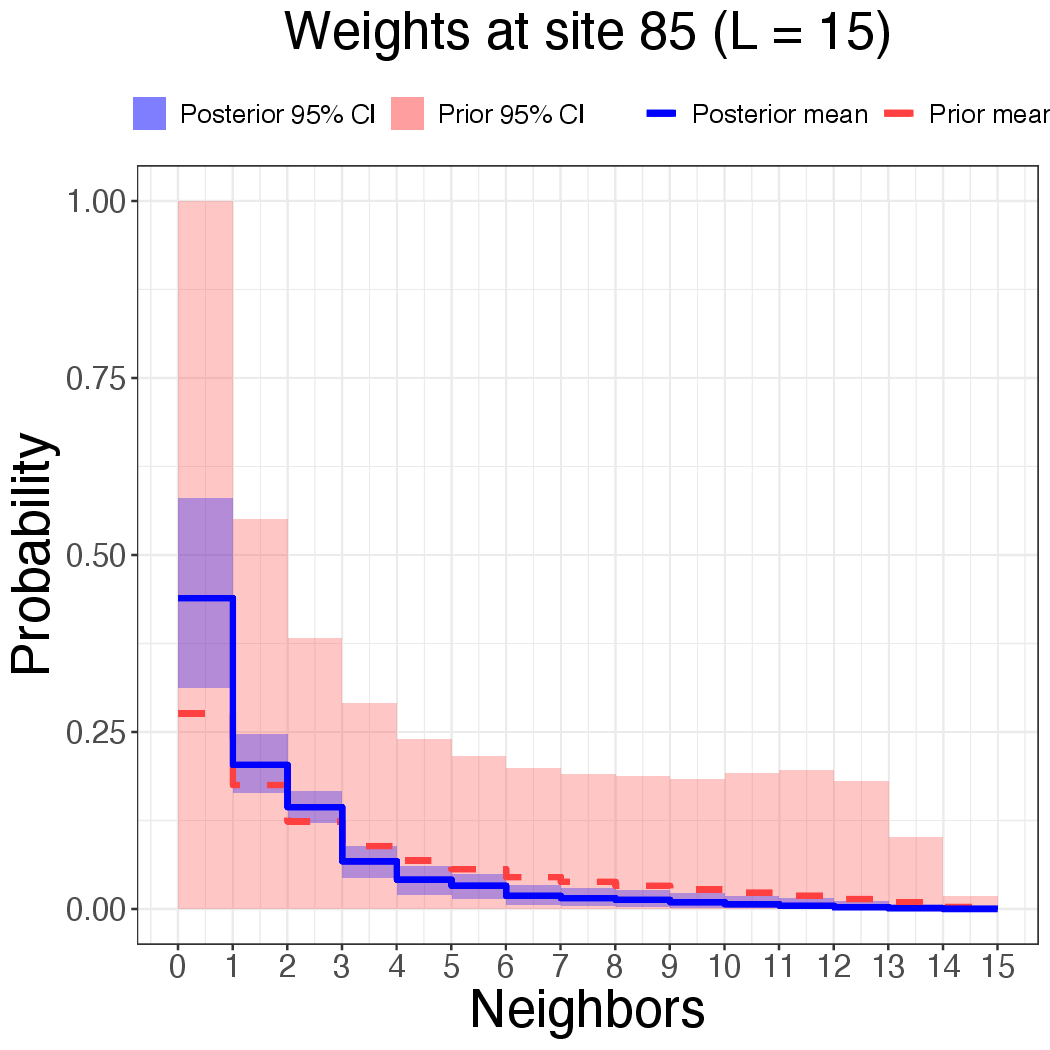}
    \includegraphics[width=.24\textwidth]{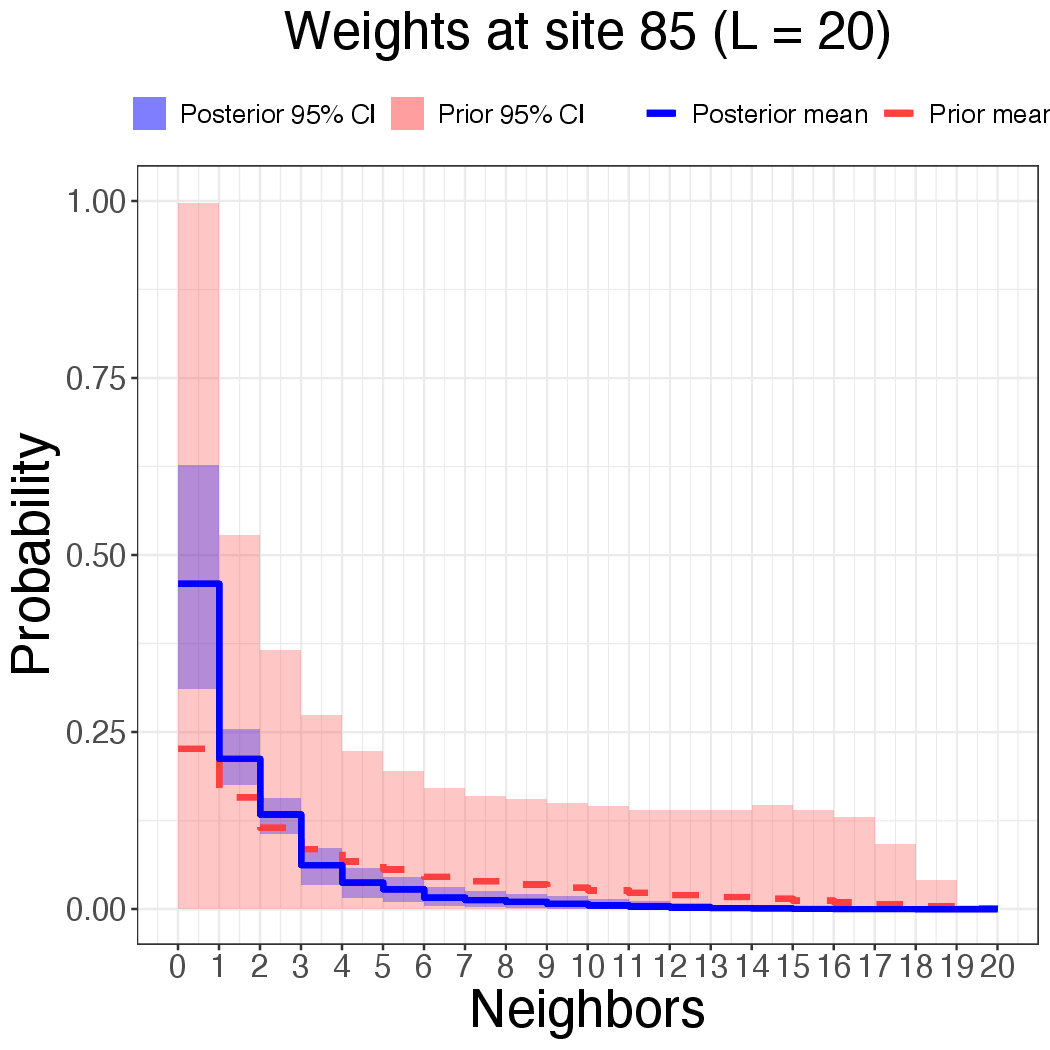}\\
    \bigskip
    \includegraphics[width=.24\textwidth]{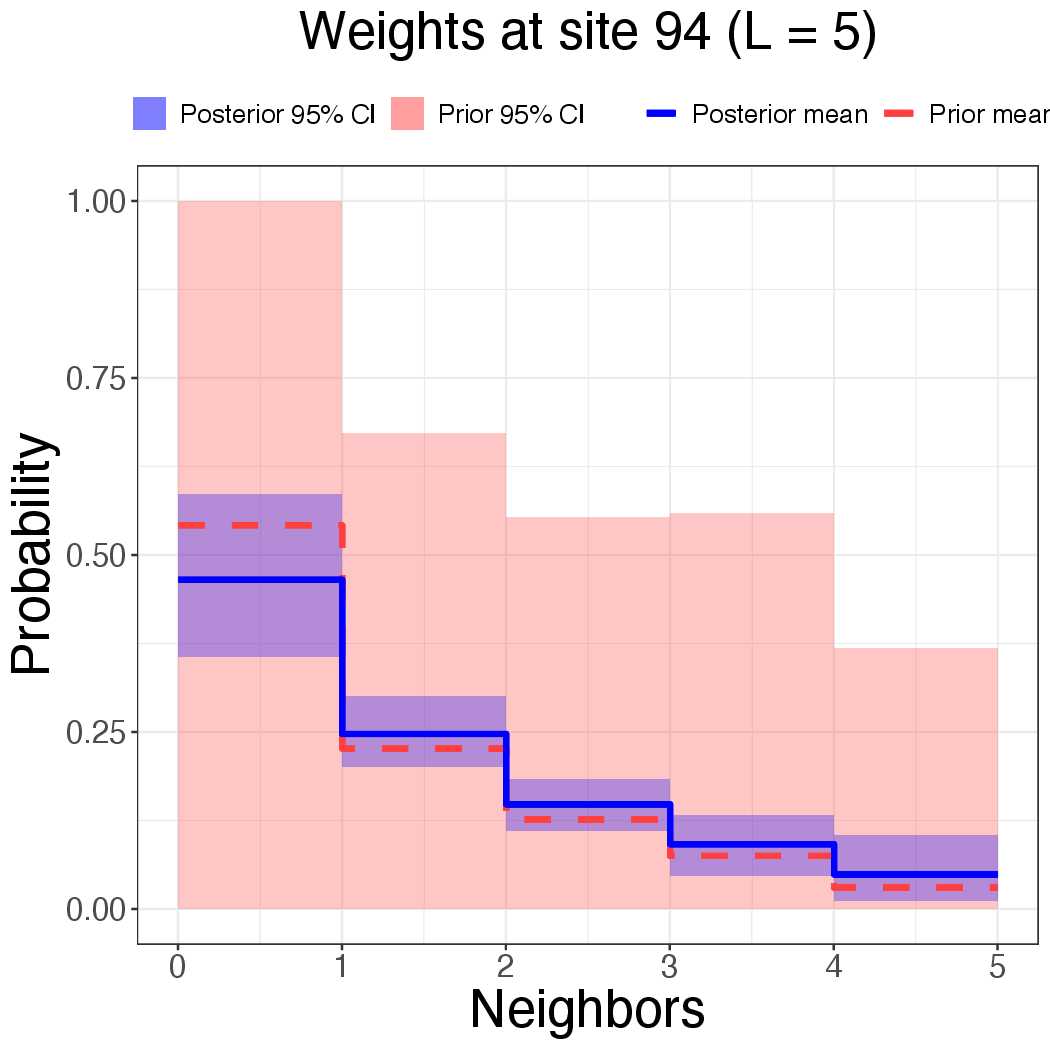}
    \includegraphics[width=.24\textwidth]{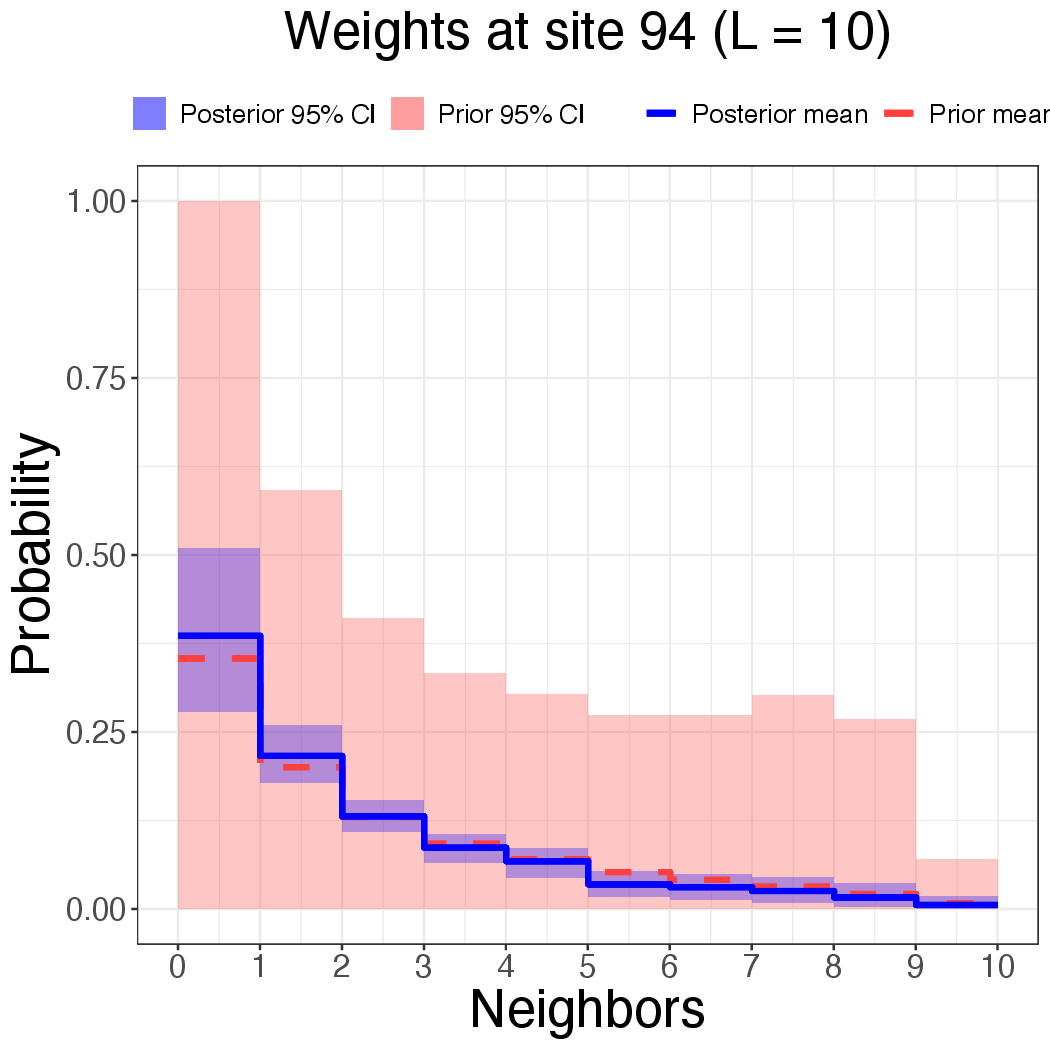}
    \includegraphics[width=.24\textwidth]{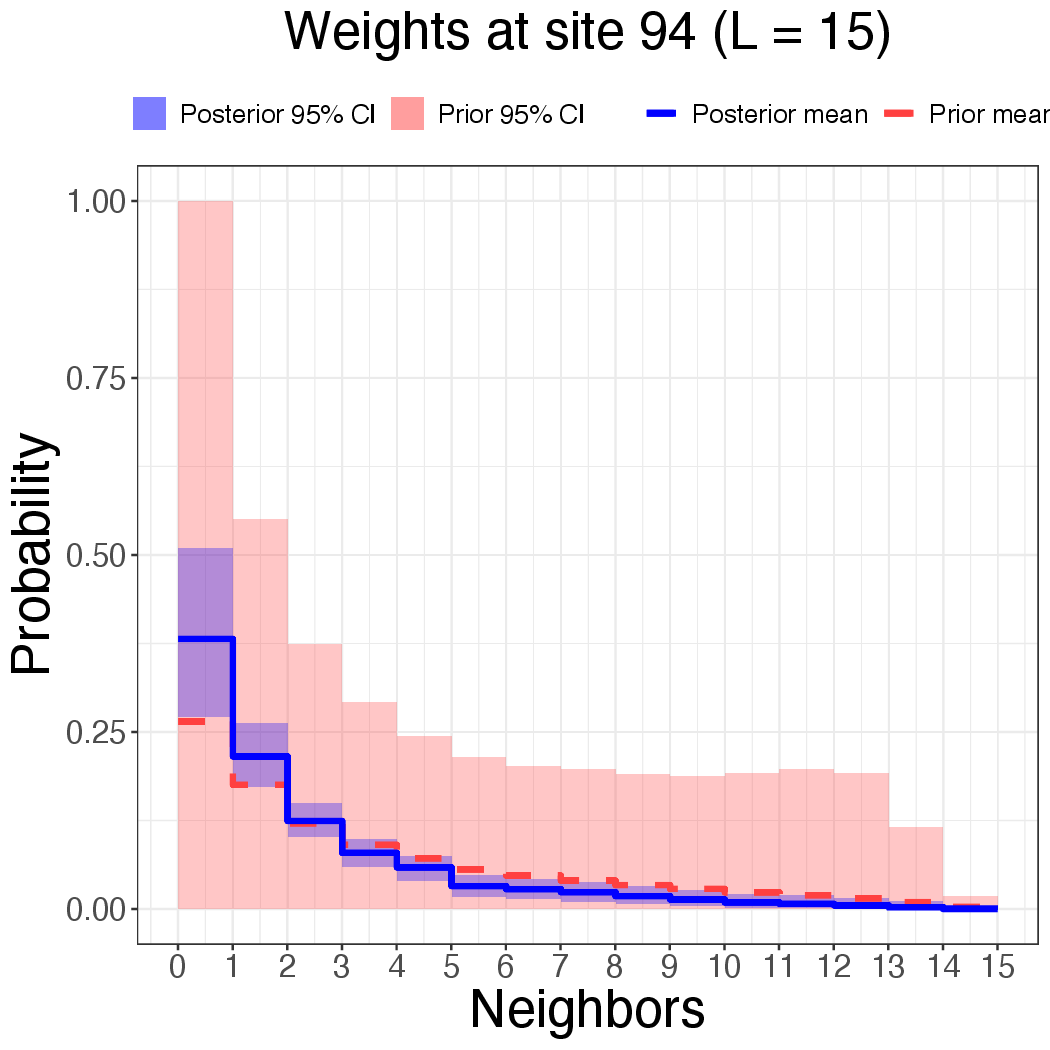}
    \includegraphics[width=.24\textwidth]{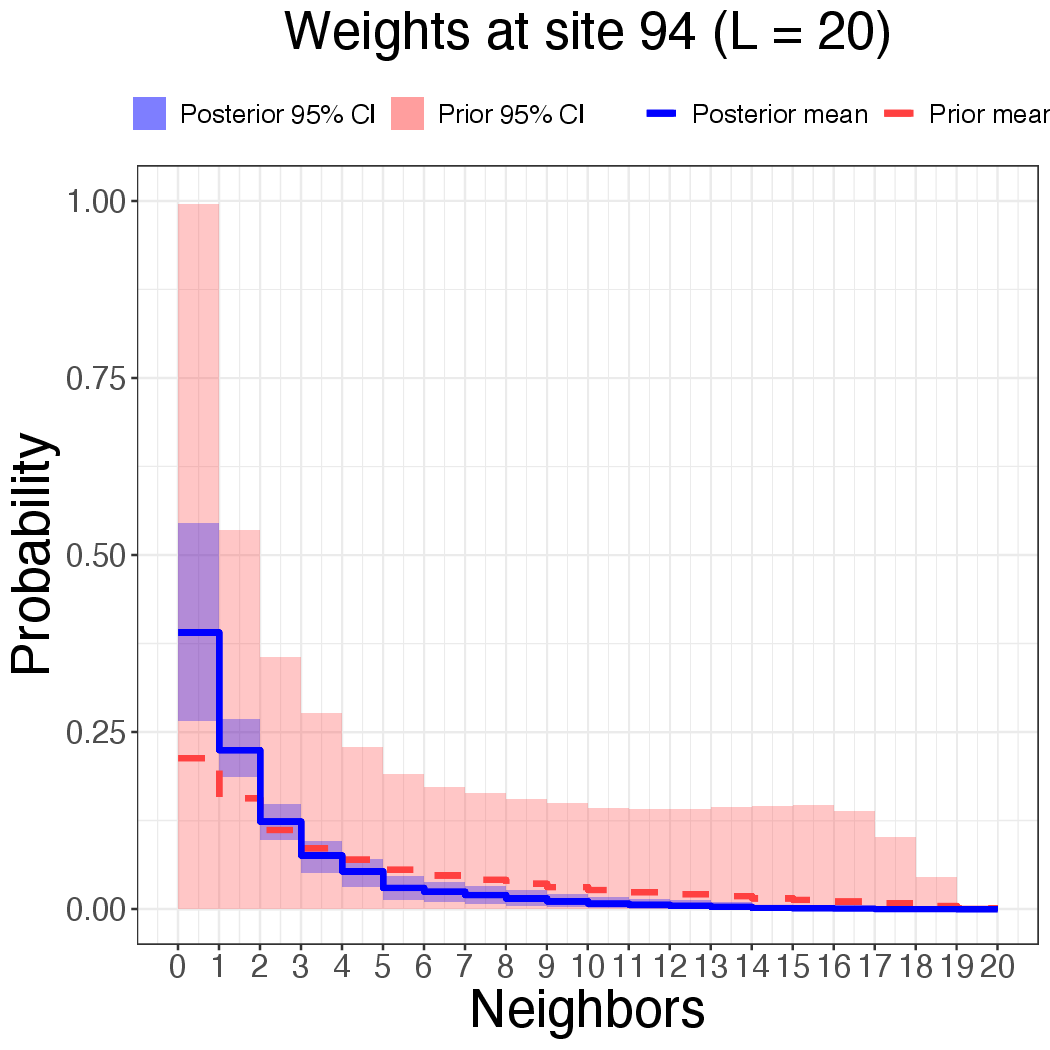}\\
    \bigskip
    \includegraphics[width=.24\textwidth]{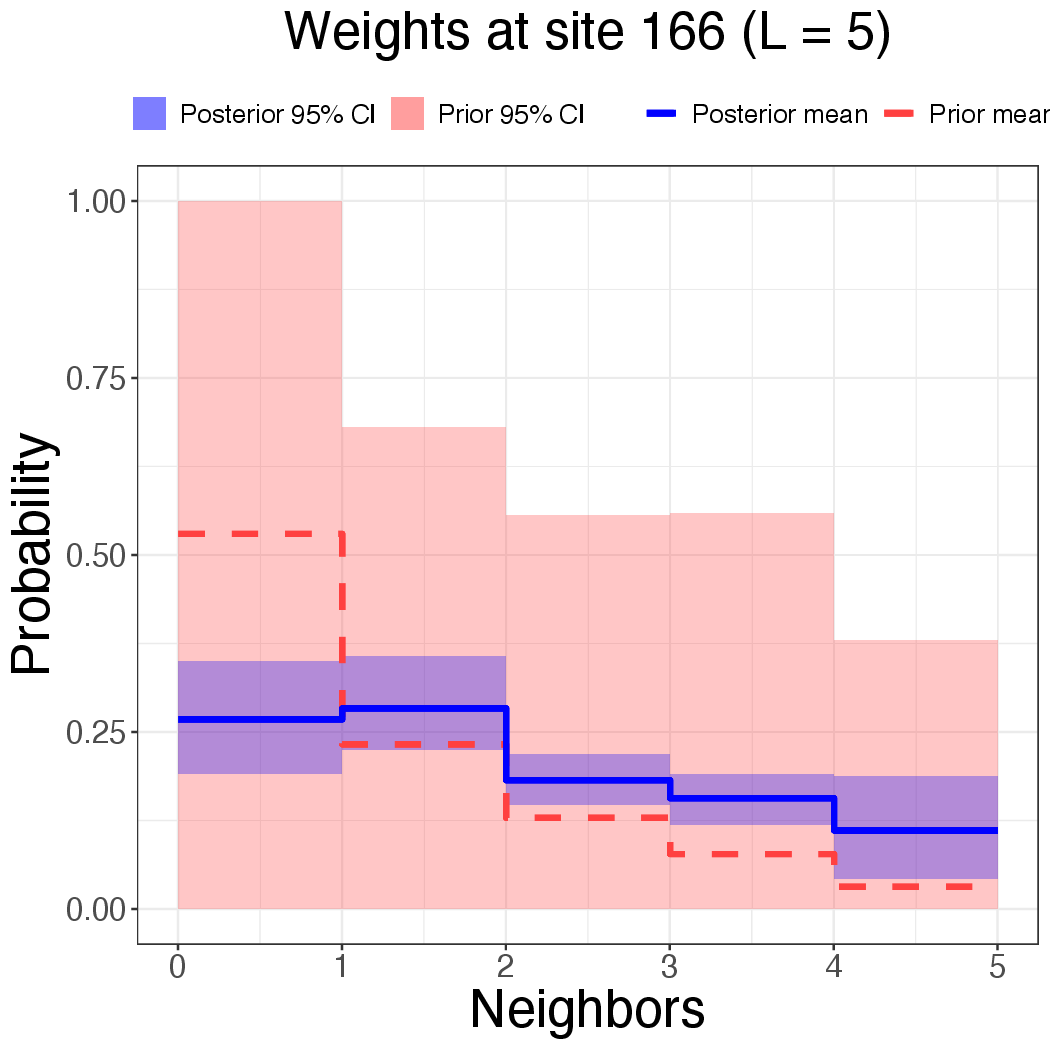}
    \includegraphics[width=.24\textwidth]{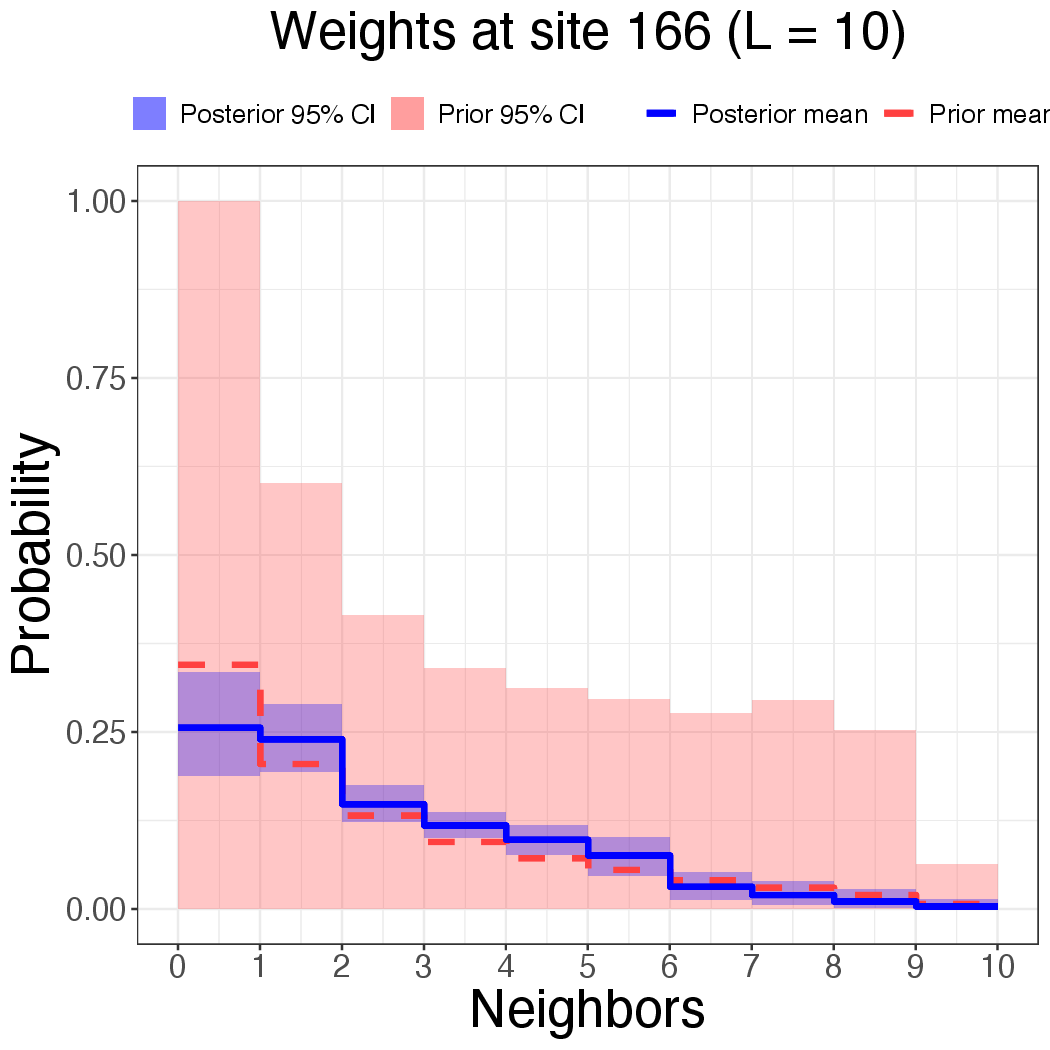}
    \includegraphics[width=.24\textwidth]{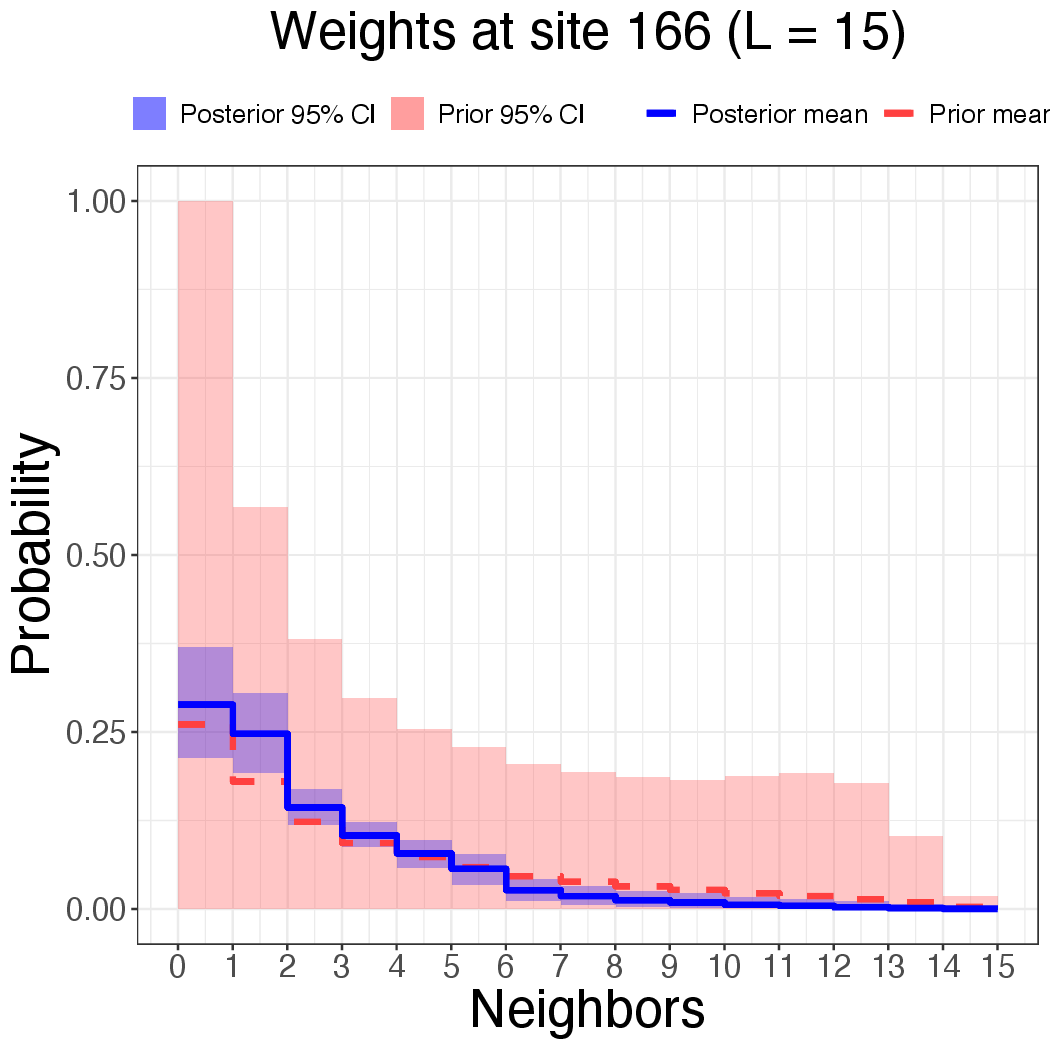}
    \includegraphics[width=.24\textwidth]{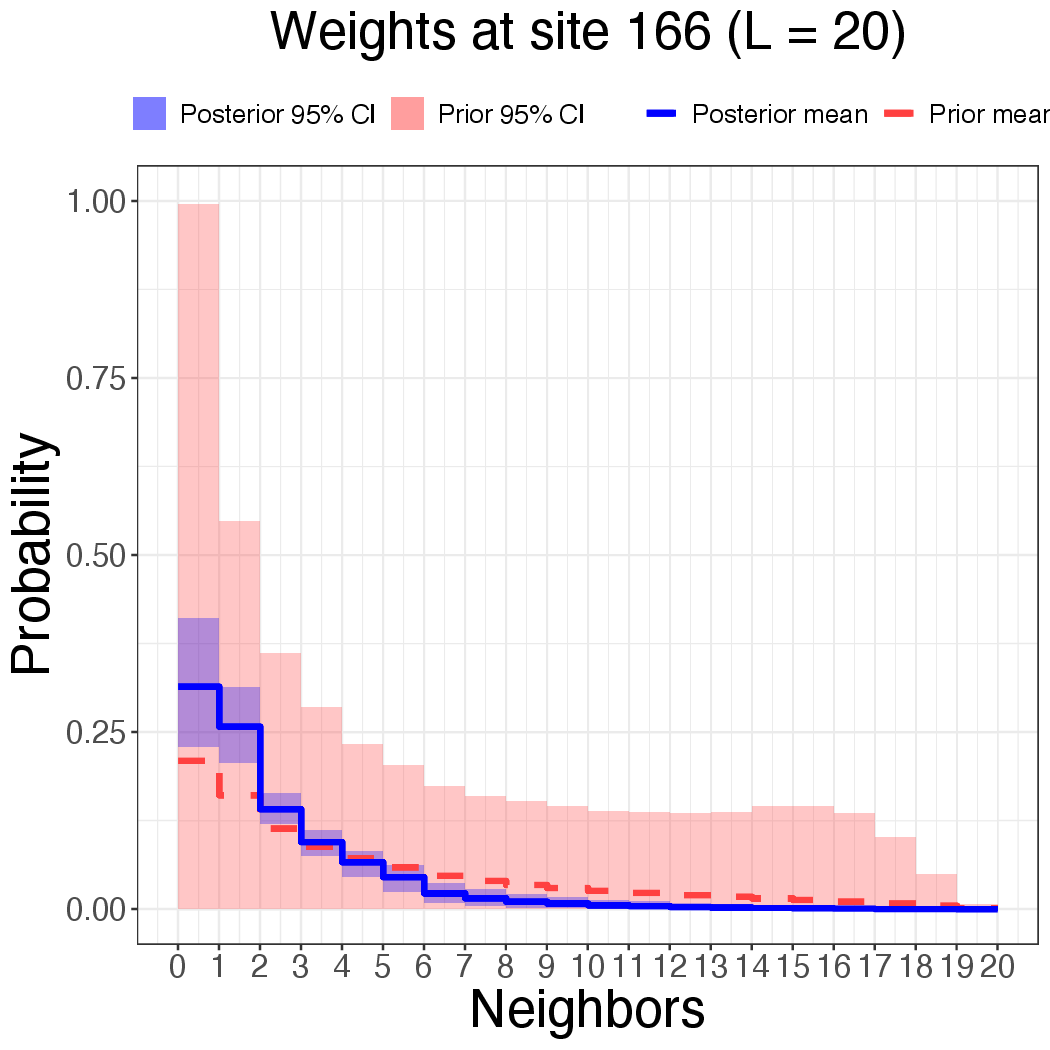}\\
    \bigskip
    \includegraphics[width=.24\textwidth]{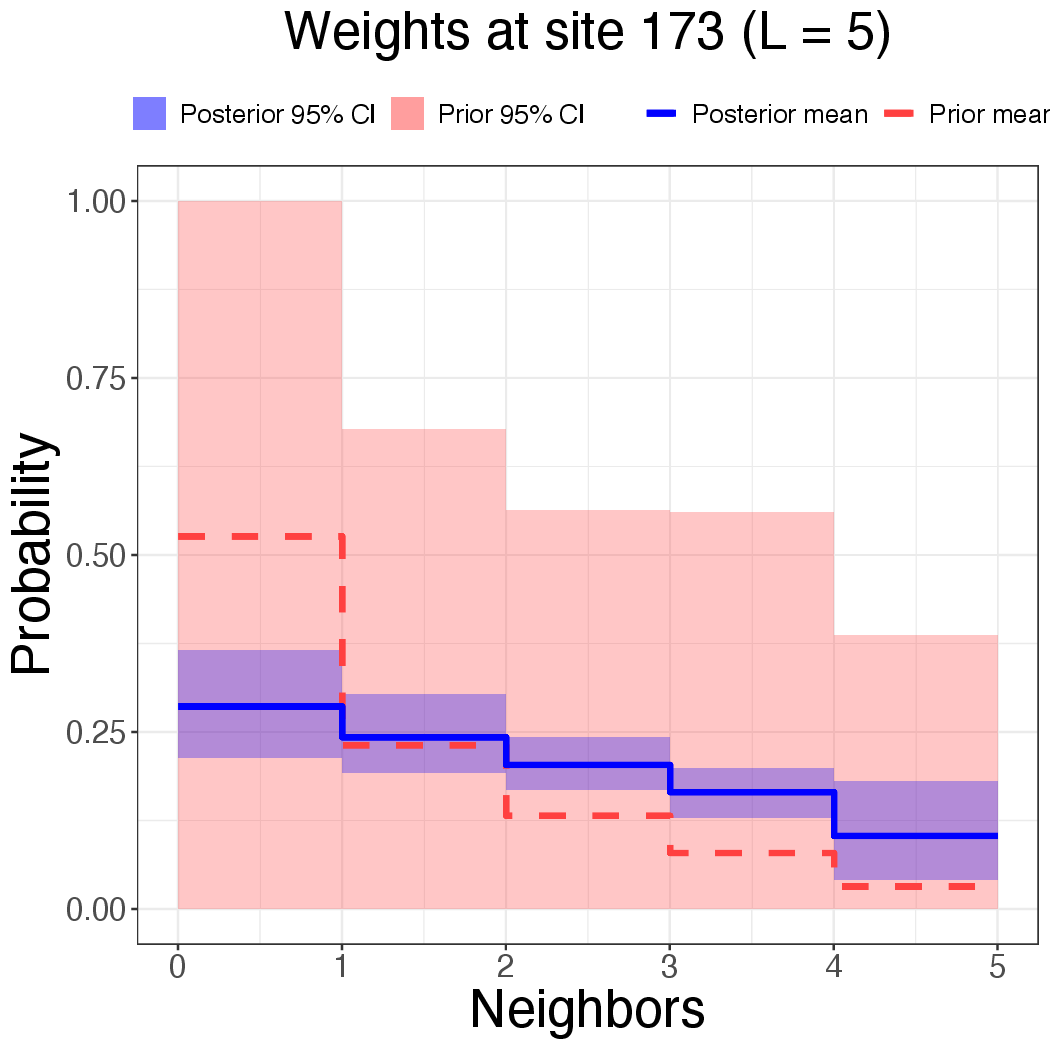}
    \includegraphics[width=.24\textwidth]{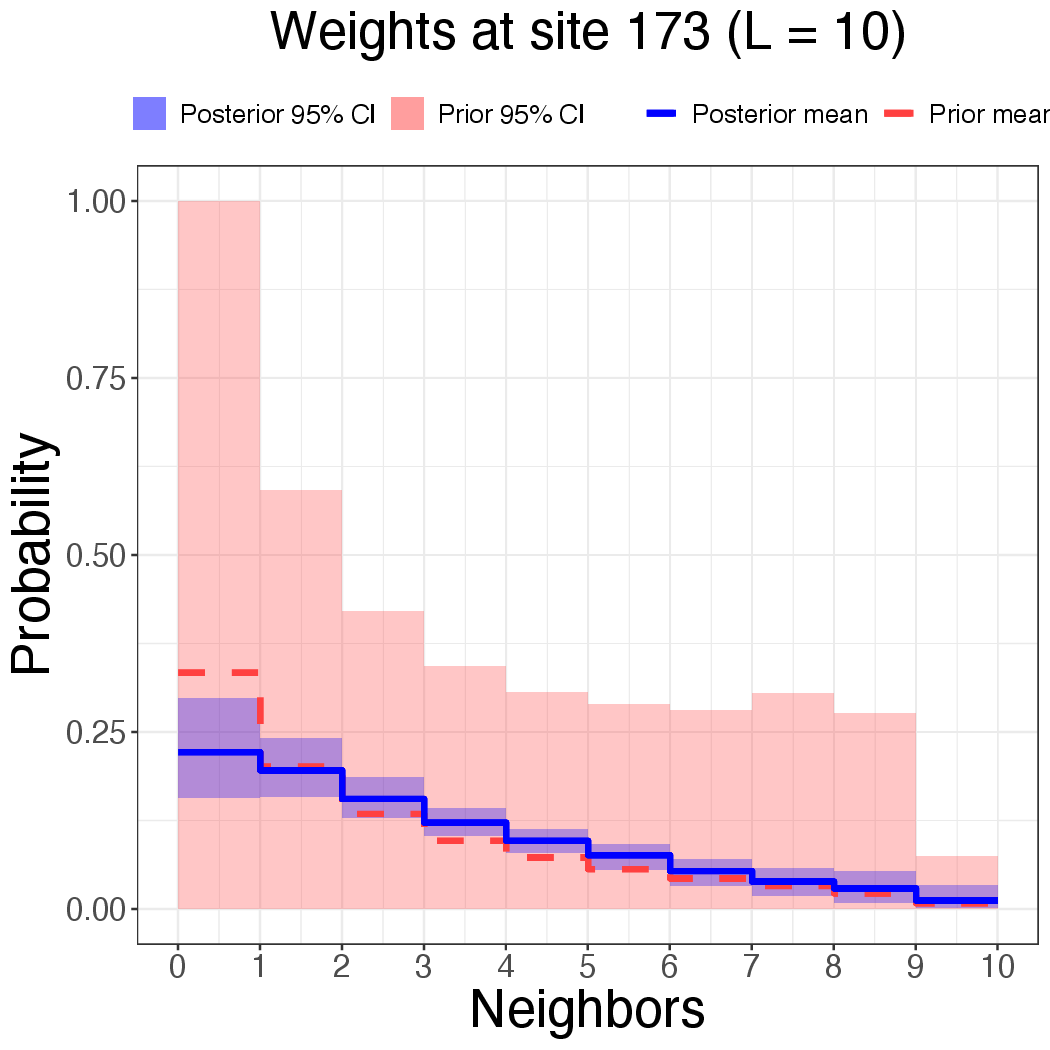}
    \includegraphics[width=.24\textwidth]{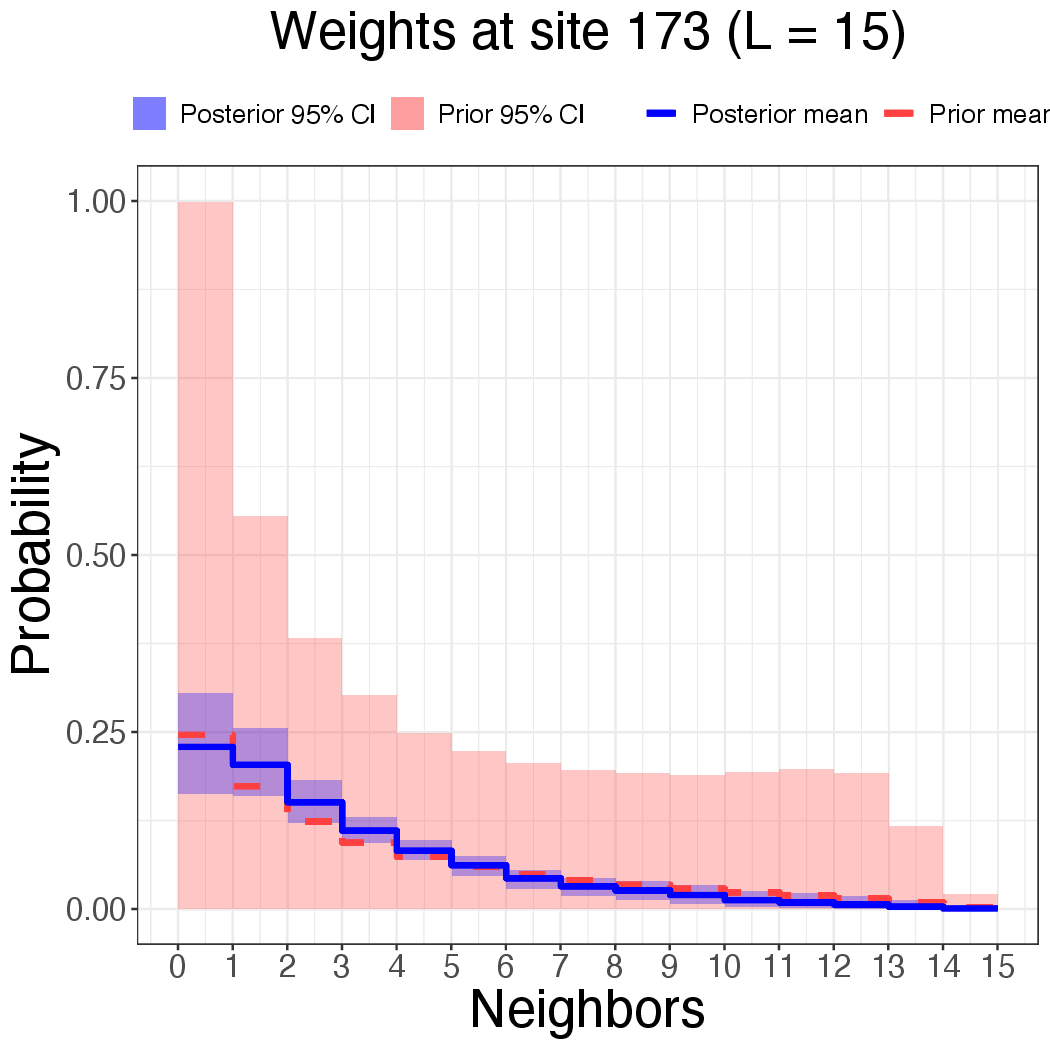}
    \includegraphics[width=.24\textwidth]{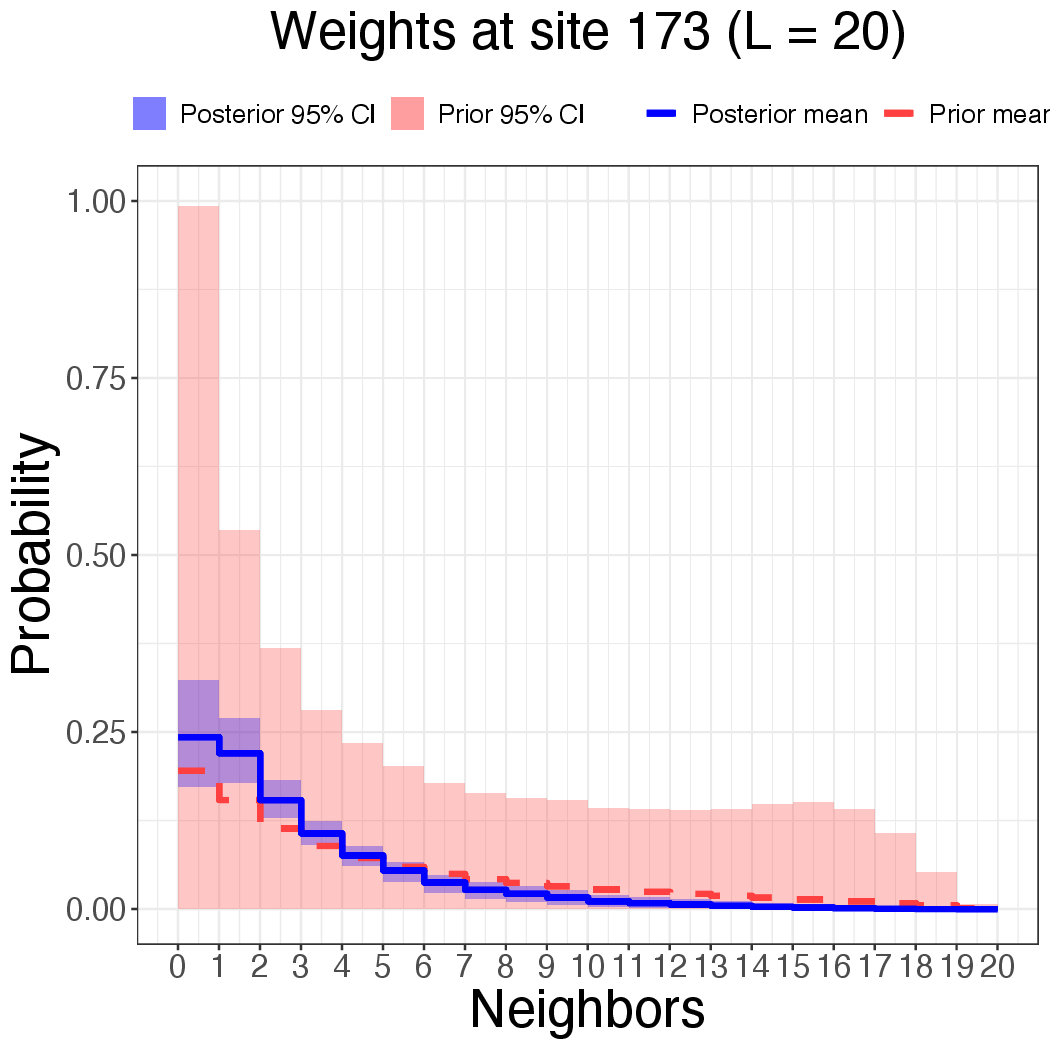}\\
    \bigskip
    \caption{BBS data analysis: Posterior means and 95\% CI estimates of the weights of the first five locations. }
    \label{fig:weights1}
\end{figure}

\begin{figure}[ht]
    \centering
    \includegraphics[width=.24\textwidth]{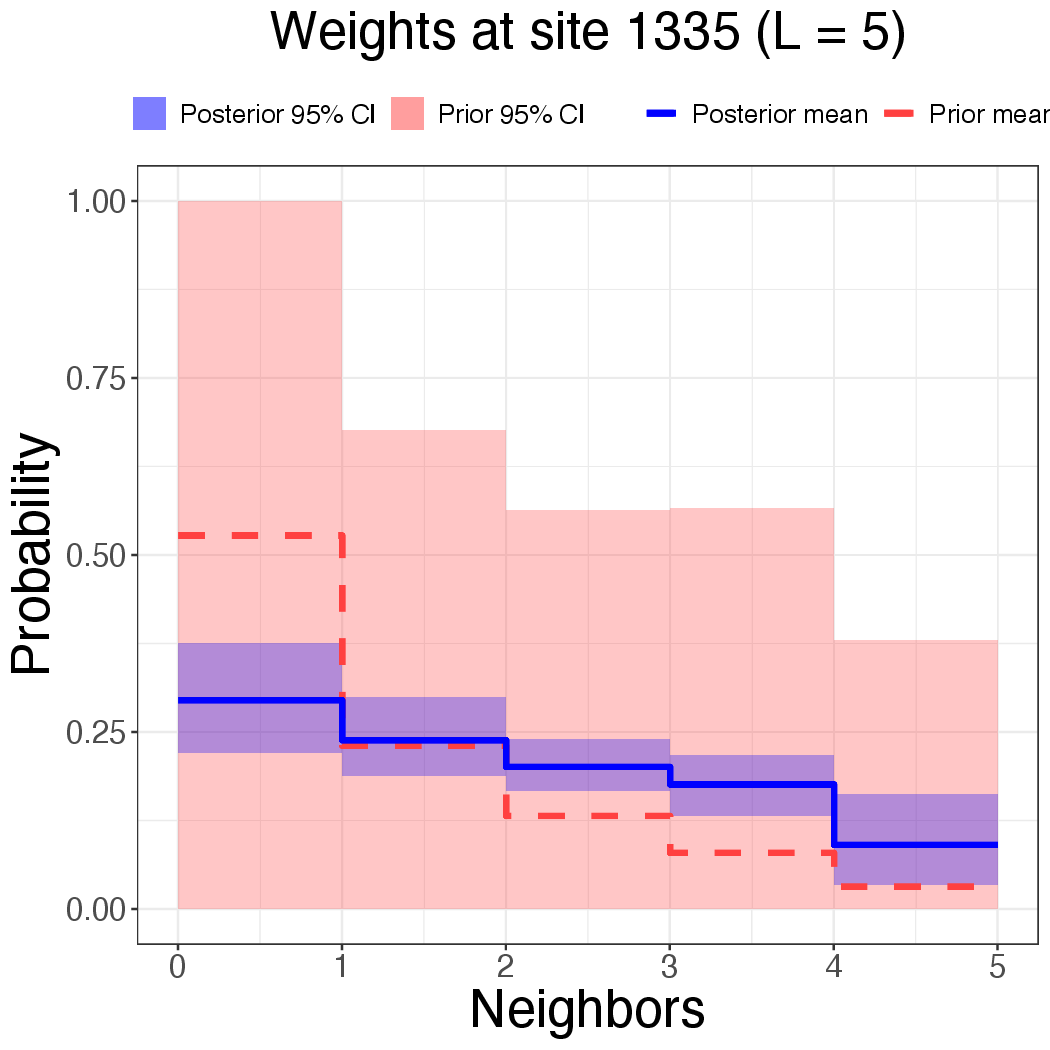}
    \includegraphics[width=.24\textwidth]{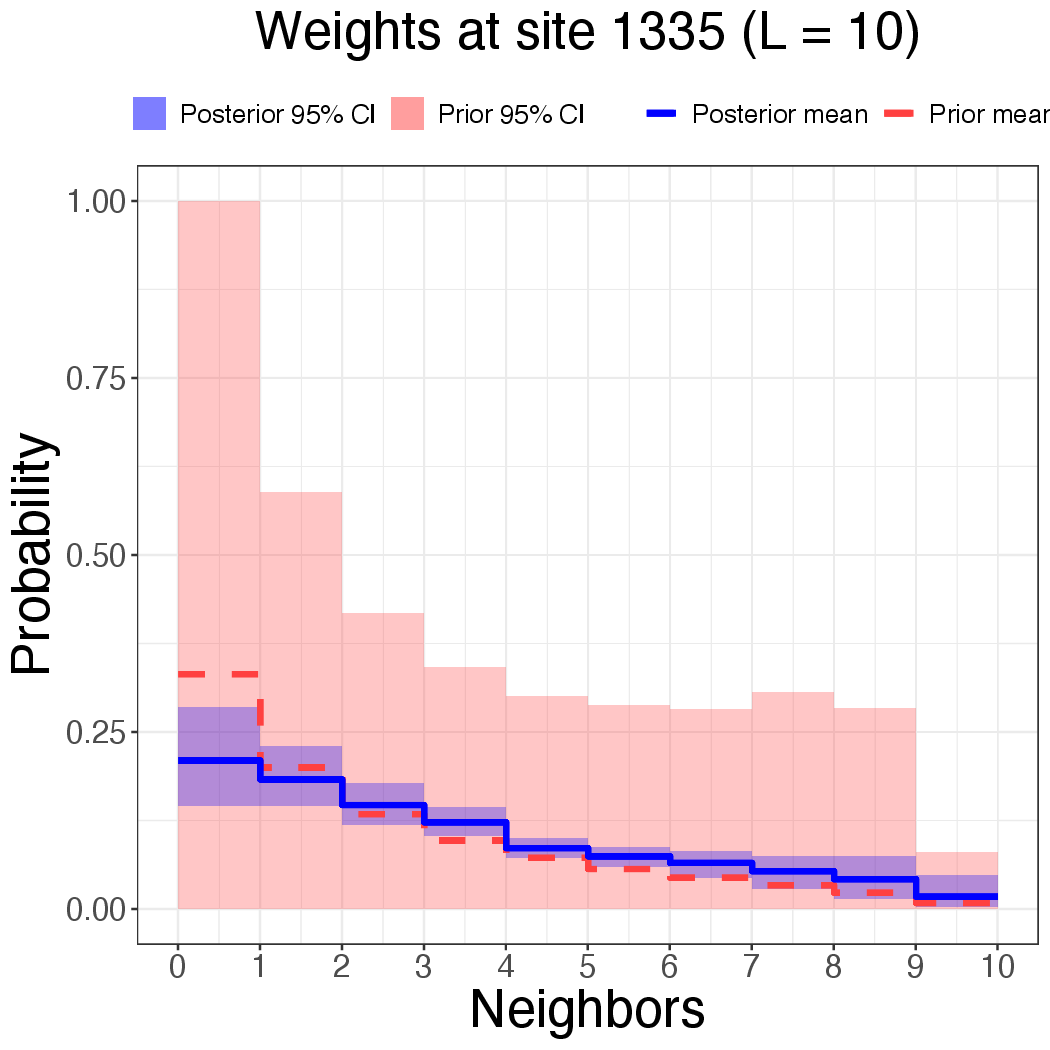}
    \includegraphics[width=.24\textwidth]{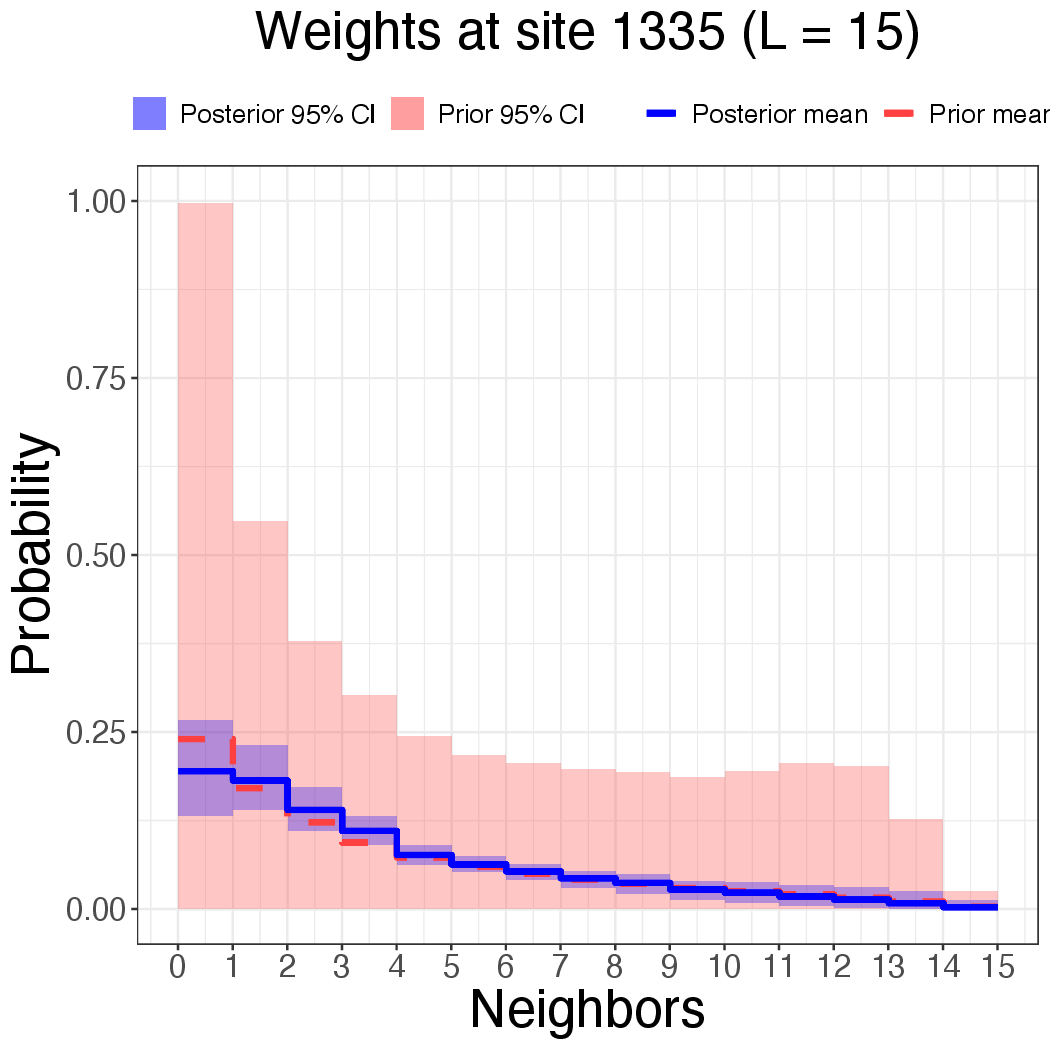}
    \includegraphics[width=.24\textwidth]{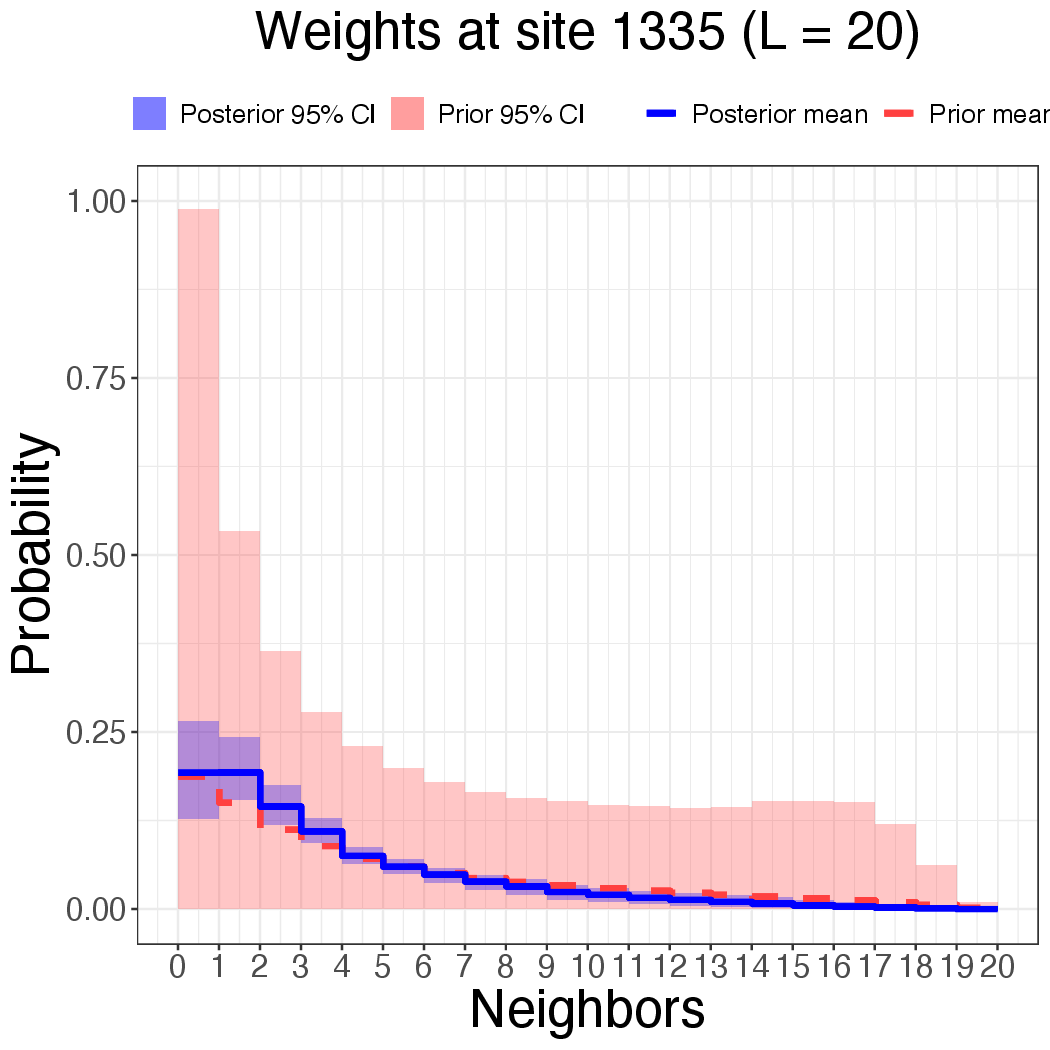}\\
    \bigskip
    \includegraphics[width=.24\textwidth]{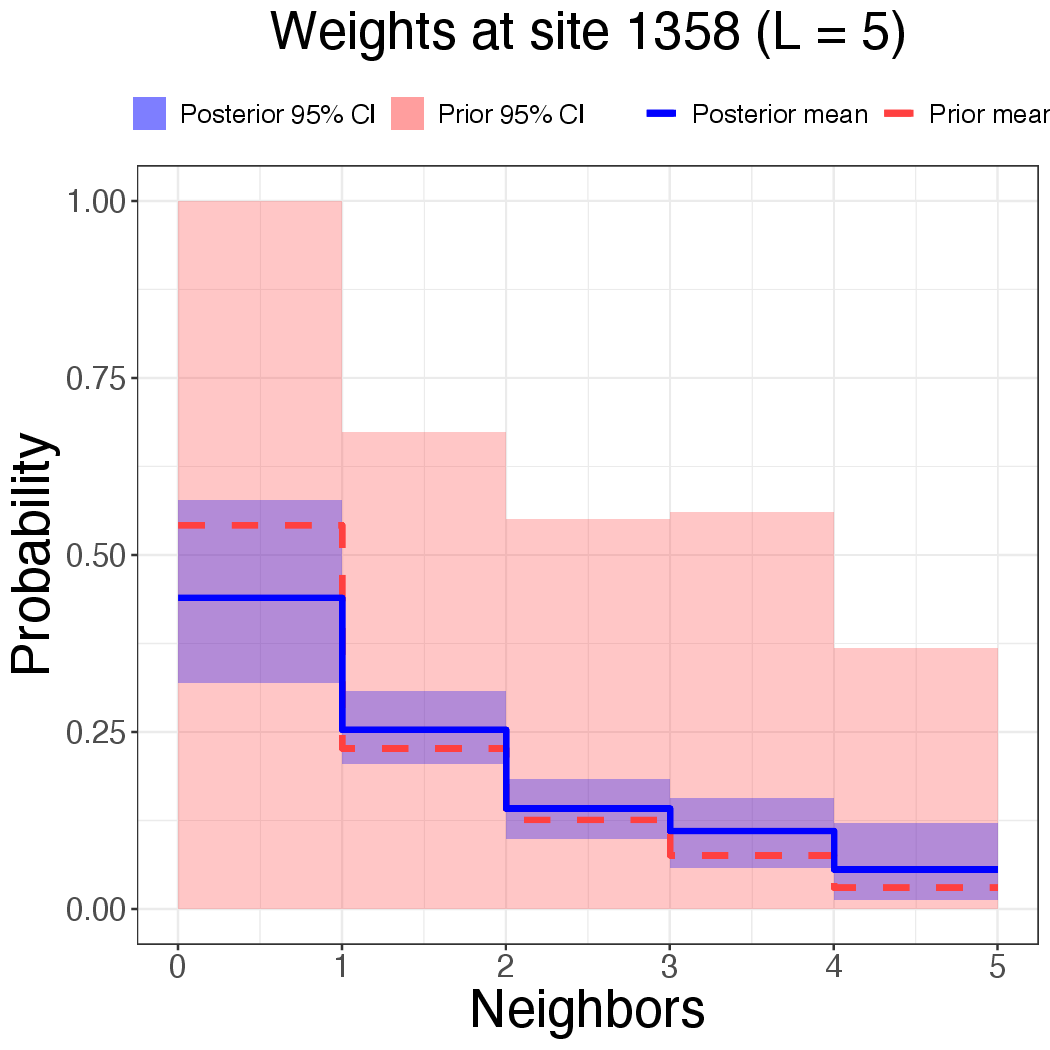}
    \includegraphics[width=.24\textwidth]{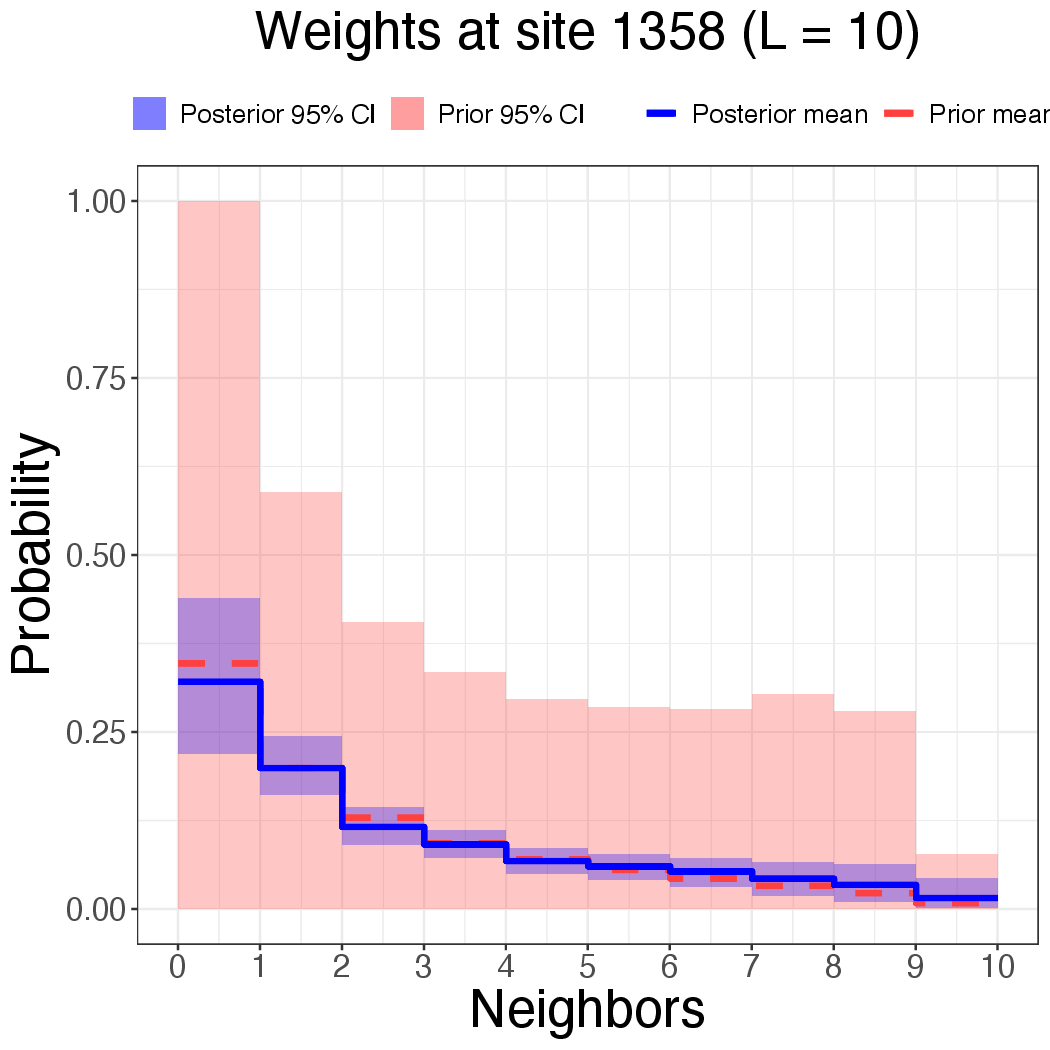}
    \includegraphics[width=.24\textwidth]{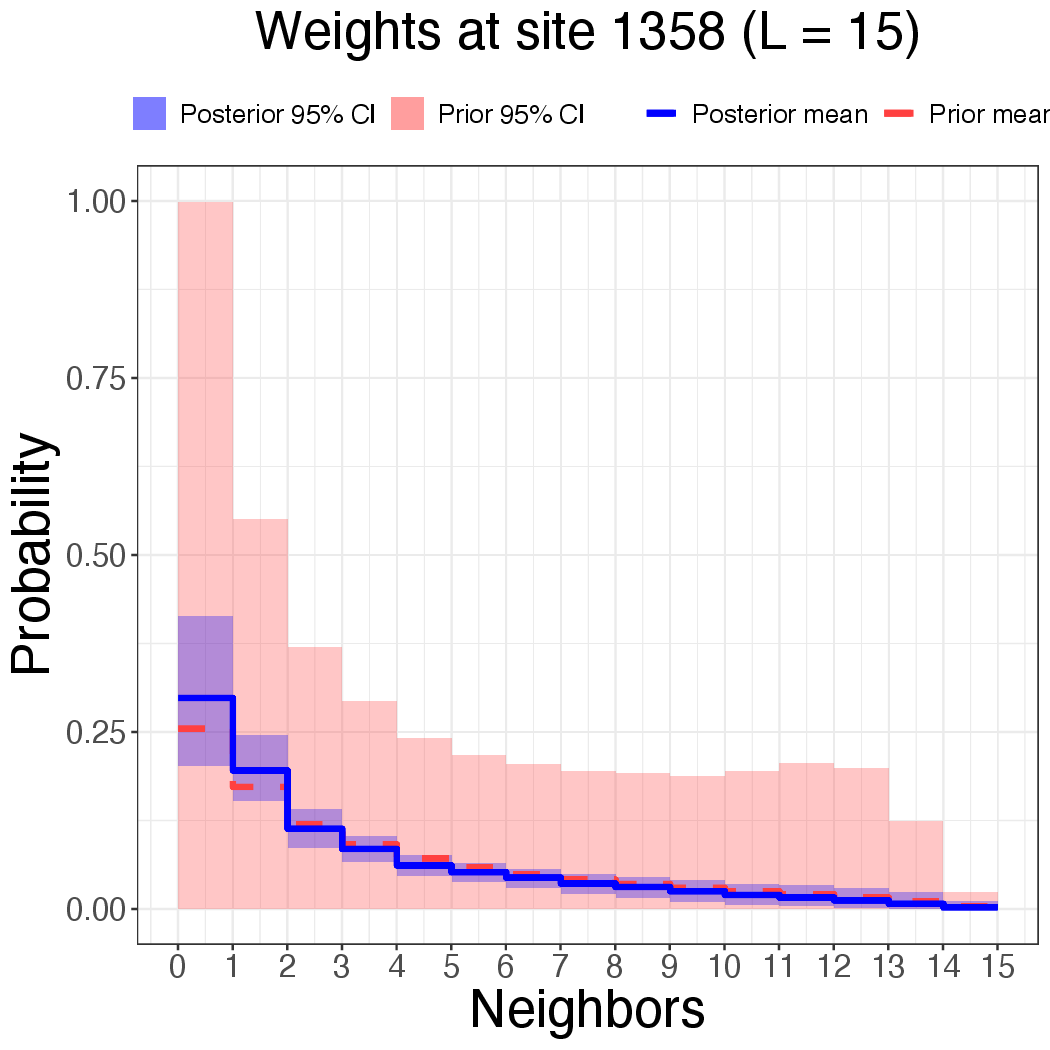}
    \includegraphics[width=.24\textwidth]{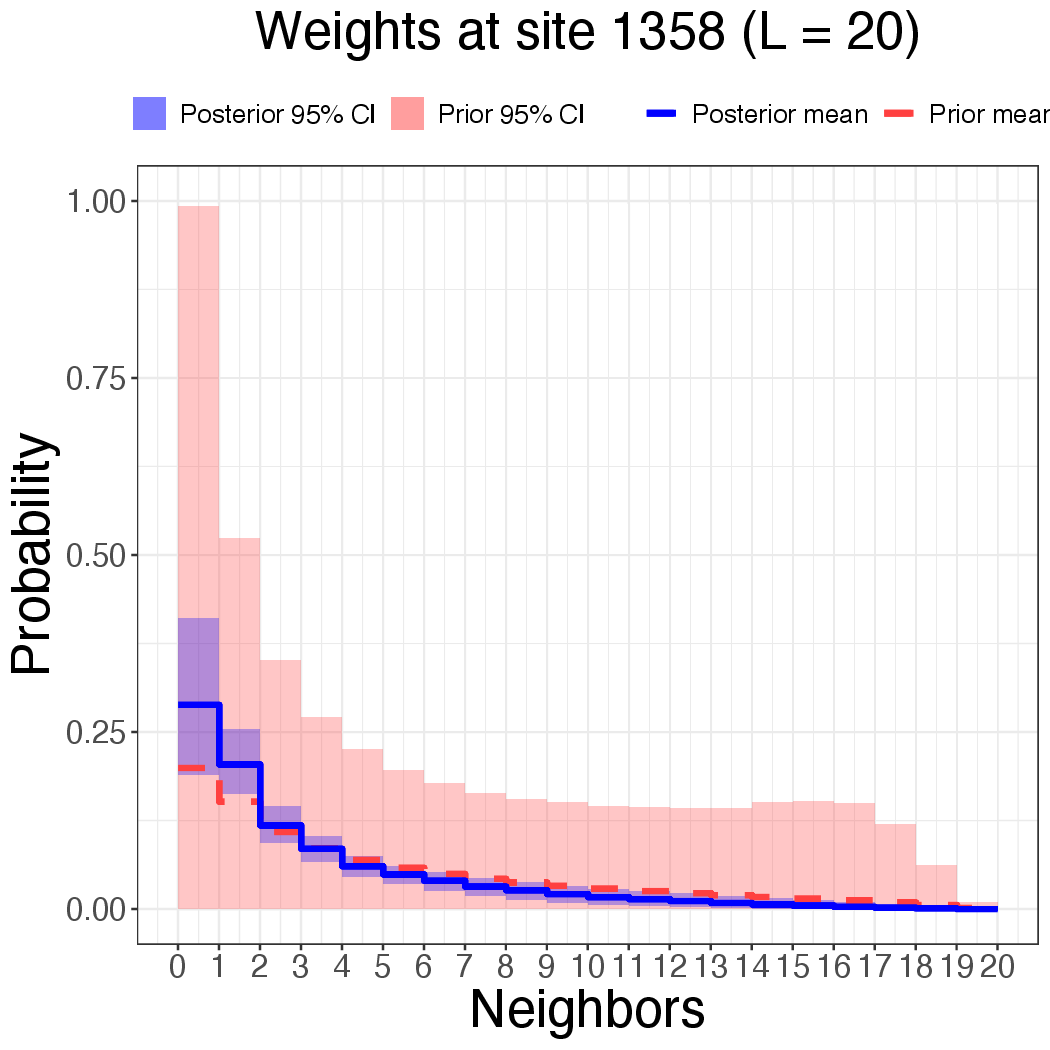}\\
    \bigskip
    \includegraphics[width=.24\textwidth]{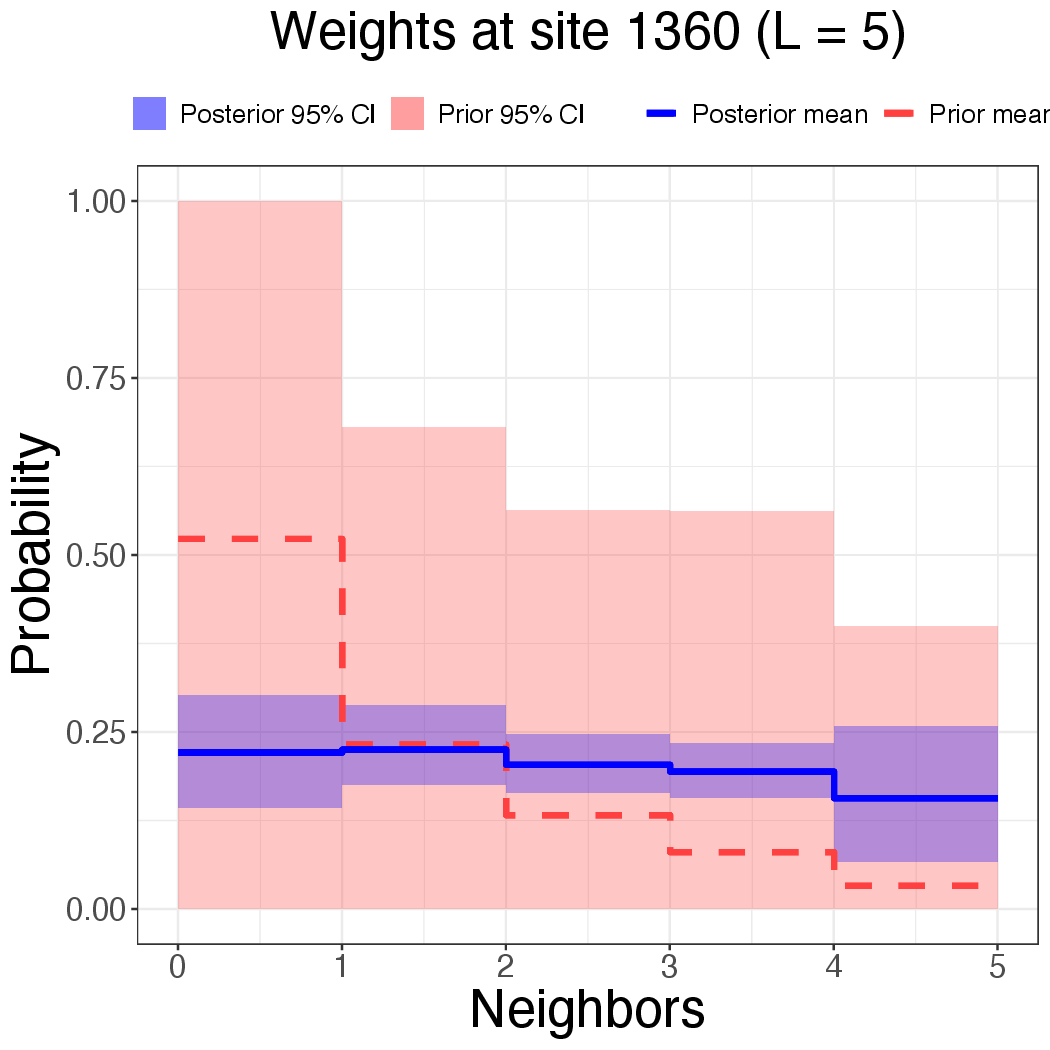}
    \includegraphics[width=.24\textwidth]{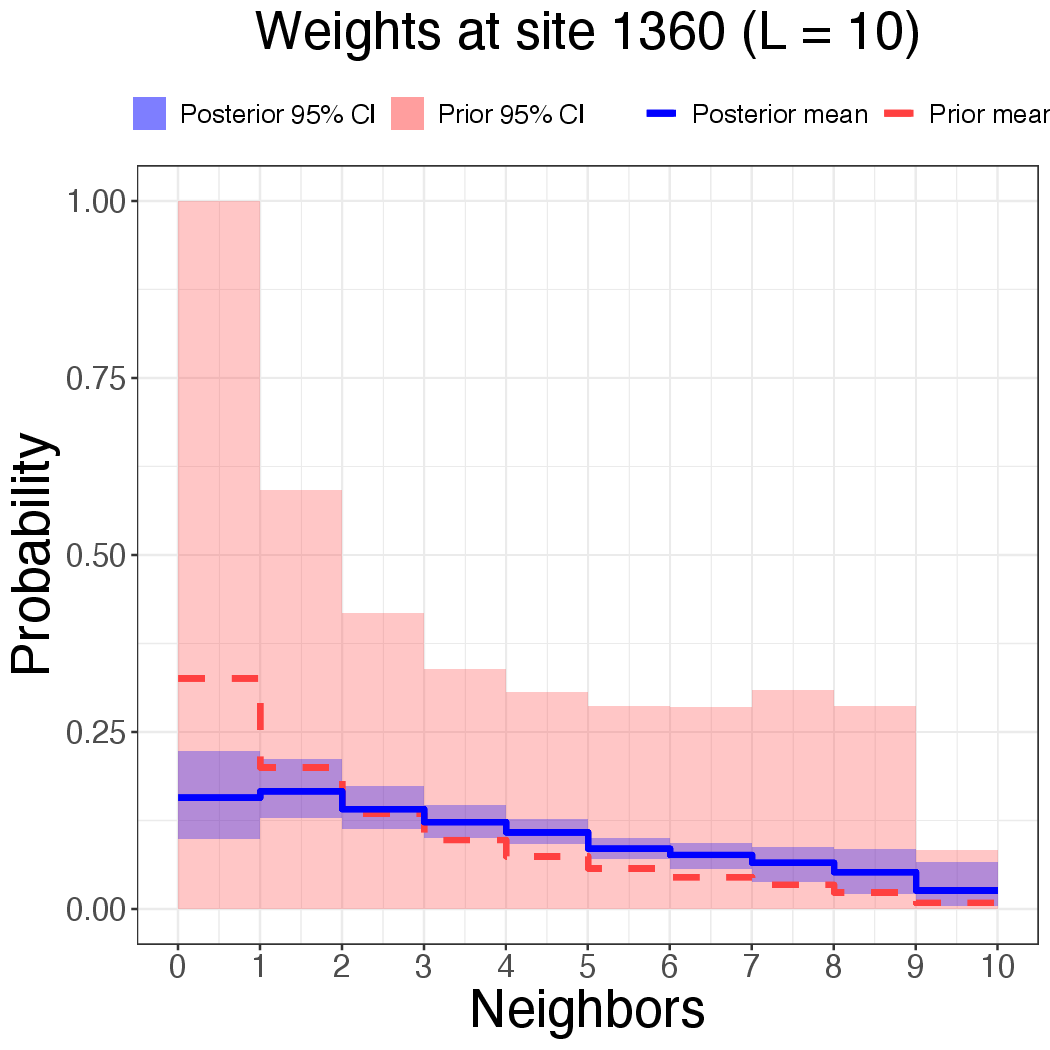}
    \includegraphics[width=.24\textwidth]{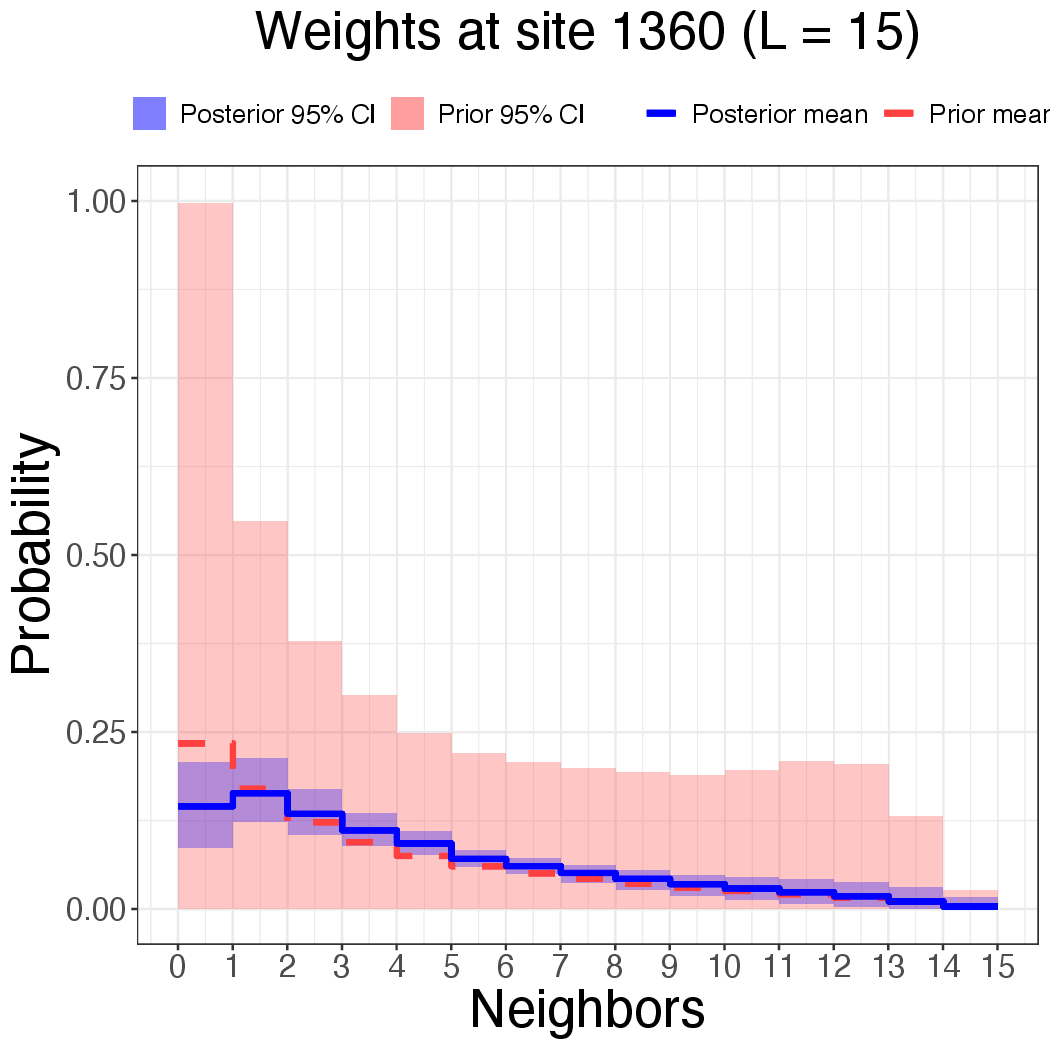}
    \includegraphics[width=.24\textwidth]{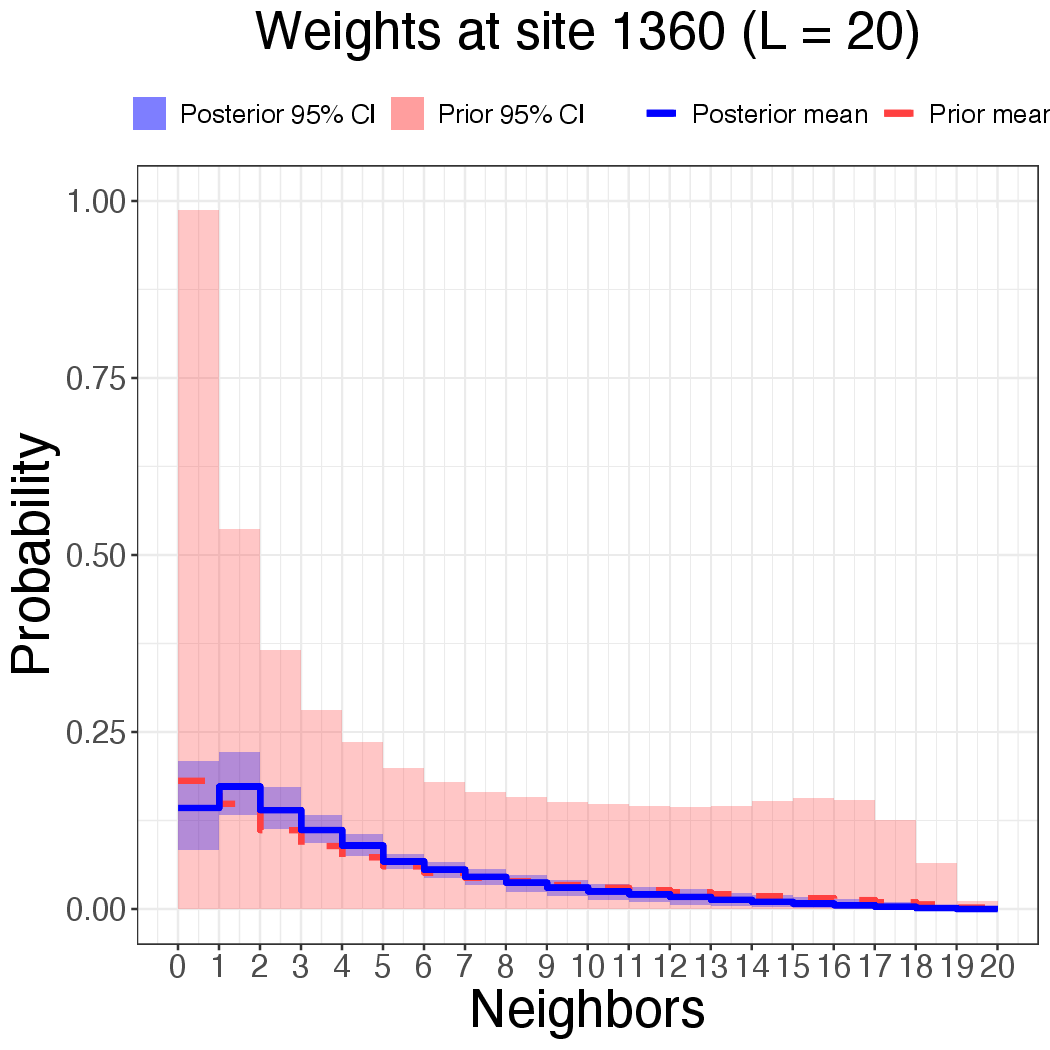}\\
    \bigskip
    \includegraphics[width=.24\textwidth]{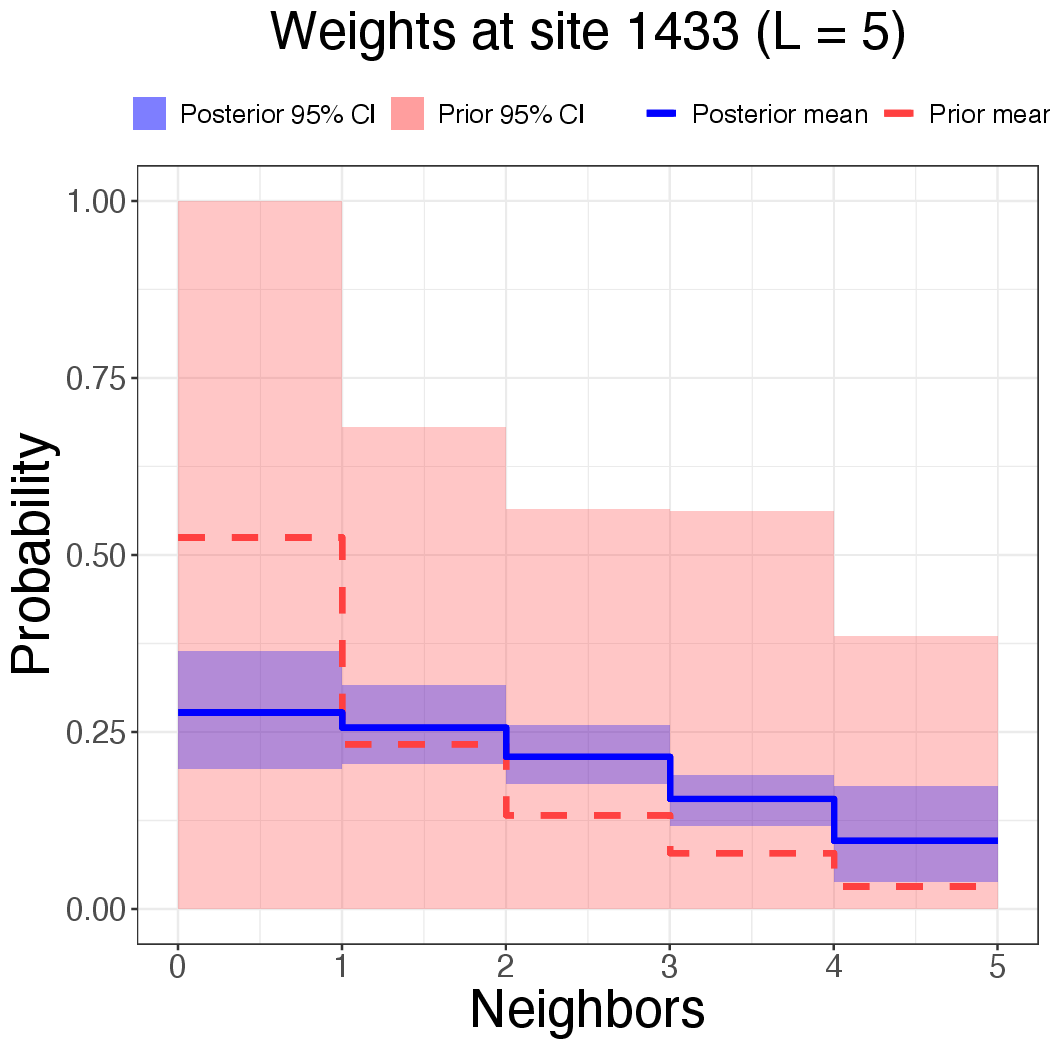}
    \includegraphics[width=.24\textwidth]{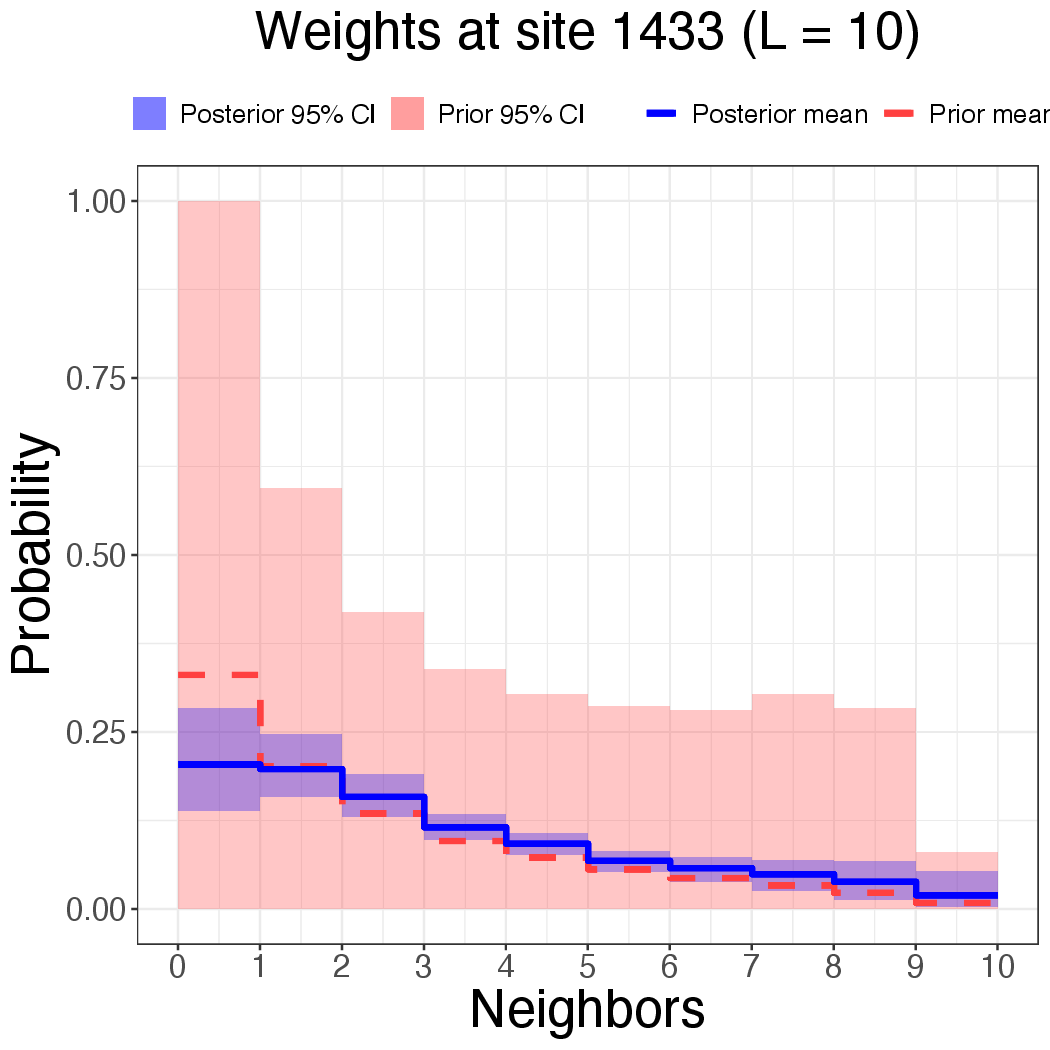}
    \includegraphics[width=.24\textwidth]{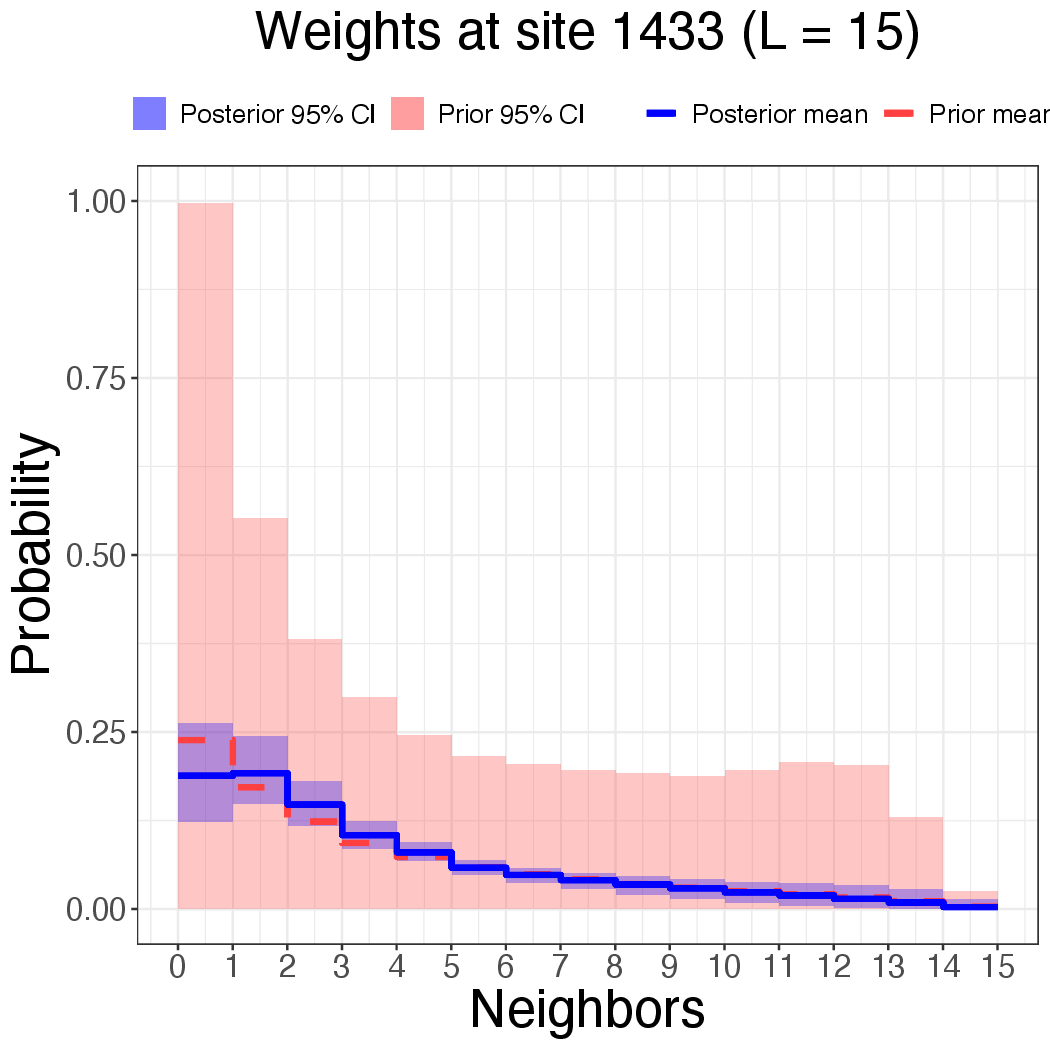}
    \includegraphics[width=.24\textwidth]{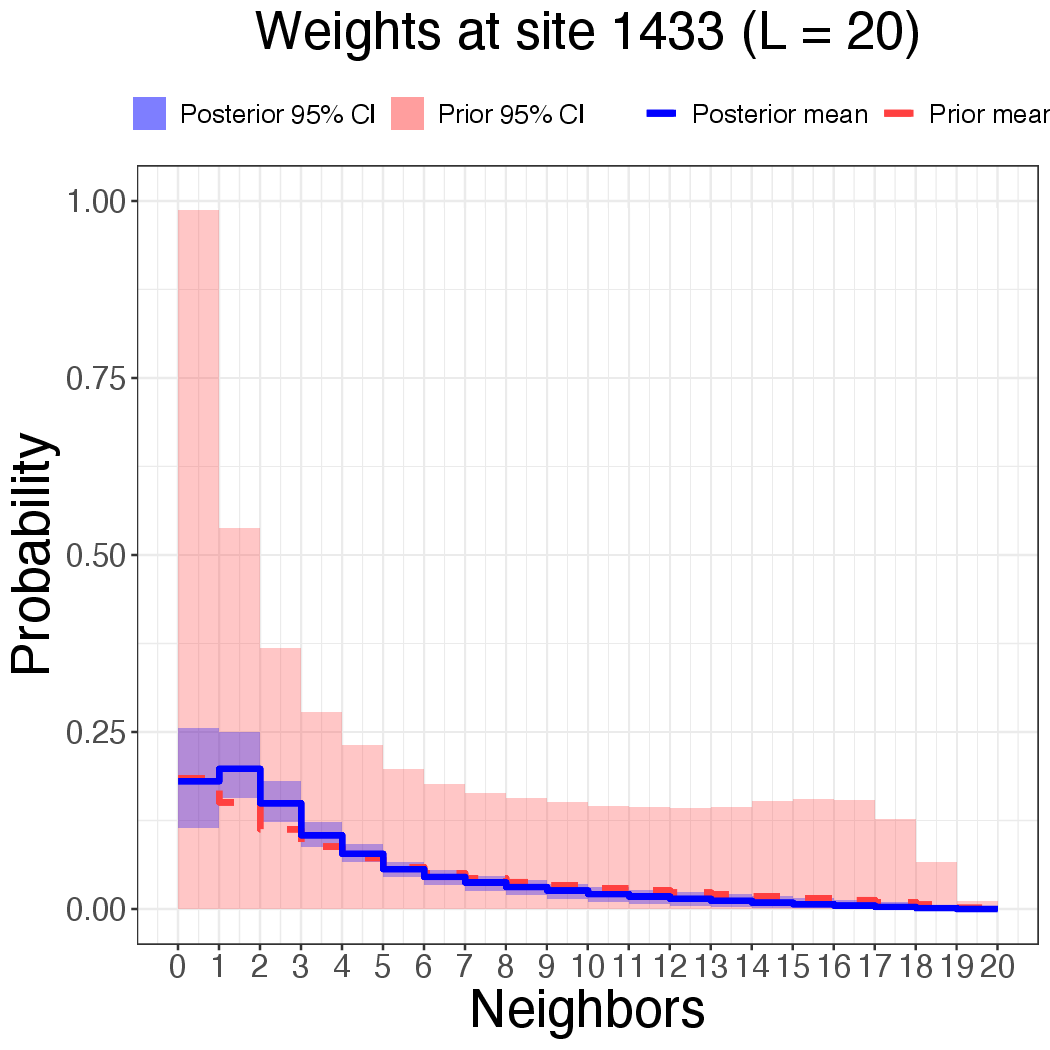}\\
    \bigskip
    \includegraphics[width=.24\textwidth]{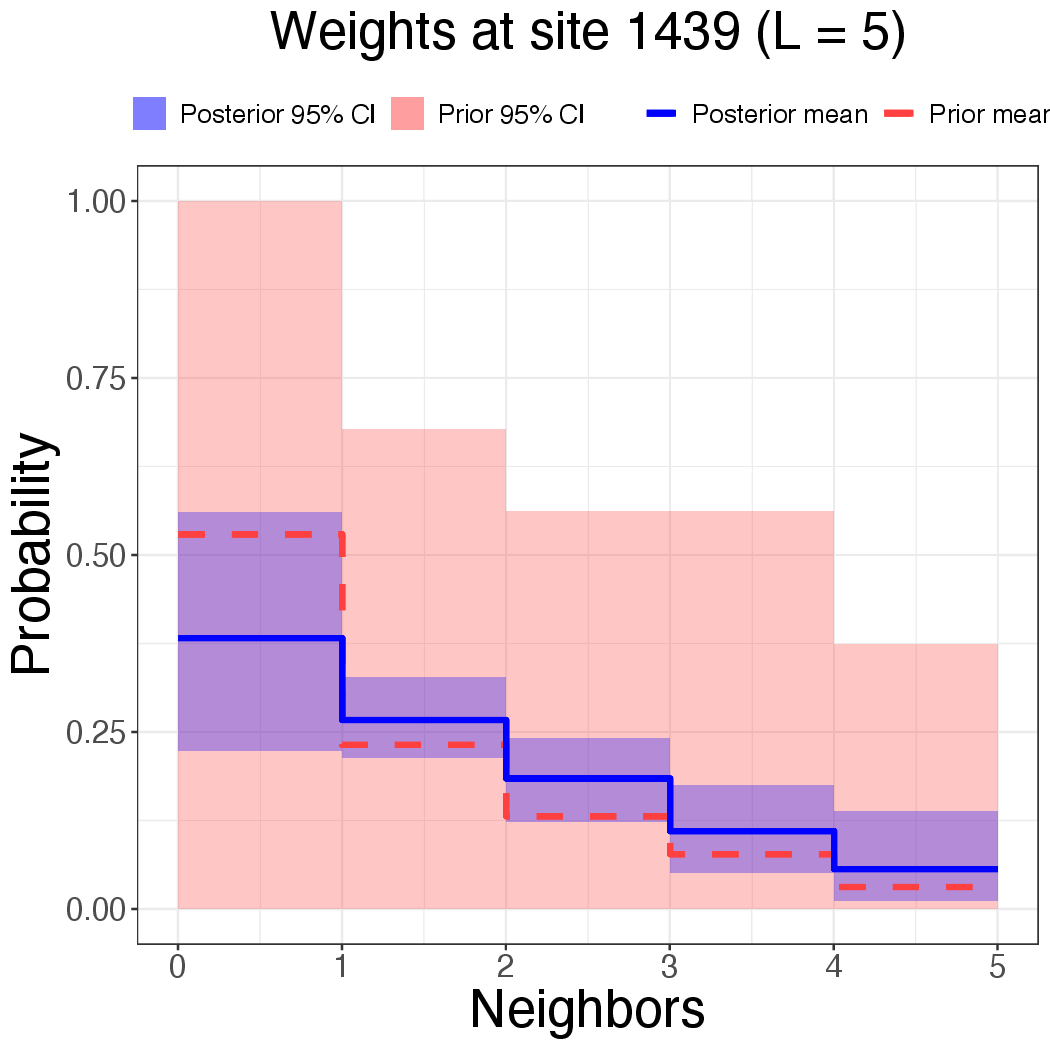}
    \includegraphics[width=.24\textwidth]{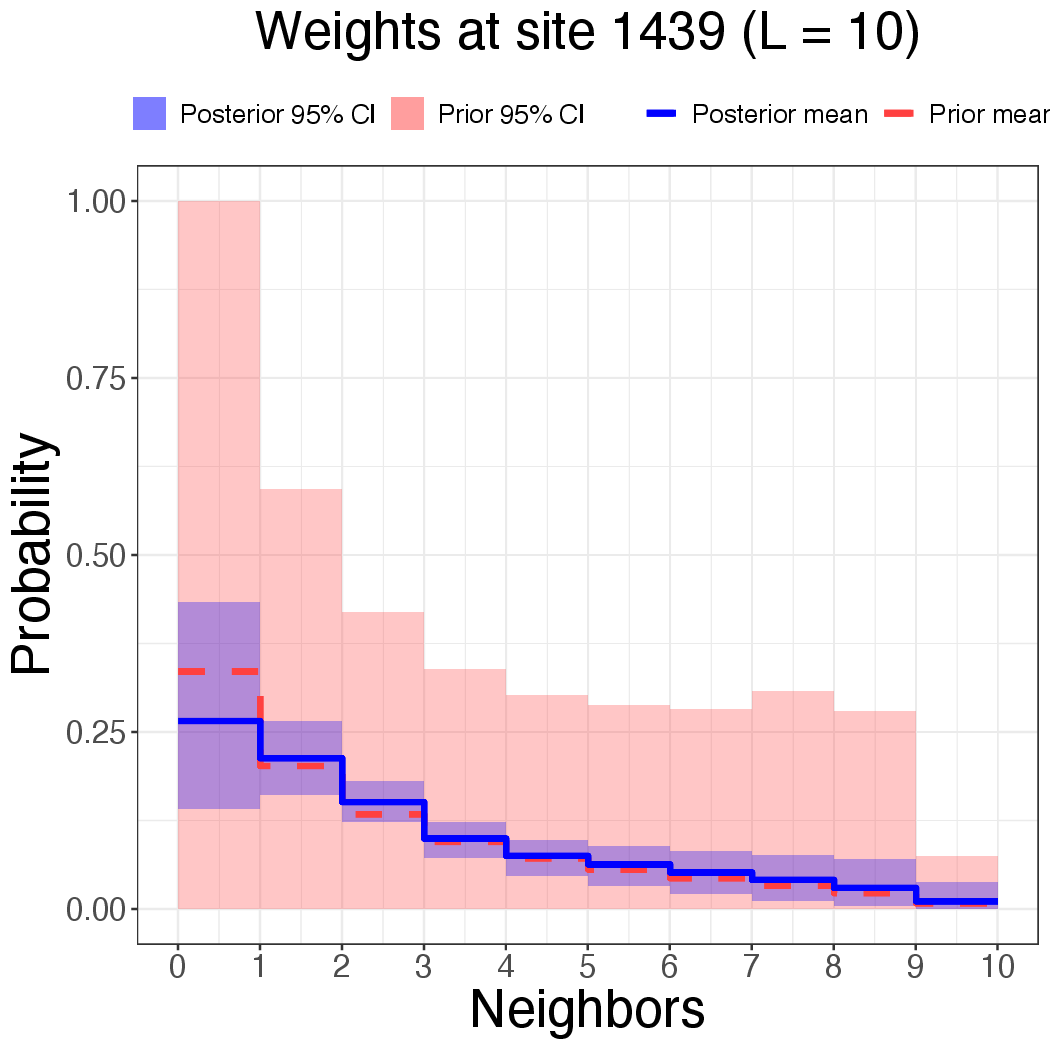}
    \includegraphics[width=.24\textwidth]{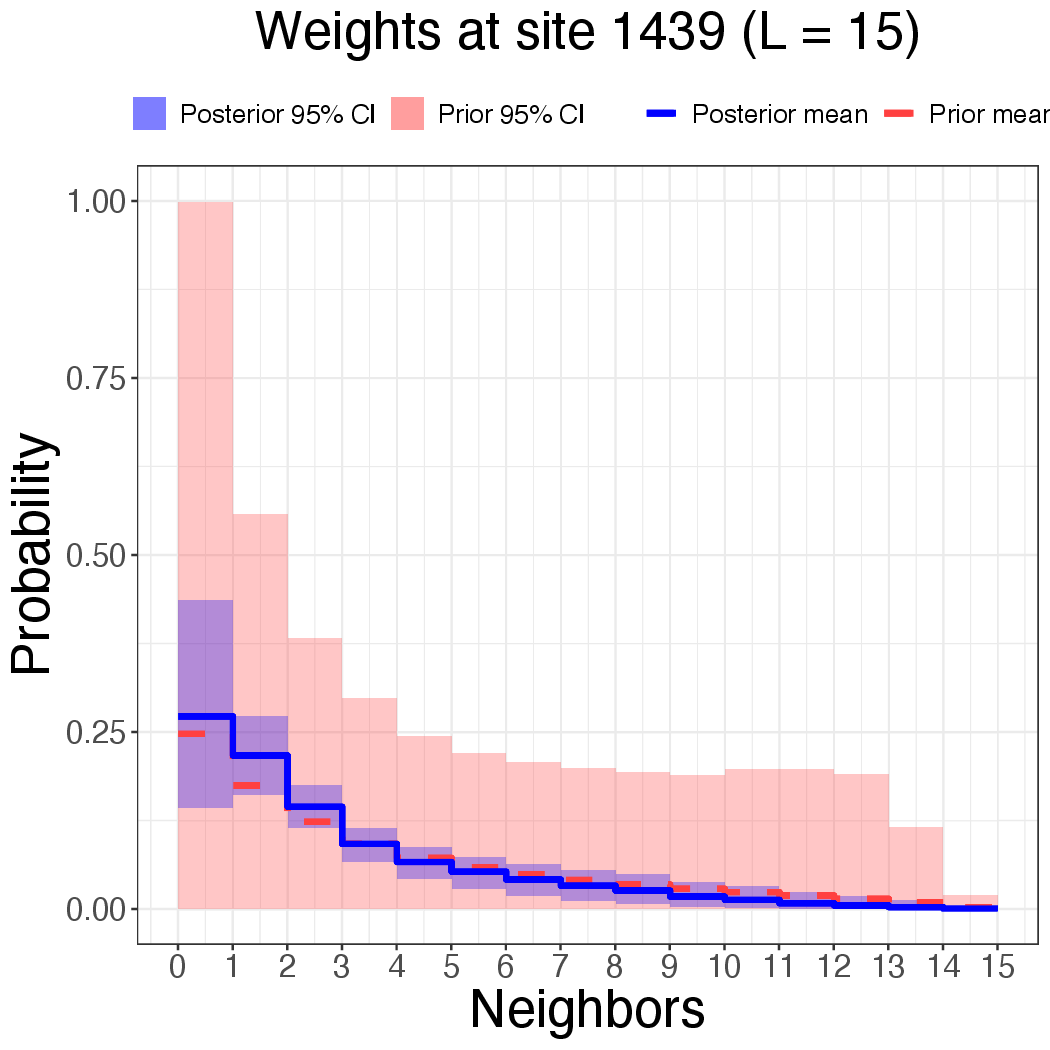}
    \includegraphics[width=.24\textwidth]{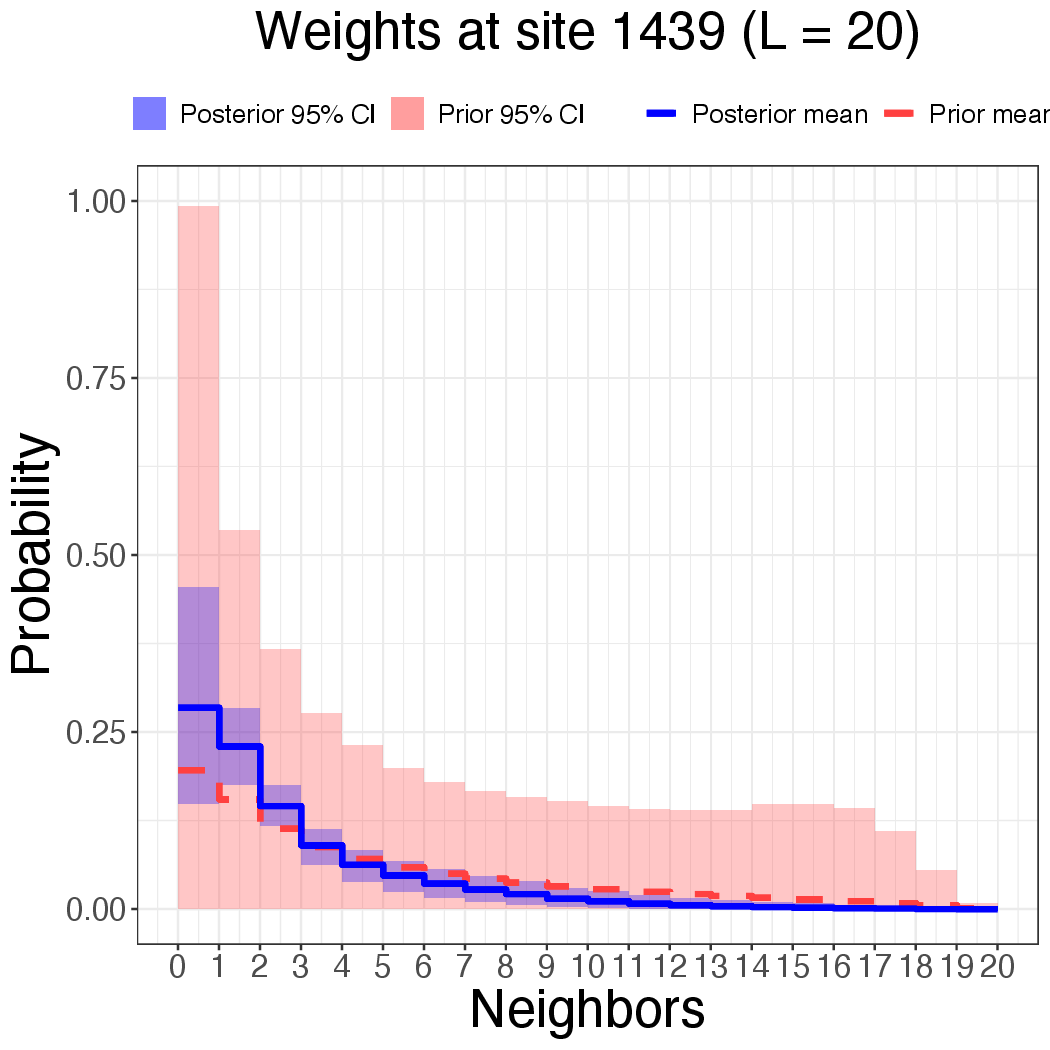}\\
    \bigskip
    \caption{BBS data analysis: Posterior means and 95\% CI estimates of the weights of the last five locations. }
    \label{fig:weights2}
\end{figure}

To examine the model performance on estimating the weights, 
we randomly selected ten locations $(\bs_{j_1},\dots,\bs_{j_{10}})$ such that 
$21\leq j_k\leq 200$ for $k = 1,\dots,5$ and $1312\leq j_k \leq 1512$ for $k = 6,\dots,10$.
Since we used random ordering to assign indices to the locations, 
the neighbors of $\bs_{j_k}$, $k = 1,\dots,5$, may consist of distant locations, 
whereas the neighbors of $\bs_{j_k}$, $k = 6, \dots, 10$, were expected to be all nearby.
Figures \ref{fig:weights1} and \ref{fig:weights2} illustrate the posterior means and 
95\% CI estimates of the weights at 
these ten locations. From the figures, we see that the model provided 
estimates of the weights that adjust to different neighborhood structures. 
The effective number of neighbors varied across locations. 
In addition, the estimates of the weights 
were quite robust as the value of $L$ increased. We can observe that the model
was able to penalize irrelevant neighbors by assigning very small probabilities.
While $L = 5$ seems too small to work as an upper bound, we observe that when 
$L$ ranged from  $10$ to $20$, the effective number of weights for each location 
was quite  consistent.

\begin{table}[t!]
\caption{BBS data analysis: posterior means and 95\% CI estimates for the parameters and computing time, 
under the Gaussian copula NBNNMP models with different values of $L$.}
    \centering
    \begin{tabular*}{\hsize}{@{\extracolsep{\fill}}lcccc}
    % \\[-5pt]
\hline
  & L = 5 & L = 10 & L = 15 & L = 20\\
\hline
$\beta_0$ & 6.52 (5.88, 7.33) & 6.56 (5.69, 7.22) & 6.48 (5.72, 7.28) & 6.48 (5.62, 7.29)\\
\hline
$\beta_1$ & -0.09 (-0.11, -0.07) & -0.09 (-0.11, -0.06) & -0.09 (-0.11, -0.07) & -0.09 (-0.11, -0.06)\\
\hline
$\phi$ & 1.61 (1.26, 2.04) & 2.51 (1.80, 3.47) & 2.65 (1.93, 3.59) & 2.62 (1.81, 3.68)\\
\hline
$\zeta$ & 0.82 (0.45, 1.82) & 1.10 (0.63, 2.15) & 1.37 (0.77, 2.70) & 1.71 (0.87, 3.80)\\
\hline
$r$ & 1.94 (1.65, 2.22) & 1.86 (1.51, 2.19) & 1.87 (1.54, 2.21) & 1.88 (1.53, 2.22)\\
\hline
$\gamma_0$ & -1.28 (-3.60, 0.96) & -1.29 (-3.49, 1.01) & -1.51 (-3.77, 0.66) & -1.69 (-3.85, 0.41)\\
\hline
$\gamma_1$ & 0.00 (-0.02, 0.03) & 0.00 (-0.02, 0.02) & 0.00 (-0.02, 0.02) & 0.00 (-0.02, 0.02)\\
\hline
$\gamma_2$ & 0.03 (-0.01, 0.08) & 0.02 (-0.02, 0.06) & 0.01 (-0.02, 0.06) & 0.01 (-0.02, 0.05)\\
\hline
$\kappa^2$ & 2.39 (1.48, 3.65) & 2.23 (1.46, 3.31) & 1.93 (1.24, 2.95) & 1.63 (1.09, 2.30)\\
\hline
Time (mins) & 29.17 & 32.71 & 38.49 & 50.91\\
\hline
    \end{tabular*}
    \label{tbl:real2}
\end{table}

Finally, a sensitivity analysis  was carried out to study 
the impact of $L$ on the model performance. We randomly split the data into two sets, 
a training set with 1212 observations and a testing set with 300 observations. 
We then applied the Gaussian copula NBNNMP with $L$ from 
$5$ to $20$, and evaluated the model performance based on out-of-sample predictive 
performance as shown in Table \ref{tbl:sa}. There were no discernible 
differences among the models with $L$ between 9 and 20. The conclusion from the
robustness analysis of the choice of $L$ is that  $L = 20$ 
works as a reasonable upper bound for this particular data example.

\begin{table}[t!]
    \caption{BBS data analysis: performance metrics of the Gaussian copula NBNNMP models 
             with different values of $L$.}
    \centering
    \begin{tabular*}{\hsize}{@{\extracolsep{\fill}}lcccccc}
    % \\[-5pt]
    \hline
     & RMSPE & 95\% CI & 95\% CI width & CRPS & ES & VS\\
    \hline
     L = 5 & 19.90 & 0.93 & 66.07 & 9.79 & 235.34 & 39759593\\
    \hline
     L = 6 & 19.82 & 0.94 & 65.91 & 9.75 & 234.50 & 39446330\\
    \hline
     L = 7 & 19.83 & 0.94 & 66.04 & 9.75 & 234.73 & 39464801\\
    \hline
     L = 8 & 19.80 & 0.94 & 66.19 & 9.75 & 234.36 & 39345232\\
    \hline
     L = 9 & 19.75 & 0.94 & 66.33 & 9.72 & 233.42 & 39073447\\
    \hline
     L = 10 & 19.72 & 0.94 & 66.27 & 9.72 & 233.50 & 39066501\\
    \hline
     L = 11 & 19.74 & 0.94 & 66.40 & 9.73 & 233.75 & 39179711\\
    \hline
     L = 12 & 19.73 & 0.95 & 66.67 & 9.70 & 233.10 & 38919544\\
    \hline
     L = 13 & 19.73 & 0.94 & 66.50 & 9.71 & 233.29 & 38978258\\
    \hline
     L = 14 & 19.70 & 0.95 & 66.69 & 9.71 & 233.20 & 38920854\\
    \hline
     L = 15 & 19.72 & 0.95 & 66.70 & 9.71 & 233.26 & 38865662\\
    \hline
     L = 16 & 19.73 & 0.94 & 66.70 & 9.72 & 233.50 & 38998533\\
    \hline
     L = 17 & 19.72 & 0.95 & 66.67 & 9.72 & 233.55 & 38982480\\
    \hline
     L = 18 & 19.72 & 0.94 & 66.80 & 9.72 & 233.63 & 39013058\\
    \hline
     L = 19 & 19.74 & 0.94 & 66.67 & 9.72 & 233.94 & 39111633\\
    \hline
     L = 20 & 19.79 & 0.94 & 66.75 & 9.74 & 234.30 & 39194713\\
    \hline
    \end{tabular*}
    \label{tbl:sa}
\end{table}

\subsubsection{Comparison of three copula NBNNMP models}

We compare three discrete copula NBNNMP models with $L = 20$. Each model used
either the spatial Gaussian, Gumbel or Clayton copulas, with negative binomial marginals
$\mathrm{NB}(\exp(\x(\sv)^\top\bbeta),r)$.
We used the same link functions and prior
specifications for copulas as in Section 5.1 of the main paper and the same priors
for other parameters as in Section 5.2 of the main paper.
We fitted the models to 1215 randomly selected observations and used the remaining 
300 for model comparison. 
For each model, we ran the MCMC algorithm for 30000 iterations, 
discarding the first 10000 iterations, and collected posterior samples every 5th iteration.
Table \ref{tbl:real} shows the comparison based on out-of-sample predictive performance.
Overall, the Gaussian copula outperformed the other two.

\begin{table}[t!]
    \caption{BBS data analysis: performance metrics for NBNNMPs based on different copulas.}
    \centering
    \begin{tabular*}{\hsize}{@{\extracolsep{\fill}}lcccccc}
    % \\[-5pt]
\hline
    & RMSPE & 95\%CI cover & 95\%CI width & CRPS & ES & VS\\
\hline
Gaussian & 19.75 & 0.94 & 66.62 & 9.72 & 233.91 & 39136486\\
\hline
Gumbel & 19.71 & 0.96 & 68.77 & 9.81 & 236.18 & 39665090\\
\hline
Clayton & 19.97 & 0.93 & 71.51 & 9.91 & 237.21 & 39566563\\
\hline
    \end{tabular*}
    \label{tbl:real}
\end{table}

\subsubsection{Comparison with the SGLMM method}

We also assessed the model performance by comparison with
the SGLMM-GP model. Again, we randomly split the data into a training set with 1212 observations and a testing set with 300 observations. We ran the MCMC algorithm for 
the Gaussian copula NBNNMP ($L=20)$ for 30000 iterations, discarding the first 
10000 iterations, and collecting posterior 
samples every 5th iteration. Since the MCMC for SGLMM-GP involves sampling 
the spatial random effects, we ran the algorithm for 50000 iterations and collected posterior samples every five iterations, with the first 30000 as burn-in.

Table \ref{tbl:bbs_model_comp} shows the parameter estimates and predictive performances
by the two models. The regression coefficient estimates were quite similar.
Both models indicate an increasing trend in the counts as the latitude decreases.
Regarding the out-of-sample predictive performance, the
NBNNMP model performed uniformly better than the SGLMM-GP, with a huge
gain in computing time. 

\begin{table}[b!]
    \caption{BBS data analysis: parameter estimates and performance metrics of the Gaussian copula NBNNMP and the SGLMM-GP models.}
    \centering
    \begin{tabular*}{\hsize}{@{\extracolsep{\fill}}lcc}
    % \\[-5pt]
\hline
  & NBNNMP & SGLMM-GP\\
\hline
$\beta_0$ & 6.57 (5.83, 7.19) & 6.67 (6.55, 6.81)\\
\hline
$\beta_1$ & -0.09 (-0.10, -0.07) & -0.10 (-0.10, -0.09)\\
\hline
RMSPE & 19.79 & 20.41\\
\hline
95\% CI & 0.94 & 0.94\\
\hline
95\% CI width & 66.56 & 76.56\\
\hline
CRPS & 9.74 & 10.10\\
\hline
ES & 234.22 & 239.02\\
\hline
VS & 39204378.76 & 40185343.15\\
\hline
Time (mins) & 37.56 & 4208.33\\
\hline
    \end{tabular*}
    \label{tbl:bbs_model_comp}
\end{table}

\subsection{Randomized quantile residual analysis}

\begin{figure}[t!]
    \centering
    \includegraphics[width=\textwidth]{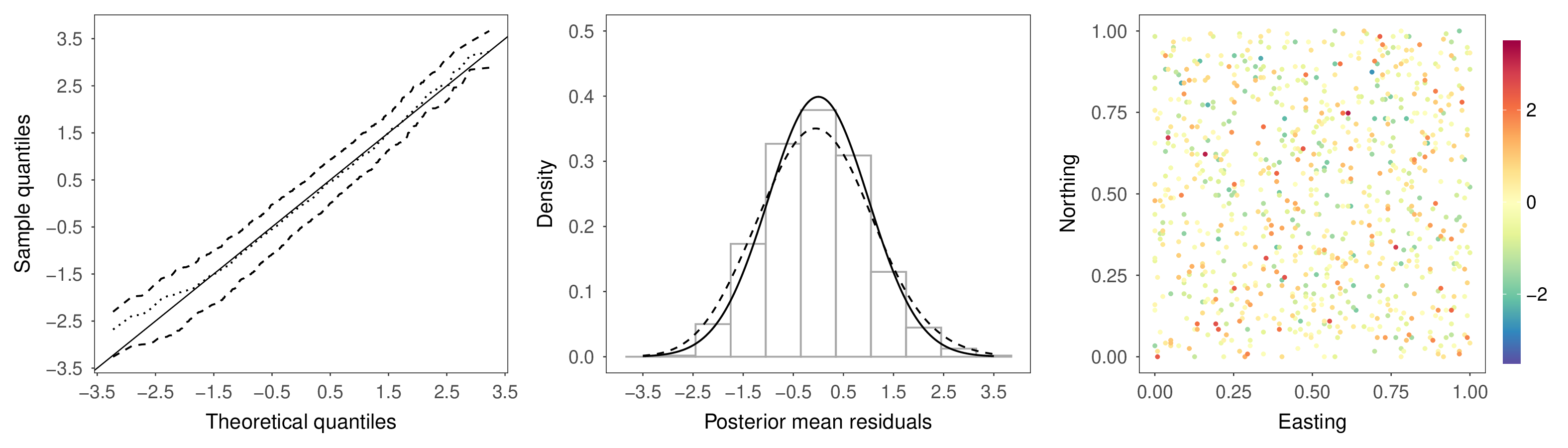}
    \medskip
    \includegraphics[width=\textwidth]{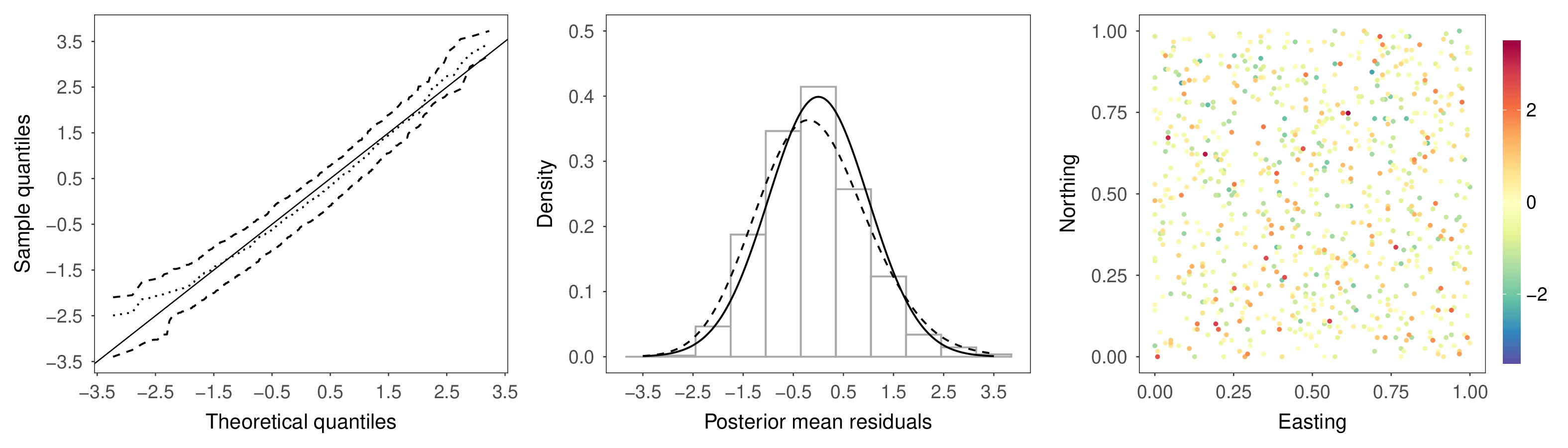}
    \medskip
    \includegraphics[width=\textwidth]{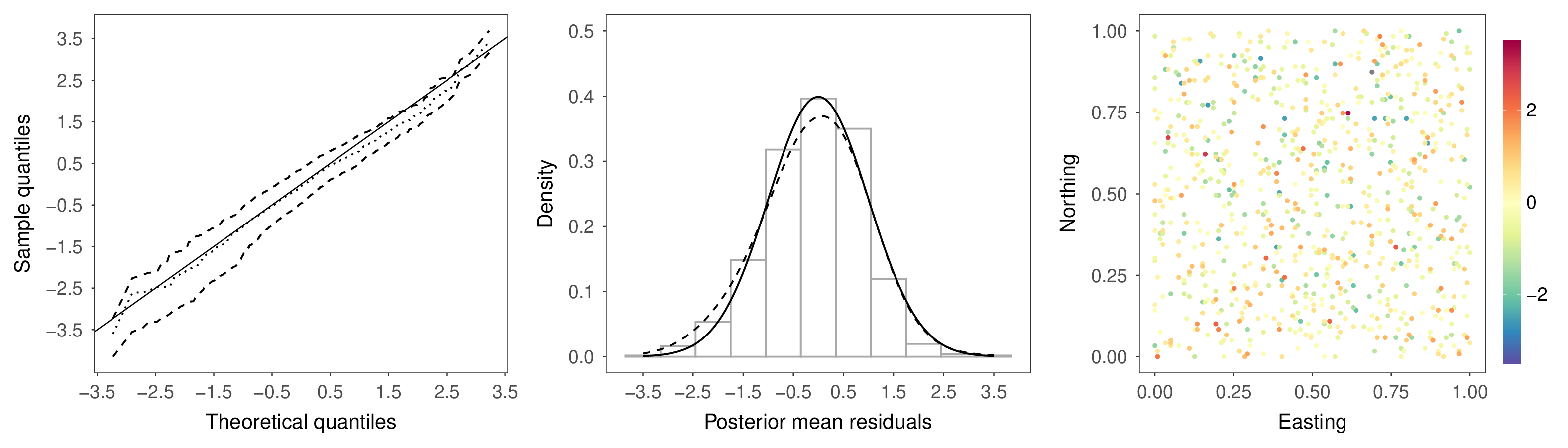}
    \caption{
    Simulated data example 1 - randomized quantile residual analysis for Scenario 1 ($\sigma_1 = 1)$.
    Left column: Gaussian quantile-quantile plots. 
    Dotted and dashed lines correspond to the posterior mean and 95\% interval bands, respectively.
    Middle column: Histograms of the posterior means of the residuals.
    Solid and dashed lines are the standard Gaussian density and the kernel density estimate of the
    posterior means of the residuals, respectively. 
    Right column: spatial plots of the posterior means of the residuals.
    Rows from top to bottom correspond to the Gaussian, Gumbel, and Clayton models, respectively.
    }
    \label{fig:sim1-rqr1}
\end{figure}

\begin{figure}[t!]
    \centering
    \includegraphics[width=\textwidth]{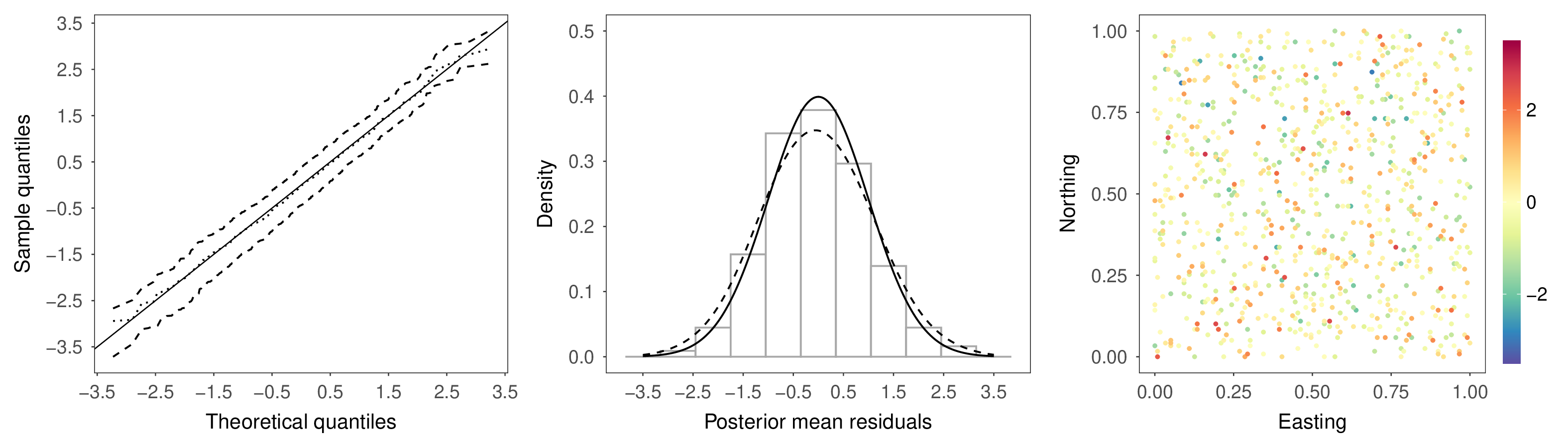}
    \medskip
    \includegraphics[width=\textwidth]{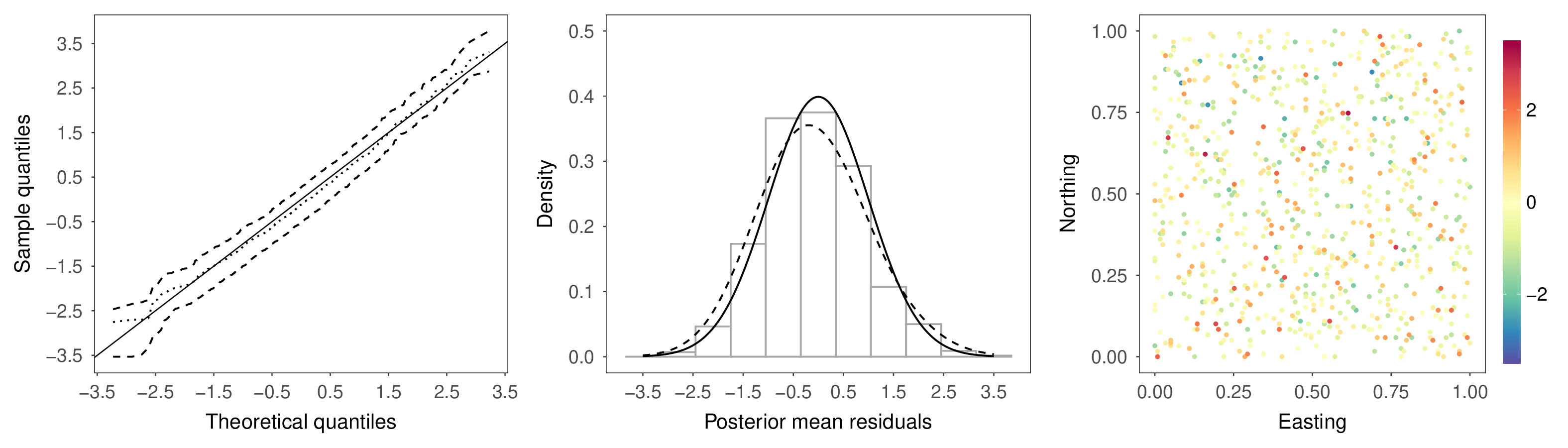}
    \medskip
    \includegraphics[width=\textwidth]{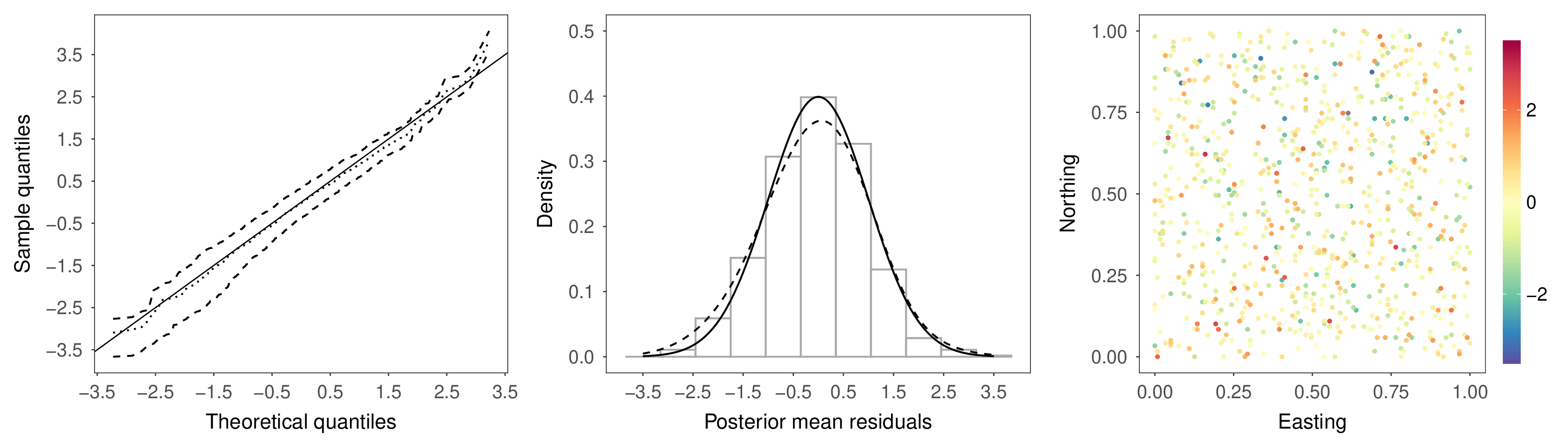}
    \caption{
    Simulated data example 1 - randomized quantile residual analysis for Scenario 2 ($\sigma_1 = 3)$.
    Left column: Gaussian quantile-quantile plots. 
    Dotted and dashed lines correspond to the posterior mean and 95\% interval bands, respectively.
    Middle column: Histograms of the posterior means of the residuals.
    Solid and dashed lines are the standard Gaussian density and the kernel density estimate of the
    posterior means of the residuals, respectively. 
    Right column: spatial plots of the posterior means of the residuals.
    Rows from top to bottom correspond to the Gaussian, Gumbel, and Clayton models, respectively.
    }
    \label{fig:sim1-rqr2}
\end{figure}

\begin{figure}[t!]
    \centering
    \includegraphics[width=\textwidth]{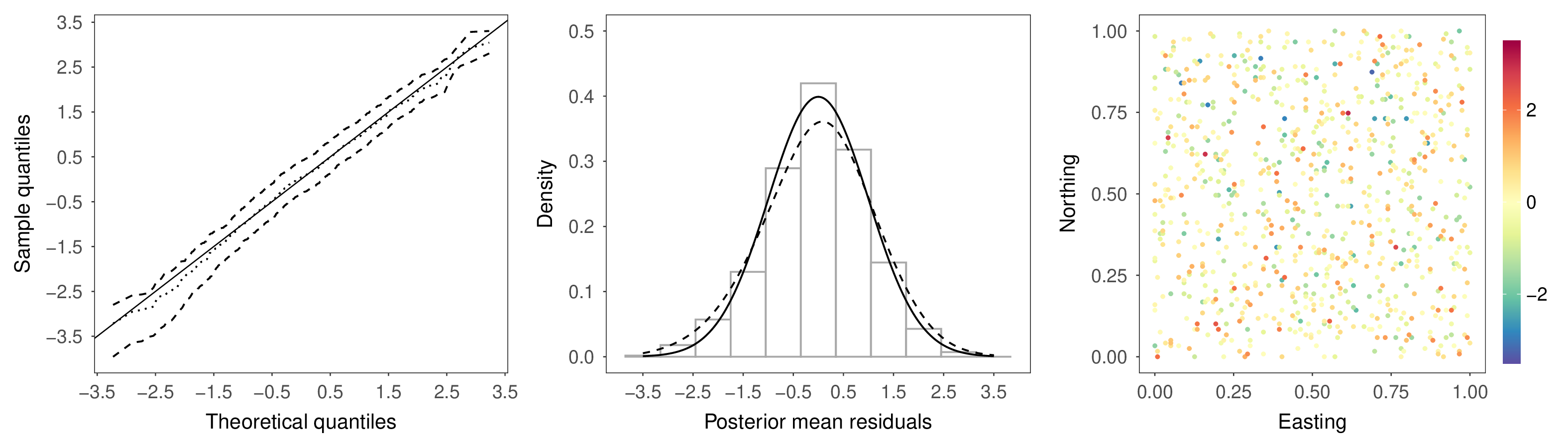}
    \medskip
    \includegraphics[width=\textwidth]{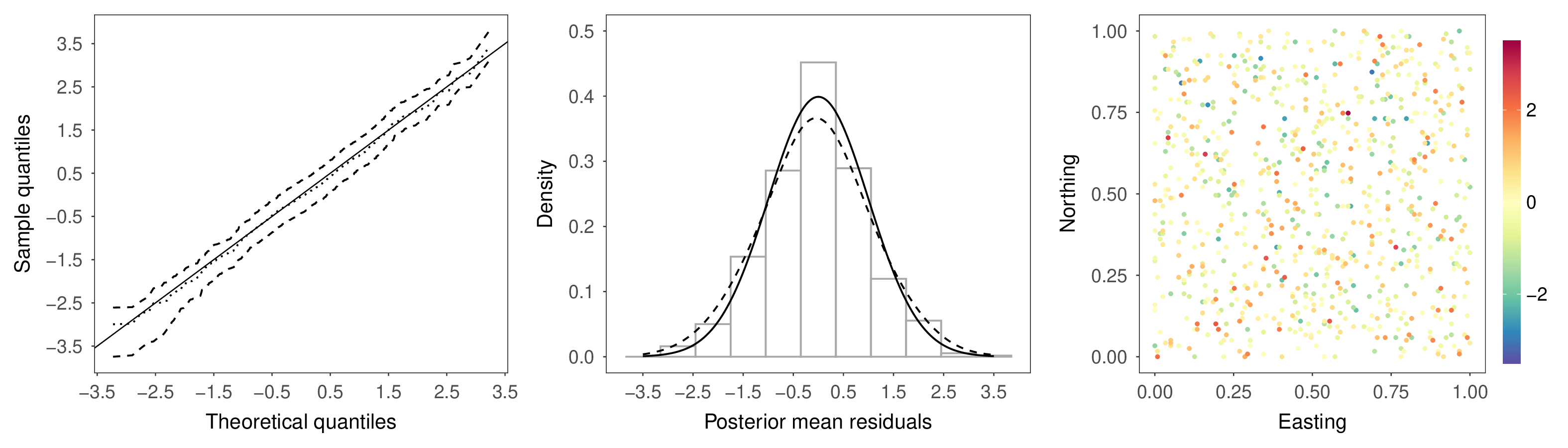}
    \medskip
    \includegraphics[width=\textwidth]{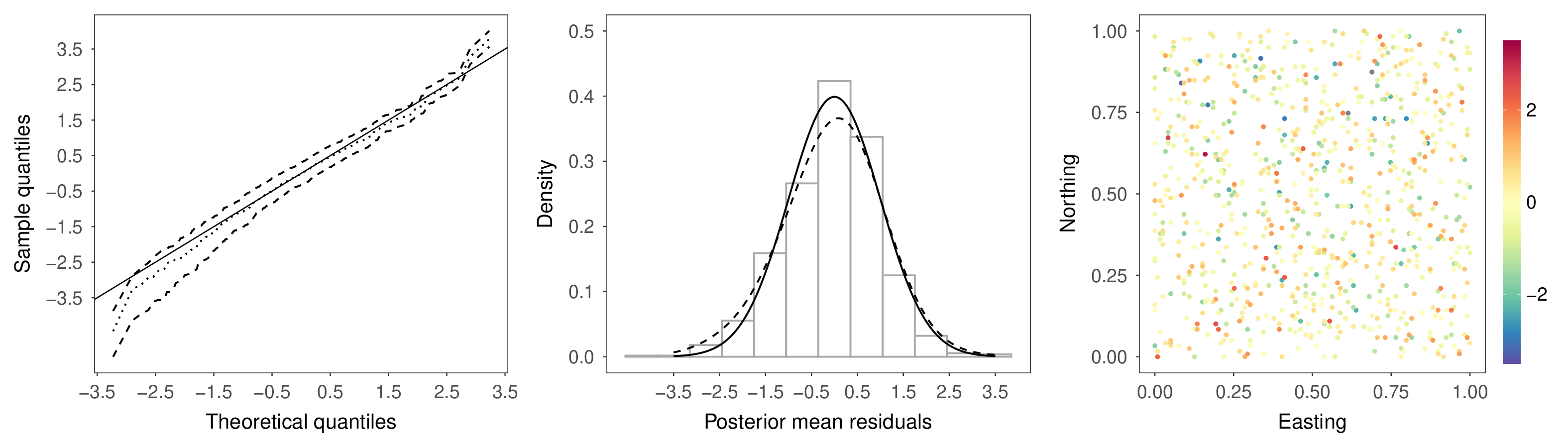}
    \caption{
    Simulated data example 1 - randomized quantile residual analysis for Scenario 3 ($\sigma_1 = 10)$.
    Left column: Gaussian quantile-quantile plots. 
    Dotted and dashed lines correspond to the posterior mean and 95\% interval bands, respectively.
    Middle column: Histograms of the posterior means of the residuals.
    Solid and dashed lines are the standard Gaussian density and the kernel density estimate of the
    posterior means of the residuals, respectively. 
    Right column: spatial plots of the posterior means of the residuals.
    Rows from top to bottom correspond to the Gaussian, Gumbel, and Clayton models, respectively.
    }
    \label{fig:sim1-rqr3}
\end{figure}

\begin{figure}[t!]
    \centering
    \includegraphics[width=\textwidth]{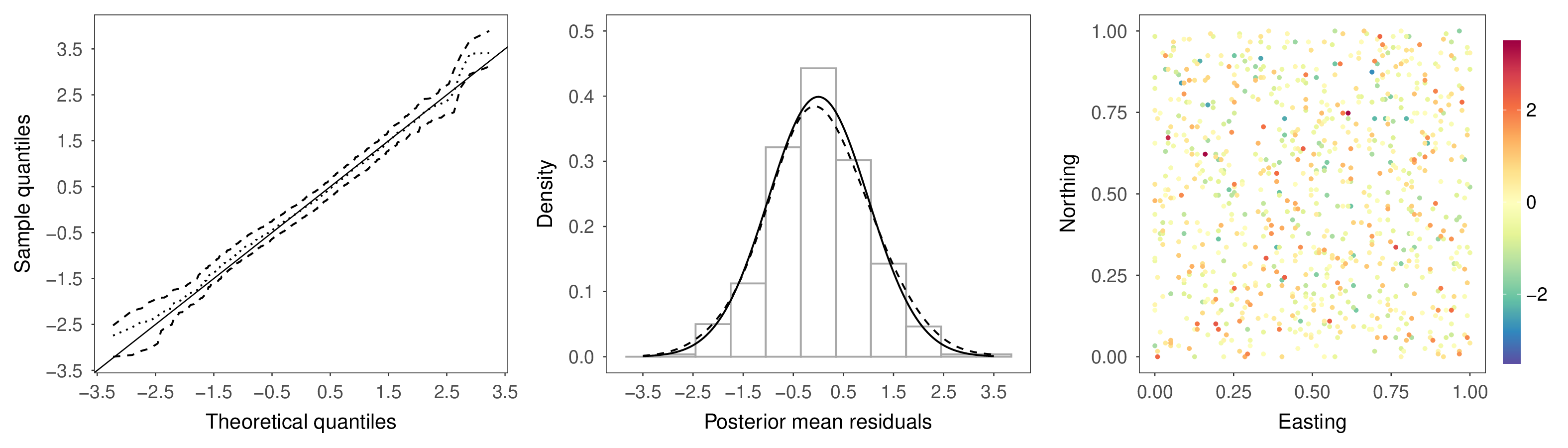}
    \caption{
    Simulated data example 2 - randomized quantile residual analysis for the NBNNMP model.
    Left panel: Gaussian quantile-quantile plot. 
    Dotted and dashed lines correspond to the posterior mean and 95\% interval bands, respectively.
    Middle panel: Histogram of the posterior means of the residuals.
    Solid and dashed lines are the standard Gaussian density and the kernel density estimate of the
    posterior means of the residuals, respectively. 
    Right panel: spatial plot of the posterior means of the residuals.
    }
    \label{fig:sim2-rqr}
\end{figure}

Model checking results using randomized quantile residuals 
for simulation examples 1 and 2 are illustrated in
Figures \ref{fig:sim1-rqr1}, \ref{fig:sim1-rqr2}, \ref{fig:sim1-rqr3}, and Figure \ref{fig:sim2-rqr},
respectively. For simulation example 1, each figure corresponds to a scenario and contains
posterior summary of the Gaussian quantile-quantile plot, the histogram and spatial plot of the 
posterior means of the residuals. We can see that in all cases, the results indicate good model fits.

\section*{Additional References}

Joe, H. (2014). \textit{Dependence modeling with copulas}. Boca Raton, FL: CRC press.

\end{document}